\documentclass[a4paper,11pt]{article}

\usepackage{jheppub}
\usepackage{amsfonts}
\usepackage{dsfont}
\usepackage{amsmath,amssymb}

\usepackage{bbm}
\usepackage{bm}
\usepackage{color}
\usepackage{slashed}
\usepackage{siunitx}
\DeclareSIUnit \kpc {kpc}
\usepackage{booktabs}
\usepackage{physics}
\usepackage{array}
\usepackage{subcaption}
\usepackage[utf8]{inputenc}
\usepackage{graphicx}
\usepackage{bbold}
\usepackage{bm}
\usepackage{todonotes}
\usepackage{enumitem}
\DeclareGraphicsExtensions{.pdf,.png,.jpg}

\newcommand*{\cf}{cf.\ }
\newcommand*{\eg}{e.\,g.\ }
\newcommand*{\ie}{i.\,e.\ }
\newcommand*{\Eq}{eq.\,}
\newcommand*{\Eqs}{eqs.\,}

\newcommand*{\evalat}[2]{\left. #1\right|_{#2}}
\renewcommand*{\vec}[1]{\bm{#1}}

\newcommand*{\sigmaann}{\sigma_{\text{ann}}}
\newcommand*{\sigmaanns}{\sigma_{\text{ann}}^{\text{s-wave}}}
\newcommand*{\sigmaannp}{\sigma_{\text{ann}}^{\text{p-wave}}}
\newcommand*{\sigmaBSF}{\sigma_{\text{\tiny BSF}}}
\newcommand*{\gammaBSD}{\Gamma_{\text{\tiny BSD}}}
\newcommand*{\Sann}{S_{\text{ann}}}
\newcommand*{\SBSF}{S_{\text{\tiny BSF}}}

\newcommand*{\vrel}{v_{\text{rel}}}
\newcommand*{\vcm}{v_{\text{cm}}}
\newcommand*{\vrelt}{\tilde{v}_{\text{rel}}}
\newcommand*{\vcmt}{\tilde{v}_{\text{cm}}}
\newcommand*{\rhochit}{\tilde{\rho}_X}
\newcommand*{\psit}{\tilde{\psi}}
\newcommand*{\fchit}{\tilde{f}_X}
\newcommand*{\et}{\tilde{\epsilon}}
\newcommand*{\vt}{\tilde{v}}
\newcommand*{\vesc}{v_{\text{esc}}}
\newcommand*{\vesct}{\tilde{v}_{\text{esc}}}

\title{Indirect detection of dark matter with (pseudo)-scalar interactions}

\author[a]{Simone Biondini,}
\author[b]{Julian Bollig}
\author[b]{and Stefan Vogl}
\affiliation[a]{Department of Physics, University of Basel, Klingelbergstrasse\ 82, CH-4056 Basel, Switzerland}
\affiliation[b]{Institute of Physics, University of Freiburg, Hermann-Herder-Straße 3, 79014 Freiburg, Germany}

\emailAdd{simone.biondini@unibas.ch}
\emailAdd{julian.bollig@physik.uni-freiburg.de}
\emailAdd{stefan.vogl@physik.uni-freiburg.de}

\abstract{Indirect detection is one of the most powerful methods to search for annihilating dark matter. In this work, we investigate the impact of non-perturbative effects in the indirect detection of dark matter. For this purpose we utilize a minimal model consisting of a fermionic dark matter candidate in the TeV mass range that interacts via scalar- and pseudo-scalar interactions with a massive scalar mediator mixing with the Higgs. The scalar interaction induces an attractive Yukawa potential between dark matter particles, such that annihilations are Sommerfeld enhanced, and bound states can form. These non-perturbative effects are systematically dealt with (potential) non-relativistic effective field theories and we derive the relevant cross sections for dark matter. We discuss their impact on the relic density and indirect detection. Annihilations in dwarf galaxies and the Galactic Center require special care and we derive generalized $J$-factors for these objects that account for the non-trivial velocity dependence of the cross sections in our model. We use limits on the gamma-ray flux based on Fermi-LAT observations and limits on the rate of exotic energy injection from Planck to derive bounds on the parameter space of the model. Finally, we estimate the impact that future limits from the Cherenkov Telescope Array are expected to have on the model.}

\begin{document}
\maketitle
\flushbottom


\section{Introduction}
\label{sec:intro}

The advent of precision cosmology, mainly driven by increasingly accurate observations of the Cosmic Microwave Background (CMB), has made stunning insights into the structure and the dynamics of our universe at the largest scales possible. Among the most startling results of this program is the realization that the energy content of the universe is heavily dominated by two components that are only poorly understood: dark matter (DM) and dark energy. The concordance cosmological model, known as $\Lambda$CDM, requires that $\sim$27\% of the energy budget of the universe is in the form of cold DM to explain the CMB anisotropies measured by the Planck satellite \cite{Planck:2018vyg}. This constitutes a serious challenge for particle physics since none of the known particles can account for it.

Both on the side of ultraviolet complete theories and simplified models, a plethora of particle physics models have been proposed that contain viable dark matter candidates (see \eg \cite{Bertone:2004pz,Feng:2010gw}): stable, or very long-lived, particles that are electromagnetically neutral and interact only (very) weakly with the visible sector. The possibility to observe DM particles with particle-physics-inspired strategies requires additional interactions between the visible and the dark sector beyond gravity. One organizing principle that allows to classify different DM candidates, is the production mechanism in the early universe. This is particularly helpful if one is interested in the phenomenology of the DM particles since different production mechanisms require vastly different masses and couplings. One of the classic solutions is the thermal production of DM via freeze-out from the SM bath \cite{Kolb:1990vq}. This has inspired a wide range of experiments, such as direct detection, indirect detection, and collider searches, see \eg \cite{MarrodanUndagoitia:2015veg,Klasen:2015uma,Gaskins:2016cha,Arcadi:2017kky,Kahlhoefer:2017dnp} for some recent reviews. Out of these approaches, indirect detection stands out as being the most immediately connected to the production process. It aims to observe the continued annihilation of DM particles in the late universe after freeze-out and picks out this signal from backgrounds of astrophysical origin. There is a vast range of indirect searches that target very different messengers such as gamma rays \cite{Fermi-LAT:2015att,Fermi-LAT:2016uux,Fermi-LAT:2017opo,HESS:2016mib,VERITAS:2017tif,Hoof:2018hyn}, antiprotons \cite{AMS:2015tnn,Cuoco:2016eej,Heisig:2020nse}, positrons \cite{PAMELA:2008gwm,AMS:2019rhg,John:2021ugy} and neutrinos \cite{ANTARES:2015vis,IceCube:2022clp,IceCube:2021xzo}. All of these have their individual strength and limitations. The comparative ease of detection, and the fact that photons travel unaffected from the source to Earth, make gamma rays particularly interesting. With the Cherenkov Telescope Array (CTA) a next-generation gamma-ray telescope is on the way \cite{CTAConsortium:2017dvg}. Therefore, it is timely to reassess the impact of existing indirect detection limits on dark matter produced by freeze-out and compare this with the potential of CTA observations.

A particularly interesting scenario for indirect detection is so-called secluded DM \cite{Pospelov:2007mp,Pospelov:2008jd}, \ie DM annihilating into states beyond the SM. This naturally allows to decouple the relic density from other observables and thus avoids stringent bounds from direct detection experiments \cite{PandaX-4T:2021bab,LZ:2022lsv,XENON:2023cxc} and the LHC \cite{ATLAS:2019wdu,ATLAS:2021kxv,CMS:2017jdm,CMS:2018mgb}. When these new states are light compared to the DM mass, special care is needed for the computation of the annihilation rates since they can lead to long-range forces between the annihilating DM particles. The presence of such a long-range interaction leads to a modification of the annihilation rate by the Sommerfeld effect \cite{Hisano:2003ec,Hisano:2004ds,Arkani-Hamed:2008hhe} and, if the interaction is attractive and strong enough, it allows for the existence of bound states \cite{March-Russell:2008klu,vonHarling:2014kha}.

On the one hand, the inclusion of Sommerfeld factors has become over the last years a somewhat standard ingredient in both the extraction of the DM relic density and the indirect detection limits \cite{Hisano:2004ds,Feng:2010zp,Slatyer:2011kg,Abazajian:2011ak,Lu:2017jrh,Ando:2021jvn}. On the other hand, accounting for bound-state formation is less established and the methods are still being developed. In this case, a large body of results focuses on one aspect at a time (DM energy density \cite{vonHarling:2014kha,Petraki:2015hla,Beneke:2016ync,Ellis:2015vna,Liew:2016hqo,Mitridate:2017izz,Cirelli:2016rnw,Beneke:2016jpw,Harz:2018csl,Biondini:2018pwp,Oncala:2019yvj,Oncala:2021tkz,Biondini:2021ycj,Garny:2021qsr,Binder:2019erp,Binder:2021vfo} or experimental signatures \cite{March-Russell:2008klu,Laha:2015yoa,Asadi:2016ybp,Coskuner:2018are,Chu:2018faw,Bottaro:2021srh} respectively), whereas there are a few studies in the literature that consistently combine the impact of such non-perturbative effects on the cross sections for the thermal freeze-out \textit{and} the indirect detection. Relevant examples are found in refs.~\cite{An:2016gad,Pearce:2015zca,Cirelli:2016rnw,Baldes:2020hwx}, where the impact of a vector mediator, \ie a dark photon which undergoes kinetic mixing with the U(1)$_Y$  of the SM, is scrutinized. In this work, we aim to fill the gap for a class of DM models that feature self-interaction as mediated by a scalar particle. More specifically, we consider both scalar and pseudo-scalar interactions with a Dirac fermion dark matter.  Using non-relativistic effective field theory, we derive the cross sections that fix the DM energy density and play a role in DM annihilations in astrophysical environments, such as the Galactic Center (GC) and dwarf spheroidal galaxies (dSphs). These are then employed to map out the cosmologically viable parameter space for the model and combined with present and prospective limits from the gamma-ray telescopes Fermi-LAT and CTA, respectively.

The paper is structured as follows. In section~\ref{sec:model} we describe the DM model and its relevant energy scales. Then, the cross sections and decay widths are presented in section~\ref{sec:non_pert_EFT} within the framework of non-relativistic effective field theories. The Boltzmann equation that we use to extract the DM energy density, together with a few highlights about the impact of Sommerfeld and bound-state formation, are given in section~\ref{sec:relic_density}. The scrutiny of various astrophysical and cosmological environments that can constrain present-day DM annihilation is addressed in section~\ref{sec:indirectdetection}. Here, the current bounds and future prospects for the model, which include the next-generation instrument CTA, are presented. Conclusions and outlook are offered in section~\ref{sec:conclusion}.


\section{DM model and energy scales}
\label{sec:model}

In this section, we describe the field content of the DM model and discuss the relevant energy scales. We assume DM to be a Dirac fermion that carries no charge under the SM gauge group, namely, it is an SM singlet. However, we introduce an interaction between DM fermions as mediated by a scalar particle via Yukawa-type interactions. The scalar mediator of the dark sector couples to the SM via a Higgs portal interaction. The Lagrangian density of the model reads
 \begin{equation}
    \mathcal{L}= \bar{X} (i \slashed{\partial} -M) X + \frac{1}{2} \partial_\mu \phi \, \partial^\mu \phi -\frac{1}{2}m_\phi^2 \phi^2 -  \bar{X} (g + ig_5 \gamma_5)  X \phi - \frac{\lambda_\phi}{4!} \phi^4  +\mathcal{L}_{\hbox{\scriptsize portal}} \, ,
    \label{lag_mod_relativistic}
\end{equation}
where $X$ is the DM Dirac field and $\phi$ is a real scalar. The scalar self-coupling is denoted with $\lambda_\phi$, whereas the scalar and pseudo-scalar couplings with the fermion are $g$ and $g_5$ respectively. For simplicity, we assume the scalar self-coupling to be negligible such that plays no role in our analyses.\footnote{Such an interaction would be responsible for the generation of a thermal mass for the scalar mediator in the early universe, $m_{\phi,\textrm{thermal}} = T \sqrt{\lambda_\phi/12}$.} The mass of the scalar mediator $m_\phi$ is assumed to be smaller than the DM mass $M$, in order to allow long-range effects. In this work, we restrict to the case where the scalar coupling is larger than the pseudo-scalar coupling, namely $\alpha \equiv g^2/(4 \pi) \gg \alpha_5 \equiv g_5^2/(4 \pi)$. In doing so, we ensure that the dominant non-perturbative effects originate from a scalar-type interaction, that induces an attractive potential, and we can largely neglect the mixed scalar-pseudo-scalar and pure-pseudo-scalar induced contributions \cite{Kahlhoefer:2017umn,Biondini:2021ycj}.

For the main scope of this work, namely exploring non-perturbative effects in indirect detection of dark matter, it is sufficient to work within a simplified model realization with a scalar particle mediating interactions between DM particles~\cite{Kaplinghat:2013yxa,DeSimone:2016fbz}. The mass parameters and couplings of the Lagrangian given in \Eq\eqref{lag_mod_relativistic} are taken as free and we do not address here the fermion and scalar mass generation of the dark sector (see \eg\cite{Kahlhoefer:2015bea,Duerr:2016tmh} for a simplified model with two mediators, a vector and a scalar, where the gauge invariance and spontaneous symmetry breaking in the dark sector is fully accounted for). It is worth mentioning that the mediator sector can be extended to have a richer set of interactions~\cite{Wise:2014jva,Kahlhoefer:2017umn,Oncala:2018bvl}. For example, an interaction of the form $\rho_\phi \phi^3$ with a dimensional coupling could exist, which enables additional bound state formation processes~\cite{Oncala:2018bvl,Oncala:2019yvj}.

The dark and visible sectors are connected via Higgs-portal interactions, which read as follows 
\begin{equation}
	\mathcal{L}_{\text{portal}}=-\mu_{\phi h}\phi\left(H^\dagger H -\frac{v^2}{2}\right)-\frac{1}{2}\lambda_{\phi h}\phi^2\left(H^\dagger H -\frac{v^2}{2}\right) \, ,
 \label{L_portal}
\end{equation}
where $H$ is the SM Higgs doublet. We have shifted the scalar mediator field such that it has no vacuum expectation value. In the following, we shall take the quartic coupling $\lambda_{\phi h}$ to be negligible and only retain the dimensional coupling to induce the mixing between $\phi$ and the Higgs boson.\footnote{This choice implies the absence of sizable contributions to the Higgs mass through mediator loops. In order to achieve thermal contact between the SM and dark sector in the early universe, rather small couplings are typically needed , see Sec.~\ref{sec:thermalization}.} We decided to induce the mixing between the scalar and the Higgs through a dimensional coupling $\mu_{\phi h}$ (and with $v_\phi=0$) in order to keep the number of relevant free parameters at a minimum. Note that we could have also set $\mu_{\phi h}=0$ and kept the quartic mixing $\lambda_{\phi h}$ alongside the quartic dark-scalar coupling $\lambda_\phi$. This choice would introduce a vev for the scalar $v_\phi\neq0$ in analogy to the Standard Model Higgs sector. The mixing term in the Lagrangian after EWSB as well as the definition of the mixing angle $\delta$ remain the same if one substitutes $\mu_{\phi h}\to v_\phi \lambda_{\phi h}$. We have checked that the phenomenology of both scenarios, together with the resulting dark scalar and Higgs boson masses, is equivalent up to $\order{\sin^2\delta}$ of the scalar mixing angle, which we will neglect in the following due to its smallness. The scalar mixing and the implications for the indirect detection are addressed in section~\ref{sec:BR_phi}.

Fermion-antifermion pair annihilation into dark scalars is the key process both for fixing the DM energy density via the freeze-out mechanism and for providing the astrophysical signatures of indirect detection.  $X \bar{X}$ annihilations in the early universe, as well as later-stage annihilations, either during the formation of the CMB or today, occur in the non-relativistic regime. Here fermions and antifermions move with relative velocities $\vrel \ll 1 $. In this kinematical region, also referred to as near-threshold, repeated scalar exchange between DM particles with $\vrel \lesssim \alpha$ induce ladder diagrams that are not suppressed and need to be resummed. This resummation manifests itself in the dynamical generation of bound-state poles of order $M \alpha^2$ at negative energies (below threshold) and a continuum spectrum of scattering states at positive energies (above threshold). The typical momentum exchanged between the fermion-antifermion pairs for $\vrel \sim \alpha$ is $M \alpha$, which is of the order of the inverse Bohr radius of the bound states. The dark matter mass and the dynamically generated scales are the more separated the smaller $\alpha$ is, namely $M \gg M \alpha \gg M \alpha^2$. In the following, we refer to these energy scales as hard, soft, and ultrasoft respectively.
\begin{figure}[t!]
    \centering
    \includegraphics[scale=0.6]{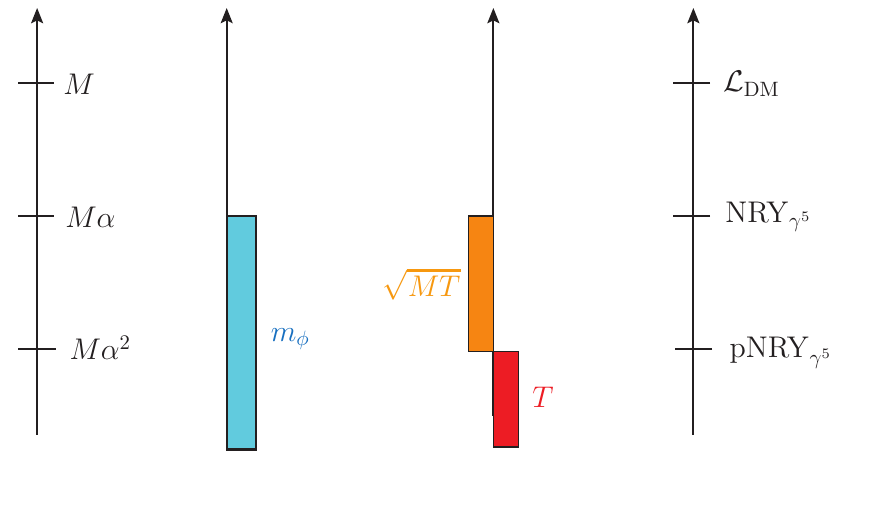}
    \caption{
      Hierarchy of energy scales and effective field theories considered in this work for the DM model Lagrangian density defined in \Eq\eqref{lag_mod_relativistic}. In-vacuum and thermal scales are shown.}
    \label{fig:scales_EFT}
\end{figure}

Another relevant energy scale is the mediator mass $m_\phi$. We will take it to be smaller than the DM mass $M$ and at most as large as the soft scale $M \alpha$. Thus, long-range interactions typically still occur, however, there are sharp deviations from the Coulombic regime that manifest in various ways. From a phenomenological point of view, we are mainly interested in $m_\phi\geq 2 m_{\pi^{0}}$. Lighter mediators do not produce an appreciable amount of $\gamma$-rays, which is the main observation signature considered in this work. However, lighter mediators can have implications for other observables such as the positron flux or the DM mass distribution in dense halos, see \eg \cite{Kahlhoefer:2017umn,Kainulainen:2015sva}. The main difference to massless mediators is that DM self-interactions are screened when the soft scale $M \alpha$ is of the order of the mediator mass $m_\phi$. For scattering states subject to an attractive potential, the Sommerfeld enhancement is typically reduced and does not monotonically grow at small velocities. Moreover, a rich resonance structure is found that depends on the mediator and dark matter mass ratio $m_\phi/M$ \cite{Hisano:2003ec,Hisano:2004ds,Arkani-Hamed:2008hhe}. As far as bound states are concerned, there is even a condition that has to be satisfied for their existence of at least one bound state, namely $1.6\,m_\phi < \alpha M$ \cite{Shepherd:2009sa}. In order to make contact with previous work on the subject, we define the two dimensionless variables for later use
\begin{equation}
    \zeta =\frac{\alpha}{\vrel} \, \quad \mbox{and} \, \quad  \xi=\frac{M \alpha}{2 m_\phi} \, .
\end{equation}
The latter variable involves the Bohr momentum $M \alpha /2$ or, equivalently, the inverse Coulomb Bohr radius $a_0 \equiv 2/ M \alpha$. The Coulombic limit is recovered for large velocities with respect to $\alpha$, namely $\zeta \ll 1$, and for mediator masses much smaller than the soft scale $M \alpha$, \ie $\xi \gg 1$.

Finally, thermal scales are relevant for the freeze-out of DM particles in the early universe. Most notably, there is the temperature of the thermal plasma $T$, which at the thermal freeze-out of massive relics is smaller than the DM mass, $T \ll M$. However, the temperature can still be of the order of the dynamically generated soft and ultrasoft scales $M \alpha$ and $M \alpha^2$. In this work, we include thermal effects due to the medium assuming that the temperature is about the ultrasoft scale $M \alpha^2$ or smaller. This implies that thermal effects do not enter the potential, which may be taken as an in-vacuum Yukawa potential. For thermalized dark matter fermions, one finds $\vrel \sim \sqrt{T/M}$, so that their typical momentum is $M \vrel \sim \sqrt{MT} \lesssim M \alpha$, which implies $\vrel \lesssim \alpha$. For the temperature range considered in this work, it also holds that $\sqrt{MT}\gtrsim M\alpha^2$. These conditions qualify $M \vrel$ as a soft scale and $M \vrel^2 \sim T$ as an ultrasoft scale. Therefore, the hierarchy of energy scales, that we show in figure~\ref{fig:scales_EFT}, is 
\begin{equation}
    M \gg M\alpha \gtrsim \sqrt{MT} \gg M\alpha^2 \gtrsim T \,.
    \label{scale_arrang_0}
\end{equation}
The mediator mass is taken to be $m_\phi \lesssim M \alpha$. Because of the assumption of the scalar sector interactions, namely no trilinear interactions and $\lambda_\phi \ll 1$, thermal masses for the mediator are not considered in the hierarchy of scales (the condition $T \lesssim M \alpha^2$ would make it small in any case). In the following section~\ref{sec:non_pert_EFT}, we discuss the non-relativistic EFTs that we use to compute near-threshold observables. Indeed, the use of the Lagrangian \Eq\eqref{lag_mod_relativistic} is in general unpractical since all energy scales are entangled in the amplitudes.  


\section{Non-perturbative effects in NREFTs}
\label{sec:non_pert_EFT}

In the following, we find it convenient to express pairs in a scattering state with $(X \bar{X})_p$, where $p=M \vrel/2$ denotes the momentum of the relative motion, whereas a fermion-antifermion pair in a bound state is indicated with $(X \bar{X})_n$. In order to shorten the notation, $n \equiv | nlm \rangle$ stands for the whole set of quantum numbers of a given bound state. In this section, we summarise the non-relativistic effective field theories that are useful to compute the main observables of interest. These include DM pair annihilations into light scalar mediators, which can occur both for above-threshold scattering states $(X \bar{X})_p$ and bound states $(X \bar{X})_n$, and bound-state formation.  We shall compute the relevant cross sections, as well as decays widths, in the framework of a potential non-relativistic EFT, dubbed as pNRY$\gamma_5$, which is obtained from the model Lagrangian \eqref{lag_mod_relativistic} via a two-step matching procedure \cite{Luke:1997ys,Biondini:2021ccr,Biondini:2021ycj}. First, the modes with energy/momentum of the order of the hard scale $M$ are integrated out, which results in NRY$_{\gamma_5}$. In this theory, the soft and the ultrasoft scales are still intertwined. In the second stage, the typical scale of soft momentum exchanges between non-relativistic fermions and antifermions is integrated out, together with the mediator mass $m_\phi$, that results in the lowest-lying EFT, namely pNRY$_{\gamma_5}$.  Here, the degrees of freedom are fermion-antifermion pairs that interact only with ultrasoft scalar mediators and obey a Schr\"odinger equation with a Yukawa potential. The towers of EFTs are shown in figure~\ref{fig:scales_EFT}.

Here we discuss the non-relativistic EFT framework and its main ingredients for a self-contained discussion. The details on the derivation of the EFTs have been addressed in refs.\cite{Luke:1996hj,Luke:1997ys,Biondini:2021ccr,Biondini:2021ycj} and more recently in ref.~\cite{Biondini:2023zcz} for the inclusion of thermal scales.


\subsection[NRY$_{\gamma^5}$ and pNRY$_{\gamma^5}$]{\protect\boldmath NRY$_{\gamma^5}$ and pNRY$_{\gamma^5}$}
\label{sec:NRYandpNRY}

NRY$_{\gamma_5}$  describes non-relativistic fermions and antifermions that interact with a scalar. The set of interaction vertices is comprised of the bilinear sector of the corresponding effective Lagrangian.  Moreover, fermion-antifermion annihilations into light mediators are accounted for by four-fermion operators \cite{Bodwin:1994jh}. The NRY$_{\gamma_5}$ Lagrangian reads schematically 
\begin{eqnarray}
    \mathcal{L}_{\hbox{\tiny NRY$_{\gamma_5}$}} = \mathcal{L}^{\textrm{bilinear}}_{\psi} + \mathcal{L}^{\textrm{bilinear}}_{\chi} + \mathcal{L}_{\textrm{4-fermions}}  + \mathcal{L}_{\textrm{scalar}} \, ,
    \label{NRY_scheme}
\end{eqnarray}
where $\psi$ is the two-component Pauli spinor that annihilates a dark matter fermion, whereas $\chi^\dagger$ is the Pauli spinor that annihilates an antifermion. Here, scalar particles have energy and momenta smaller than $M$. The fermion bilinear at order $1/M$ with the matching coefficients set to their tree-level values reads \cite{Luke:1996hj,Biondini:2021ycj} 
\begin{equation}
    \mathcal{L}^{\textrm{bilinear}}_{\psi} = \phantom{+} \psi^\dagger \left( i \partial_0  - \,  g\phi +  g_5 \frac{\sigma \cdot [\bm{\nabla} \phi]}{2M} - g_5^2 \frac{\phi^2}{2M} + \frac{\bm{\nabla}^2}{2 M} \right) \psi \, ,
     \label{bilinear_psi_5}
\end{equation}
for the non-relativistic fermion field, whereas 
\begin{equation}
    \mathcal{L}^{\textrm{bilinear}}_{\chi} = \phantom{+} \chi^\dagger \left( i \partial_0  + \,  g\phi +  g_5 \frac{\sigma \cdot [\bm{\nabla} \phi]}{2M} + g_5^2 \frac{\phi^2}{2M}  - \frac{\bm{\nabla}^2}{2 M} \right) \chi \, ,
     \label{bilinear_chi_5}
\end{equation}
for the antifermion. We notice that the leading fermion-pseudo-scalar interaction is suppressed with respect to the leading fermion-scalar interaction by $k/M$, where $k$ is the soft or ultra-soft momentum carried by the field $\phi$.  The operators proportional to one power of $g_5$ are parity-violating, and the notation $[\bm{\nabla} \phi]$ stands for the derivative acting on the scalar field only.

The four-fermion Lagrangian $\mathcal{L}_{\textrm{4-fermions}} $ in \Eq\eqref{NRY_scheme} is especially relevant because it accounts for DM fermion-antifermion annihilations into light scalars. Through the optical theorem, pair annihilations correspond to the imaginary parts of four-point Green's functions and, at leading order in the couplings, describe the $2 \to 2$ process $X \bar{X} \to \phi \phi$. We adopt the spectroscopy notation from NRQED/NRQCD \cite{Caswell:1985ui,Bodwin:1994jh} so that one can classify the annihilations in terms of the total spin $S$ of the pair, the relative angular momentum $L$ and the total angular momentum $J$. The operators and matching coefficients of the four-fermion Lagrangian are then labeled with $^{2S+1}L_J$ (\cf\Eqs\eqref{dimension_6_lag} and \eqref{dimension_8_lag}). This will turn out useful for assigning the corresponding Sommerfeld factors. In the following, we list the matching coefficients of the four-fermion operators by retaining the full dependence on the mass ratio $m_\phi/M$, and hence we generalize the results presented in  ref.~\cite{Biondini:2021ycj}. In our setting of the hierarchy of scales, such corrections are modest (the largest value for the mediator mass is taken to be the soft scale $M \alpha$). The general result with the $m_\phi$ dependence could turn out to be useful if one aims to consider larger mediator masses for this model.\footnote{For mediator masses comparable with $M$ the construction of the non-relativistic EFTs changes. In particular, having $m_\phi \sim M \gg M \alpha$ would affect the potential between the fermion-antifermion pairs as the mediator scale has to be integrated out together with the DM mass.}

In the following, we explicitly list the full set of four-fermion operators that can be written up to order $1/M^4$. The corresponding matching coefficients have to be determined and depend on the DM model. There are two independent dimension-6 operators that can account for velocity-independent contributions to DM annihilation, and they read~\cite{Bodwin:1994jh}
\begin{eqnarray}
    (\mathcal{L}_{\textrm{4-fermions}} )_{d=6} = \frac{f(^1S_0)}{M^2} \, \mathcal{O}(^1S_0) + \frac{f(^3S_1)}{M^2} \, \mathcal{O}(^3S_1) \, ,
    \label{dimension_6_lag}
\end{eqnarray}
where $f(^1S_0)$ and $f(^3S_1)$ are the matching coefficients of the spin singlet and spin triplet operators 
\begin{eqnarray}
    \mathcal{O}(^1S_0)=\psi^\dagger \chi \, \chi^\dagger \psi \, , \quad \mathcal{O}(^3S_1)=  \psi^\dagger \, \bm{\sigma} \, \chi \cdot \chi^\dagger \, \bm{\sigma} \, \psi \, .
    \label{dimension_6_operators}
\end{eqnarray}
Dimension-8 operators comprise derivatives acting on the DM fermion and antifermion fields and hence have a dependence on the DM velocity.  Schematically one may write the corresponding Lagrangian in terms of operators, which are indicated with $\mathcal{O}(^{2S+1}L_J)$ and $\mathcal{P}(^{2S+1}L_J)$, and the corresponding matching coefficients $f(^{2S+1}L_J)$ and $g(^{2S+1}L_J)$, respectively, as follows~\cite{Bodwin:1994jh} 
\begin{eqnarray}
    (\mathcal{L}_{\textrm{4-fermions}} )_{d=8} &=& \frac{f(^1P_1)}{M^4} \mathcal{O}(^1P_1) + \frac{f(^3 P_0)}{M^4} \mathcal{O}(^3 P_0) + \frac{f(^3 P_1)}{M^4} \mathcal{O}(^3 P_1)  \nonumber 
    \\
    &+& \frac{f(^3 P_2)}{M^4} \mathcal{O}(^3 P_2) + \frac{g(^1 S_0)}{M^4} \mathcal{P}(^1 S_0) + \frac{g(^3 S_1)}{M^4} \mathcal{P}(^3 S_1) \nonumber 
    \\
    &+& \frac{g(^3 S_1,^3 D_1 )}{M^4} \mathcal{P}(^3 S_1, ^3 D_1)  \, , 
    \label{dimension_8_lag}
\end{eqnarray}
where the operators are  
\begin{eqnarray}
    \mathcal{O}(^1P_1) &=& \psi^\dagger \left( - \frac{i}{2} \overset{\leftrightarrow}{\bm{\nabla}} \right) \chi \cdot \chi^\dagger \left( -\frac{i}{2} \overset{\leftrightarrow}{\bm{\nabla}} \right) \psi \, ,  
    \\
    \mathcal{O}(^3P_0) &=& \frac{1}{3} \psi^\dagger \left( -\frac{i}{2} \overset{\leftrightarrow}{\bm{\nabla}} \cdot \bm{\sigma} \right) \chi\,  \chi^\dagger \left( -\frac{i}{2} \overset{\leftrightarrow}{\bm{\nabla}} \cdot \bm{\sigma} \right) \psi\, , 
    \\
    \mathcal{O}(^3P_1) &=& \frac{1}{3} \psi^\dagger \left( -\frac{i}{2} \overset{\leftrightarrow}{\bm{\nabla}} \times \bm{\sigma} \right) \chi  \cdot  \chi^\dagger \left( -\frac{i}{2} \overset{\leftrightarrow}{\bm{\nabla}} \times \bm{\sigma} \right) \psi\, , 
    \\
    \mathcal{O}(^3P_2) &=&  \psi^\dagger \left( -\frac{i}{2} \overset{\leftrightarrow}{\nabla}  \phantom{s}^{(i} \sigma^{j)} \right) \chi\,  \chi^\dagger \left( -\frac{i}{2} \overset{\leftrightarrow}{\nabla}  \phantom{s}^{(i} \sigma^{j)} \right) \psi \, ,
    \\
    \mathcal{P}(^1 S_0) &=& \frac{1}{2} \left[ \psi^\dagger \chi \, \chi^\dagger \left( \frac{i}{2} \overset{\leftrightarrow}{\bm{\nabla}} \right)^2 \psi + \textrm{h.c.} \right] \, ,
    \\
    \mathcal{P}(^3 S_1) &=& \frac{1}{2} \left[ \psi^\dagger \bm{\sigma} \chi \, \cdot \chi^\dagger \bm{\sigma} \left( \frac{i}{2} \overset{\leftrightarrow}{\bm{\nabla}} \right)^2 \psi + \textrm{h.c.} \right] \, ,
    \\
    \mathcal{P}(^3 S_1, ^3 D_1) &=& \frac{1}{2} \left[ \psi^\dagger \sigma^i \chi \, \cdot \chi^\dagger \sigma^j  \left( \frac{i}{2} \right)^2 \overset{\leftrightarrow}{\nabla}  \phantom{x}^{(i} \overset{\leftrightarrow}{\nabla}\,^{j)}  \psi + \textrm{h.c.} \right] \, .
\end{eqnarray}
Here $\overset{\leftrightarrow}{\bm{\nabla}}$  is the difference between the derivative acting on the spinor to the right and on the spinor to the left, namely $\chi^\dagger \overset{\leftrightarrow}{\bm{\nabla}} \psi \equiv \chi^\dagger (\bm{\nabla} \psi)- (\bm{\nabla} \chi)^\dagger \psi$. The notation $T^{(ij)}$ for a rank 2 tensor stands for its traceless symmetric components, $T^{(ij)}=(T^{ij}+T^{ji})/2-T^{kk}\delta^{ij}/3$.

The non-vanishing matching coefficients are found to be
\begin{eqnarray}
    && {\rm{Im}}[f(^1S_0)] = 2  \pi \alpha \alpha_5 \mathcal{F}(\tilde{r}) \, , 
     \label{match_coeff_1}
     \\
    &&  {\rm{Im}}[g(^1S_0)] = - \frac{8 \pi \alpha \alpha_5}{3} \frac{\left( 1-\frac{13}{8}\tilde{r}^2+\frac{5}{8}\tilde{r}^4 - \frac{3}{32} \tilde{r}^6\right)}{(1-\tilde{r}^2/2)^2(1-\tilde{r}^2)} \mathcal{F}(\tilde{r}),
    \label{match_coeff_2}
    \\
    &&  {\rm{Im}}[f(^3P_0)] = \frac{\pi}{6} \left[ 3 \alpha \left( 2 - \frac{\mathcal{G}(\tilde{r})}{3} \right) - \alpha_5 \mathcal{G}(\tilde{r})  \right]^2  \mathcal{F}(\tilde{r})  \, , \quad  
    \label{match_coeff_3}
    \\
    && {\rm{Im}}[f(^3P_2)] = \frac{\pi}{15} (\alpha+\alpha_5)^2 \mathcal{G}(\tilde{r})^2   \mathcal{F}(\tilde{r}) \, ,
    \label{match_coeff_4}
\end{eqnarray}
where the auxiliary functions $\mathcal{F}(\tilde{r})$ and $\mathcal{G}(\tilde{r})$ with $\tilde{r} \equiv m_\phi/M$ are 
\begin{eqnarray}
    \mathcal{F}(\tilde{r}) = \frac{\sqrt{1-\tilde{r}^2}}{\left( 1-\frac{\tilde{r}^2}{2} \right)^2}  \, , \quad  \mathcal{G}(\tilde{r}) =  \frac{1-\tilde{r}^2}{1-\frac{\tilde{r}^2}{2}}  \, .
\end{eqnarray}
The vanishing imaginary parts for $f(^3S_1)$ and $f(^1P_1)$ can be understood in terms of symmetry arguments \cite{Biondini:2021ycj}. The pseudo-scalar interaction in \Eq\eqref{lag_mod_relativistic} violates parity, however it still preserves the charge conjugation symmetry. Accordingly, only some combinations of the spin and angular momentum of the annihilating fermion-antifermion pairs are allowed.\footnote{The $C$ quantum number for a fermion-antifermion pair is $(-1)^{L+S}$, whereas a scalar field is a parity eigenstate, such that we have $C=1^n$ for $n$ scalar particles in the final state. This would select only $C=1$ for $(X\bar{X})_p$ and  $(X\bar{X})_n$, which are ${}^1 S_0$ and ${}^3 P_J$. Additional fermion species with $m_f<M$, that couple to the scalar field, can lift this condition and allow non-vanishing matching coefficients $f(^3S_1)$ and $f(^1P_1)$. An interaction among $\phi$ and SM fermions can be triggered by the mixing between $\phi$ and the SM Higgs boson which results in a $p$-wave process $X \bar{X} \to \phi \to f \bar{f}$ (see \eg ref.~\cite{DeSimone:2016fbz}). A further suppression comes from the mixing angle, in the form of an effective vertex $y_{\textrm{SM}} \sin \delta \phi \, f \bar{f}$, where $\delta$ is the mixing angle, which makes such contribution fairly smaller than the ones we account for (an exception in some parts of the parameter space is for the top-quark case, that we neglect here).}

To obtain NRY$_{\gamma_5}$ in \Eq\eqref{NRY_scheme}, we have integrated out the hard scale $M$, which is reflected by $1/M$ powers appearing in the effective Lagrangian. The next scale one has to integrate out according to the scale hierarchy in \Eq\eqref{scale_arrang_0} is the soft scale $M \alpha$, or the typical inverse distance between $X$ and $\bar{X}$ in a pair. In this work, we allow the mass of the scalar mediator to be as large as the soft scale $ M \alpha$, which also corresponds to the momentum transfer of potential exchanges among fermion-antifermion pairs (this is, in general, true for bound states, whereas it depends on the value of $\vrel$ for the scattering states).  Hence, $m_\phi$ is integrated out at the same time with the soft scale $M \alpha$ which results in a Yukawa potential experienced by $X\bar{X}$ pairs. In the situation where $m_\phi \ll M \alpha$ one naturally recovers a Coulomb-like regime and the corresponding potential. As mentioned earlier, we can set the thermal scales to zero in the matching between NRY$_{\gamma_5}$ and pNRY$_{\gamma_5}$ because of the assumption $T \lesssim M \alpha^2$.

To obtain pNRY$_{\gamma_5}$ one implements a projection of the NRY$_{\gamma_5}$ Hamiltonian onto the particle-antiparticle sector via 
\begin{equation}
    \int d^3 \bm{x}_1 d^3 \bm{x}_2 \, \varphi_{ij}(t,\bm{x}_1, \bm{x}_2) \psi^\dagger_i (t,\bm{x}_1) \chi_j (t,\bm{x}_2) | \phi_{\hbox{\tiny US}}\rangle \, ,
    \label{proj_fock}
\end{equation}  
where $i,j$ are Pauli spinor indices, while the state $| \phi_{\hbox{\tiny US}}\rangle$ contains an arbitrary number of scalars with energies much smaller than $M\alpha$, including thermal scales, and no heavy fermions/antifermions. Here $\varphi_{ij}(t,\bm{x}_1, \bm{x}_2)$ is a bilocal wave function field representing the $X\bar{X}$ system. The Lagrangian is then written in terms of the relative distance of the pair $\bm{r}=\bm{x}_1-\bm{x}_2$ and its center-of-mass coordinate $\bm{R}=(\bm{x}_1+\bm{x}_2)/2$  
\cite{Biondini:2021ccr,Biondini:2021ycj}
\begin{align}
    \mathcal{L}_{\hbox{\tiny pNRY$_{\gamma_5}$}}  &= \int d^3 \bm{r}   \, \varphi^\dagger(\bm{r},\bm{R},t) \left\lbrace  i \partial_0 +\frac{\bm{\nabla}^2_{\bm{r}}}{M} +\frac{\bm{\nabla}^2_{\bm{R}}}{4M} + \frac{\bm{\nabla}^4_{\bm{r}}}{4 M^3} - V(\bm{p},\bm{r},\bm{\sigma}_1,\bm{\sigma}_2)\right. \nonumber 
    \\
    & \left. - 2 g \phi(\bm{R},t) -g\frac{ r^i r^j }{4}  \left[ \nabla_R^i \nabla_R^j \, \phi (\bm{R},t)   \right] -   g \phi(\bm{R},t) \frac{\bm{\nabla}^2_{\bm{r}}}{M^2}    \right\rbrace \varphi(\bm{r},\bm{R},t) \nonumber
    \\
    &+  \frac{1}{2} (\partial^\mu \phi(\bm{R},t))^2  -  \frac{m_\phi^2}{2} \phi(\bm{R},t)^2 - \frac{ \lambda_\phi}{4!} \phi(\bm{R},t)^4  ,
    \label{pNREFT_sca_0}
 \end{align}
where the square brackets in the second line of \Eq\eqref{pNREFT_sca_0} indicate that the spatial derivatives act on the scalar field only, which is multipole expanded. To avoid cluttering the notation, we suppress the spin indices of the bilocal fields that are contracted with each other. Each term in the Lagrangian \eqref{pNREFT_sca_0} has a well-defined scaling. The time derivative scales as $\partial_0 \sim M \alpha^2 \sim T$, the inverse relative distance and the corresponding derivative obey ${\bm{r}}^{-1}, \bm{\nabla}_{\bm{r}} \sim M \alpha$ (the mediator mass can as well define the soft scale whenever $m_\phi \sim M \alpha$), whereas the scalar field and the center-of-mass derivative satisfy $g\phi, \bm{\nabla}_{\bm{R}} \sim M \alpha^2 \sim T$. Indeed, the active dynamical scales in the so-obtained EFT are the ultrasoft scale \textit{and} the temperature.

The potential is understood as a matching coefficient of pNRY$_{\gamma_5}$ and it emerges as such because the typical distance between fermion-antifermion pairs is removed from NRY$_{\gamma^5}$. In general, the potential comprises both a real and an imaginary part, where the latter is related to physical processes $X \bar{X} \to \phi \phi$ (see section~\ref{sec:pair_ann_pNRY5}). The potential is organized as an expansion in the coupling $\alpha$ and $1/M$ 
\begin{eqnarray}
    V(\bm{p},\bm{r},\bm{\sigma}_1,\bm{\sigma}_2) = V^{(0)} + \frac{V^{(1)}}{M}  + \frac{V^{(2)}}{M^2} + \cdots \, ,
\end{eqnarray}
with the static Yukawa potential 
\begin{equation}
    V^{(0)} = - \alpha \frac{e^{-m_\phi r}}{r} \, .
    \label{yuk_pot_LO}
\end{equation}
The contributions to the potential from the mixed scalar-pseudo-scalar and pure pseudo-scalar diagrams have been scrutinised in ref.~\cite{Biondini:2021ycj}, and they belong to the $1/M$-suppressed terms $V^{(1)}$ and $V^{(2)}$, respectively. Using power counting of the low-energy theory, one can show that these terms scale as $M \alpha^3 (g_5/g)$ and $M \alpha^4 (g_5/g)^2$ compared to the leading Yukawa potential in eq.~\eqref{yuk_pot_LO}, which scales as $M \alpha^2$. In this study, we work at (moderately) weak coupling, $\alpha <0.25$, and consider $g_5 < g$. Therefore, we can neglect pseudo-scalar effects in the potential to a good approximation. The same assumptions also guarantee that the corresponding pseudo-scalar-induced ultrasoft transitions among pairs are suppressed with respect to those displayed in the second line of \Eq\eqref{pNREFT_sca_0}. Pure-scalar coupling diagrams contribute to $V^{(2)}$ and are $\alpha^2$-suppressed with respect to the leading Yukawa potential. Finally, because of $T \lesssim M \alpha^2$, the potential does not get thermal contributions at any order. In what follows, we retain only the leading Yukawa potential \eqref{yuk_pot_LO} when solving the Schr\"odinger equation to find the corresponding wave functions of scattering and bound states.

The ultrasoft interactions in the second line of \Eq\eqref{pNREFT_sca_0} differ qualitatively from the potential-NREFT of vector mediators, namely pNRQED and pNRQCD \cite{Pineda:1997bj}. One finds a non-vanishing monopole term at order $r^0$ in the multiple expansion, whereas the electric-dipole vertex, which is linear in the relative coordinate $r$, is absent. To capture the leading non-trivial dynamics among scattering and bound-states, one has to consider terms at order $r^2$ and find the quadrupole contributions that scale as $M \alpha^4$. The derivative interaction with $\nabla_{\bm{r}}$ in the second line of \Eq\eqref{pNREFT_sca_0} arises from the $1/M^2$ spin-independent operator of NRY$_{\gamma^5}$, and it is of the same order as the quadrupole term.\footnote{The NREFT is displayed only at the leading order $1/M$ in \Eq\eqref{NRY_scheme}. For the complete set of operators at order $1/M^2$ see refs.\cite{Biondini:2021ccr}.}


\subsection{Heavy pair annihilations}
\label{sec:pair_ann_pNRY5}

Let us first discuss the annihilation of heavy pairs. This process is responsible for the depletion of the DM particles into lighter scalar mediators. Annihilations can equally occur for a particle-antiparticle pair in a scattering state or a bound state. At leading order in the coupling the processes are $(X \bar{X})_p \to \phi \phi$ and $(X \bar{X})_n \to \phi \phi$, and are described by the same set of operators in $V^{(2)}$, the only difference being on the two-particle state one projects them onto.

In NRY$_{\gamma_5}$, heavy pair annihilations are accounted for by local four-fermion operators, \cf\Eqs\eqref{dimension_6_lag} and \eqref{dimension_8_lag}. Accordingly, in pNRY$_{\gamma_5}$, the four-fermion operators generate local terms, which are proportional to a delta function $\delta^3(\bm{r})$ in the imaginary part of the potential $V(\bm{p},\bm{r},\bm{\sigma}_1,\bm{\sigma}_2)$. We write the annihilation term from the Lagrangian density of pNRY$_{\gamma_5}$ as follows~\cite{Brambilla:2002nu,Brambilla:2004jw,Biondini:2021ycj} 
\begin{eqnarray}
    \mathcal{L}^{\textrm{ann}}_{\hbox{\tiny pNRY}_{\gamma_5}}&=& \frac{i}{M^2} \, \int d^3 \bm{r} \varphi^\dagger (\bm{r}) \delta^3(\bm{r}) \left[ 2 {\rm{Im}}[f(^1S_0)] - \bm{S}^2 \left( {\rm{Im}}[f(^1S_0)]-  {\rm{Im}}[f(^3S_1)] \right) \right] \varphi (\bm{r}) \nonumber
    \\
    &-&\frac{i}{M^4} \, \int d^3 \bm{r} \varphi^\dagger (\bm{r}) \mathcal{T}_{SJ}^{ij} \nabla_{\bm{r}}^i \delta^3(\bm{r})  \nabla_{\bm{r}}^j \,  {\rm{Im}}  [f(^{2 S+1}P_{J})] \varphi \, (\bm{r}) \nonumber
    \\
    &-&\frac{i}{2M^4} \, \int d^3 \bm{r} \varphi^\dagger (\bm{r}) \,  \Omega_{SJ}^{ij} \left\lbrace \delta^3(\bm{r}),\nabla_{\bm{r}}^i \nabla_{\bm{r}}^j  \right\rbrace  {\rm{Im}} [g(^{2 S+1}S_{J})] \varphi \, (\bm{r}) \, ,
    \label{pNRY_ann_lag}
\end{eqnarray}
where $\bm{S}$ is the total spin of the pair, while $\mathcal{T}_{SJ}^{ij}$ and $\Omega_{SJ}^{ij}$ are spin projector operators (\cf\eg\cite{Brambilla:2002nu,Brambilla:2004jw}). The matching coefficients are given in \Eqs\eqref{match_coeff_1}-\eqref{match_coeff_4}.

To compute the annihilation cross section we use the optical theorem and consider the scattering amplitude from an initial to a final scattering state $| \bm{p}, \bm{0} \rangle$, where we have set the center of mass momentum to zero. By inserting the vertices that appear in \Eq\eqref{pNRY_ann_lag} in the self-energy of a scattering state, the spin-averaged annihilation cross section reads
\begin{eqnarray}
    (\sigma_{\hbox{\scriptsize ann}} v_{\hbox{\scriptsize rel}})(\bm{p}) &=& \frac{1}{2} \langle \,  \bm{p}, \bm{0}| \int d^3r \, \varphi^\dagger(\bm{r},\bm{R},t)\, \left[-\rm{Im} \,\Sigma_{\rm ann} \right] \, \varphi(\bm{r},\bm{R},t)\, |  \bm{p},\bm{0} \rangle \nonumber 
    \\
    &=& \frac{1}{M^2} \left\lbrace  \left( {\rm{Im}}[f(^1S_0)] + {\rm{Im}}[g(^1S_0)] \frac{\vrel^2}{4}\right) \Sann^{0}(\zeta, \xi) \right.\nonumber \\
    &&\left. \phantom{ssssssssss}+ \frac{{\rm{Im}}[f(^3P_0)]+5{\rm{Im}}[f(^3P_2)]}{12 } \vrel^2 \Sann^{1}(\zeta, \xi)  \right\rbrace \, .
    \label{ann_cross_pNRY5}
\end{eqnarray}
The dimension-8 derivative operators have produced two powers of momentum, that together with two inverse powers of the DM mass, gives $p^2/M^2=\vrel^2/4$ in \Eq\eqref{ann_cross_pNRY5}. The Sommerfeld factors  $\Sann^{l}(\zeta, \xi)$ correspond to the wave function of the fermion-antifermion pairs in a scattering state evaluated at the origin, \ie $\Sann^{0}(\zeta, \xi) =|\mathcal{R}_{0}(0)|^2$ and $\Sann^{1}(\zeta, \xi) = p^2 |\mathcal{R}'_{1}(0)|^2$ \cite{Sommerfeld,Iengo:2009ni,Cassel:2009wt}, where $\mathcal{R}_{l}$ is the radial wave function with angular momentum $l$ and $p=M \vrel/2$. We stress here that, by computing the annihilation cross section for scattering states in pNRY$_{\gamma^5}$, the resummation of multiple soft-scalar exchanges (ladder diagrams) is already taken care of, since pNRY$_{\gamma^5}$ is a quantum field theory for interacting pairs. The factorization of the different energy scales thus manifests in \Eq\eqref{ann_cross_pNRY5}: the hard dynamics is contained in the matching coefficients of NRY$_{\gamma^5}$, whereas the soft dynamics is contained in the wave-function squared, namely the Sommerfeld factors. With the exception of the Coulomb limit, $m_\phi \to 0$, the Sommerfeld factors have to be numerically evaluated. We follow the procedure outlined in \cite{Iengo:2009ni} and \cite{Petraki:2015hla}.

The complementary manifestation of multiple soft scalar exchanges between non-relativistic fermions and antifermions is the appearance of bound states below threshold. Differently from the Coulomb case, there is a condition on the existence of bound states that depends on the relative sizes of the would-be Coulombic Bohr radius and the screening length due to a finite mediator mass. When this condition for the existence of bound states is met, their binding energy can be written as follows
\begin{equation}
    E_{n l} =  - \gamma^2_{n l}(\xi)\frac{M \alpha^2}{4 n^2} =   - \frac{\gamma^2_{n l}(\xi)}{M a_0^2n^2} ,
    \label{Yukawa_en_levels}
\end{equation}
where $a_0=2/(M \alpha)$ is the Coulombic Bohr radius (inverse of the Bohr momentum). The discrete energy levels for a Yukawa potential depend on both the principal quantum number $n$ as well as on $l$. To compute the annihilation width, we project the annihilation terms of pNRY$_{\gamma^5}$ onto $|n, \bm{0} \rangle$ bound states. In this case, we split the spin singlet and spin triplet bound states.

As one may read from the matching coefficients, $nS \equiv|n00\rangle$ bound states in a spin triplet configuration are absent, whereas for $nP \equiv |n1m \rangle$ states ($l=1$) the situation is the opposite. The counterpart of the annihilation cross section \Eq\eqref{ann_cross_pNRY5} read as follows
\begin{eqnarray}
    \Gamma_{\textrm{ann}}^{nS} &=& \frac{|R_{nS}(0)|^2}{\pi M^2}    {\rm{Im}}[f(^1S_0)] \, , \quad  \Gamma_{\textrm{ann}}^{nP_J} =\frac{ |R'_{nP}(0)|^2}{\pi M^4}  {\rm{Im}}[f(^3 P_J)] \, ,  
    \label{dec_width_pNRY5}
\end{eqnarray}
for $J=0,2$. $R_{nS}$ and $R_{nP}$ stand for the radial bound-state wave functions. At leading order, their Coulomb limit gives the analytical expressions $|R_{nS}(0)|^2=4/(n^3 a_0^3)$ and $|R'_{nP}(0)|^2=4(n^2-1)/(9 n^5 a_0^5)$, whereas in the Yukawa case one has to extract their values at the origin numerically. As argued later in section~\ref{sec:bsf_and_bsd_rates}, we shall capture bound-state effects by including only the ground state, namely the $1S$ state. It is therefore useful to define the absolute value of the ground-state binding energy $E_b \equiv |E_{10}|=\gamma_{10}^2M \alpha^2/4$. 


\subsection{Bound-state formation and dissociation}
\label{sec:bsf_and_bsd_rates}

Dark matter pairs may annihilate in the form of bound states according to the decay widths given in \Eq\eqref{dec_width_pNRY5}. However, bound states have to form in the first place. In order to be quantitative on how such an annihilation process contributes to the overall depletion of dark matter pairs, one has to estimate their formation and dissociation rates. Within the minimal model Lagrangian \eqref{lag_mod_relativistic} the relevant bound state formation (BSF) process occurs via radiative emission of a single scalar mediator $(X \bar{X})_p \to (X \bar{X})_n + \phi$. The reverse process, namely the bound-state dissociation, is only relevant for the freeze-out in the early universe. Here, a thermal population of scalars can break the bound states if the mediator is sufficiently energetic (roughly of the order of the binding energy of a given bound state).  The population of dark scalars is nowadays absent, therefore,  bound-state formation is the relevant process for the cross section driving the indirect detection.

In order to compute the bound-state formation cross section, we consider the ultrasoft transitions that are listed in the second line of the pNRY Lagrangian \eqref{pNREFT_sca_0}. More specifically, we are interested in transitions between a scattering state and a bound state, that trigger the formation of a bound state via the radiative emission of a mediator.\footnote{There can be an additional bound-state formation process in the form of $2 \to 2$ inelastic scatterings $\textrm{SM}_1+ (X \bar{X})_p \to \textrm{SM}_2 + (X \bar{X})_n$, that are induced by mixing between the scalar $\phi$ and the Higgs boson  (see refs.~\cite{Biondini:2017ufr,Biondini:2018pwp,Binder:2019erp} for the case of a vector mediator). However, due to the small mixing with the Higgs boson, the typical effective coupling is fairly smaller than $\alpha$ and we neglect this process in this work. Because we assume the trilinear  coupling $\rho_\phi \phi^3=0$ in \Eq\eqref{lag_mod_relativistic}, the corresponding process $\phi+ (X \bar{X})_p \to \phi + (X \bar{X})_n$ is absent.} As outlined in ref.~\cite{Biondini:2021ccr}, there are various ultrasoft vertices that may contribute to the bound-state formation. The monopole contribution, namely the term at order $r^0$ in the multiple expansion, vanishes because of the orthogonality of scattering and bound-state wave functions. Then the quadrupole and derivative interactions drive the ultrasoft process $(X \bar{X})_p \to (X \bar{X})_n + \phi$. According to the hierarchy of scales in \Eq\eqref{scale_arrang_0}, we can write the bound-state formation process at finite temperature by factoring out the stimulated emission of the thermal scalar as follows \cite{Biondini:2021ccr} 
\begin{eqnarray}
    \sigmaBSF^n v_{\textrm{rel}} \big \vert_{T} =  \sigmaBSF^n v_{\textrm{rel}} \left[ 1+ n_{\textrm{B}}(\Delta E^p_n) \right] \, ,
    \label{bsf_matrix_element_T}
\end{eqnarray}
where $n_{\textrm{B}}$ is the Bose-Einstein distribution. When extracting the DM energy density, we will take the thermal average of \Eq\eqref{bsf_matrix_element_T} to use it in the effective Boltzmann equation (\cf\Eq\eqref{cross_section_eff_BE}), whereas we take its in-vacuum limit with the appropriate average as discussed in section~\ref{sec:indirectdetection} for present-day annihilations. In the following, the thermal average of a generic cross section is indicated with $\langle \sigma \vrel \rangle$ and we take Maxwell Boltzmann distributions for the incoming DM fermion and antifermion (for a detailed discussion of the thermal averaging procedure see \cite{Gondolo:1990dk}).

The in-vacuum part of the cross section for a given bound state, or exclusive cross section,  reads
\begin{eqnarray}
    \sigmaBSF^n v_{\textrm{rel}} &=& \frac{\alpha}{120 }    \left[ (\Delta E_n^p) ^2-m_\phi^2\right]^\frac{5}{2} \left[ |\langle\bm{p} | \bm{r}^2 |n  \rangle|^2 + 2 |\langle\bm{p} | r^i r^j |n  \rangle|^2\right] \nonumber
    \\
    && + 2 \alpha  \left[ (\Delta E_n^p) ^2-m_\phi^2\right]^\frac{1}{2} \,  \Big \vert \Big \langle \bm{p} \Big \vert \frac{\nabla^2_{\bm{r}}}{M^2} \Big \vert n  \Big \rangle \Big \vert^{2} \nonumber
    \\
    &&-  \frac{\alpha}{3}  \left[ (\Delta E_n^p) ^2-m_\phi^2\right]^\frac{3}{2} \textrm{Re} \left[ \Big \langle \bm{p} \Big \vert \frac{\nabla^2_{\bm{r}}}{M^2} \Big \vert n  \Big \rangle  \langle n | \bm{r}^2 | \bm{p} \rangle \right] \, ,
    \label{bsf_matrix_elements_n}
\end{eqnarray}
where the energy difference between a scattering and bound state is 
\begin{eqnarray}
    \Delta E_n^p =  \frac{M v_{\textrm{rel}}^2}{4} + \gamma_{n l}(\xi)^2 \frac{M \alpha^2}{4 n^2} \, .
\end{eqnarray}
Within pNRY, the BSF cross section factorizes in quantum-mechanical matrix elements, that are often dubbed in the literature as overlapping integrals. In the Coulomb case, it is possible to obtain analytical expressions for such matrix elements \cite{Oncala:2019yvj,Biondini:2021ycj}, whereas they have to be numerically evaluated in the Yuakawa case. In this work, we use the wave functions of the scattering states and bound states as extracted with the leading Yukawa potential in \Eq\eqref{yuk_pot_LO}. Because the emitted mediator has a finite mass, the bound-state formation process has a threshold (at vanishing, or very small, relative velocity one has $m_\phi < M \alpha^2 \gamma_{n l}^2/4n^2)$. We remark that the total bound-state formation cross section entails a sum of all bound states, namely $\sigmaBSF v_{\textrm{rel}} = \sum_n \sigmaBSF^n v_{\textrm{rel}}$, where $n$ indicates collectively the quantum numbers of a given bound state (later in section~\ref{sec:indirectdetection} we shall use a different notation for the cross sections, where the quantum numbers $n$ and $l$ will be tracked separately, \cf\Eq\eqref{xsection_ID}). In our work we shall consider the formation and decays of the ground state only, and estimate the correction of excited states on the relic density, for more details see discussion at the end of section~\ref{sec:relic_density}.

\begin{figure}
    \centering
    \includegraphics[scale=0.75]{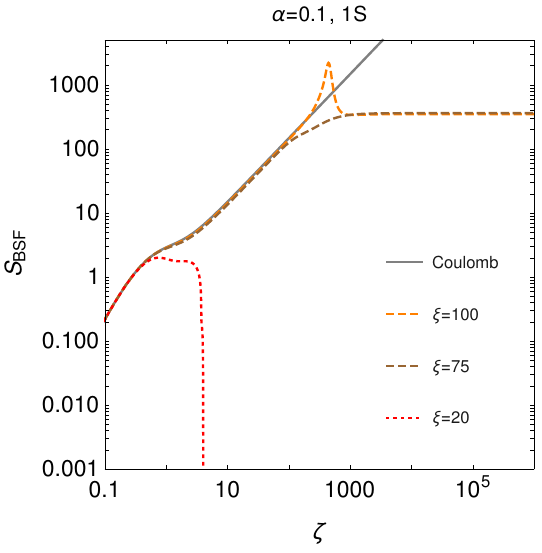}
    \hspace{0.5 cm}
     \includegraphics[scale=0.77]{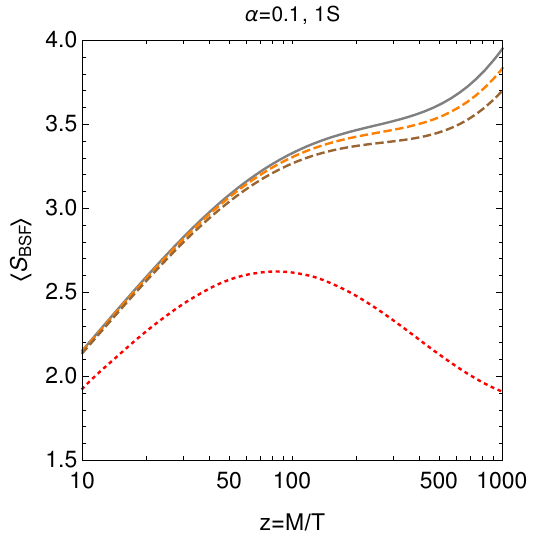}
    \caption{(Left) Bound-state formation factor of \Eq\eqref{bsf_matrix_element_1S} as a function of $\zeta$ for $\xi=20,75,100$. The Coulomb limit is shown for comparison. (Right) Thermal average of the bound-state formation factor with the inclusion of the Bose enhancement $ \left[ 1+ n_{\textrm{B}}(\Delta E^p_n) \right]$ for the same choice of the $\xi$ parameter as in the left panel.}
    \label{fig:Sbsf_sBFS_TH}
\end{figure}

We specify the general expression in \Eq\eqref{bsf_matrix_elements_n} for the ground state, and recast the bound-state formation in terms of the parameters $\zeta$ and $\xi$. The phase-space suppression $\textrm{pps}_{n l}$ due to the finite mediator mass is 
\begin{equation}
    \textrm{pps}_{10}\equiv 1-  \frac{m_\phi^2}{( \Delta E_{1S}^p)^2} = 1 - \frac{4 \zeta^4}{\alpha^2 \xi^2 \left( 1 + \gamma_{10}^2 \zeta^2 \right)^2 } \, .
    \label{pps_1S}
\end{equation}
Moreover, we factor out $\pi \alpha^4/M^2$ from the bound-state formation cross section and define a dimensionless quantity that is the analog of the Sommerfeld factors $\Sann^{l}(\zeta,\xi)$, and we obtain 
\begin{equation}
    \sigmaBSF^{\scriptsize \hbox{1S}} \vrel \equiv \frac{\pi \alpha^4}{M^2} \SBSF^{\scriptsize \hbox{1S}}(\zeta,\xi) \, ,
    \label{sigma_bsf_1S}
\end{equation}
with 
\begin{eqnarray}
    \SBSF^{\scriptsize \hbox{1S}}(\zeta,\xi)  &=&   \sqrt{\textrm{pss}_{10}} \, \frac{\left( 1 + \gamma_{10}^2 \zeta^2 \right)}{4 \pi\zeta^{10}}  \left\lbrace \frac{\left( 1 + \gamma_{10}^2 \zeta^2 \right)^4}{240 \, a_0^7 }     \left[ |\langle\bm{p} | \bm{r}^2 | \textrm{1S}  \rangle|^2 + 2 |\langle\bm{p} | r^i r^j | \textrm{1S} \rangle|^2\right] \textrm{pss}_{10}^2 \right. \nonumber
    \\
    && \left. +  a_0 \,  \zeta^8 \vert  \langle \bm{p}  \vert \nabla^2_{\bm{r}}  \vert \textrm{1S}   \rangle  \vert^{2} -  \zeta^4 \frac{\left( 1 + \gamma_{10}^2 \zeta^2 \right)^2  }{6 \, a_0^3}  \textrm{Re} \left[  \langle \bm{p}  \vert \nabla^2_{\bm{r}}  \vert \textrm{1S}  \rangle  \langle \textrm{1S} | \bm{r}^2 | p \rangle \right] \textrm{pss}_{10}  \right\rbrace \, .
    \label{bsf_matrix_element_1S}
\end{eqnarray}
For illustration, we show $\SBSF^{1S}(\zeta,\xi)$ for different values of $\xi$  in the left panel of figure~\ref{fig:Sbsf_sBFS_TH}, where we fix the scalar coupling to be $\alpha=0.1$. The Coulomb limit, which corresponds to $\xi \to \infty$, is also added for comparison. One may see various behaviors for the Yukawa bound-state factors. For $\xi =100$ and $\xi=75$ in this example, the bound-state factors flatten at large $\zeta$, or equivalently at small velocities, due to the screening of a finite mediator mass. Resonance structures appear also in the bound-state factor $\SBSF^{1S}(\zeta,\xi)$ as can be seen in the curve for $\xi=100$. For $\xi=20$ we notice that the bound-state formation drops severely and vanishes because of the threshold for the emission of the massive mediator (one enters the regime $m_\phi > M \alpha^2 \gamma_{10}^2/4$ for $\zeta \gtrsim 3$). In the right panel of figure~\ref{fig:Sbsf_sBFS_TH}, the thermally averaged bound-state formation $\langle S_{\hbox{\tiny BSF}}^{\hbox{\scriptsize 1S}} \rangle$ is performed by taking a Maxwell Boltzmann distribution for the incoming fermion-antifermion pair \cite{vonHarling:2014kha}. The bound-state factor is plotted as a function of the time variable $z=M/T$ that is used to solve the Boltzmann equation \eqref{Boltzmann_eq_eff}. In this case, the Bose enhancement factor $ \left[ 1+ n_{\textrm{B}}(\Delta E^p_n) \right]$ is included in the thermal average. One can notice how the resonance structure is averaged out and not visible anymore (compare $\xi=100$ and $\xi=75$ curves).

As for the determination of the DM energy density, one last ingredient is needed, namely the bound-state dissociation. Whenever ionization equilibrium holds, the bound-state dissociation can be simply inferred from the bound-state formation via the principle of detailed balance. For this particular case, one often refers to it as the Milne relation in the literature (see \eg \cite{Harz:2018csl,Biondini:2023zcz} for recent derivations). More specifically the ionization cross section can be written as 
\begin{eqnarray}
     \sigma^n_{\textrm{ion}} (|\bm{k}|) = \frac{g_X^2}{g_\phi \, g_{n}} \frac{M^2 \vrel^2}{4 |\bm{k}|^2} \sigmaBSF^{n} \, . 
\end{eqnarray}
Here, $g_X=2$, $g_\phi=1$, and $g_n$ stand for the d.o.f.\! of the DM particle, scalar mediator, and a given bound state, respectively. The latter is then inserted in a convolution integral in the momentum of the incoming thermal scalar $\phi$ to obtain the bound-state dissociation rate
\begin{eqnarray}
    \gammaBSD^n =  g_\phi \int_{|\bm{k}|>k_{\textrm{min}}} \frac{d^3 k}{( 2\pi)^3} n_B(|\bm{k}|) \sigma^n_{\textrm{ion}} (|\bm{k}|) \, ,
\end{eqnarray}
where $|\bm{k}|$ is the momentum of the scalar mediator and $k_{\textrm{min}} = \sqrt{|E_{nl}|^2-m_\phi^2}$. Alternatively, one can derive the bound-state dissociation rate from the thermal self-energy of a bound state in pNRY$_{\gamma^5}$ \cite{Biondini:2021ycj} (see \cite{Brambilla:2008cx} for the derivation in the context of QCD and heavy quarkonium). For the estimation of the DM energy density, we shall employ the bound-state dissociation of the ground state.

Before moving on, let us comment on the running of the scalar coupling $g$. The corresponding coupling strength $\alpha$ appears in the hard matching coefficients \Eq\eqref{match_coeff_1}-\eqref{match_coeff_4}, in the Sommerfeld factors through $\zeta=\alpha/\vrel$ and finally in bound-state formation cross section \Eq\eqref{bsf_matrix_elements_n} (through the ultrasoft vertex of pNRY$_{\gamma^5}$). A hard, soft, and ultrasoft scale, respectively, play a role. A running coupling would require tracking the different $\alpha(\mu)$ in the various observables.  One way to inspect the running of $\alpha$ is to consider the one-loop diagrams (DM fermion and scalar self-energies, and the vertex diagram) and perform the integration by regions \cite{Beneke:1997zp}, \ie expanding the integrals according to the typical momentum in the loops (soft and ultrasoft) with respect to the mass scales $M$ and $m_\phi$. Alternatively, one can perform the calculation in the non-relativistic EFTs. Both methods show that no running is induced at scales below $M$ and, therefore, $\alpha$ can be fixed at the hard annihilation scale where it remains frozen.\footnote{We focus on the running of $\alpha$ only, since $\alpha_5$ only enters the matching coefficients comprising hard energy modes. Indeed, due to the assumption $g_5 \ll g$, we neglect the effect of $\alpha_5$ for the potential and the ultrasoft vertices.} 


\section{Relic density}
\label{sec:relic_density}

In this section, we summarise the main steps and approximations to obtain the DM energy density of the model. This will serve as an input for the interplay between the cosmologically viable parameter space and the corresponding indirect detection prospects.

In this work, we rely on an effective treatment of bound states in terms of semi-classical Boltzmann equations, which allows estimating the bound-state effects on the DM energy density~\cite{vonHarling:2014kha,Ellis:2015vaa}.\footnote{Such treatment has been very recently revisited in \cite{Garny:2021qsr,Binder:2021vfo} to include transitions among bound states.} In the most general case, the situation is rather complex because there is an equation for the DM particle number density, denoted by $n_{ X}$, and an equation for the number density of each bound state, that results in a set of coupled Boltzmann equations. However, whenever the reactions that govern the rate of change of the bound states are faster than the Hubble rate, the network of Boltzmann equations for the bound states greatly simplifies and turns into algebraic equations \cite{Ellis:2015vaa}. In our case, the relevant particle rates are the bound state dissociation rate and the bound state decay width, which are both (much) larger than the Hubble rate for the mass parameters and couplings that we consider in this work.

In doing so, a single Boltzmann equation for $n_X$ is found, where the reprocessing of fermion-antifermion pairs into bound states and their decays is accounted for in an effective cross section
\begin{equation}
    \frac{d n_{X}}{d t} + 3H n_{X} = - \frac{1}{2}\langle \sigma_{\textrm{eff}} \, v_{\textrm{rel}} \rangle (n^2_{X}-n^2_{X,\textrm{eq}}) \, ,
    \label{Boltzmann_eq_eff}
\end{equation}
where the Hubble rate can be expressed in terms of the energy density $H=\sqrt{8 \pi e /3}/M_{\textrm{Pl}}$, $M_{\textrm{Pl}} = 1.22 \times 10^{19}\textrm{ GeV}$ is the Planck mass, $e=\pi^2 T^4 g_{\textrm{eff}}/30$, and $g_{\textrm{eff}}$ denotes the effective number of relativistic degrees of freedom. The factor of $1/2$ on the right-hand side of \Eq\eqref{Boltzmann_eq_eff} is needed for DM particles that are not self-conjugated \cite{Gondolo:1990dk}. The effective thermally averaged cross section when neglecting bound-to-bound transitions yields
\begin{equation}
    \langle  \sigma_{\textrm{eff}} \, v_{\textrm{rel}} \rangle  =  \langle \sigma_{\textrm{ann}} \, v_{\textrm{rel}} \rangle + \sum_{n} \langle   \sigma^n_{\textrm{\tiny BSF}} \, v_{\textrm{rel}} \rangle \, \frac{\Gamma_{\textrm{ann}}^n}{\Gamma_{\textrm{ann}}^n+\gammaBSD^n} \, , 
    \label{cross_section_eff_BE}
\end{equation}
where the sum runs over all possible bound states. As already mentioned, we capture the leading bound-state effects by including the ground state only and hence restricting the sum to the 1S state.

\begin{figure}[t!]
    \centering
    \includegraphics[scale=0.78]{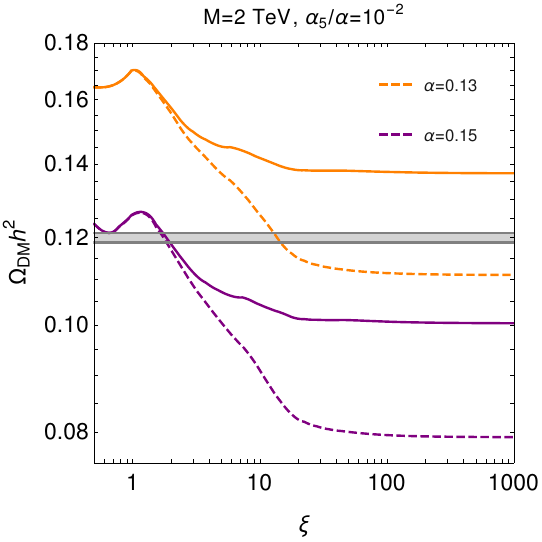}
    \hspace{0.4 cm}
      \includegraphics[scale=0.78]{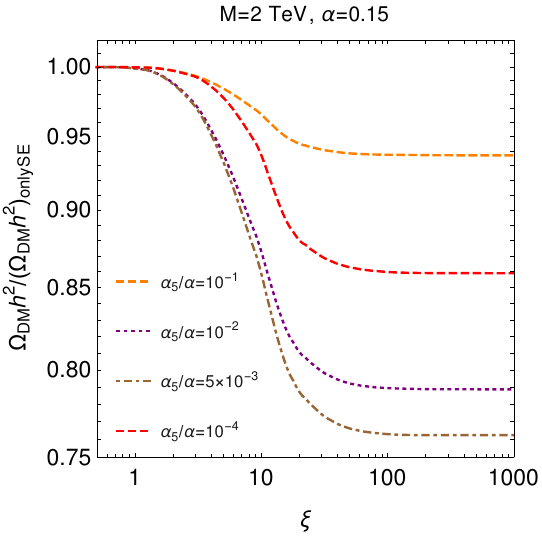}
    \caption{(Left) Predicted dark matter energy density as a function of the $\xi$ parameter for $M=2$ TeV and $\alpha_5/\alpha=10^{-2}$. Solid lines correspond to the effective cross section in \Eq\eqref{cross_section_eff_BE} without bound-state effects, whereas dashed lines include the bound-state formation and decay of the ground state. The gray lines give the range of the observed dark matter energy density $\Omega_{\hbox{\tiny DM}}h^2=0.1200 \pm 0.0012$ \cite{Planck:2018vyg}. (Right) The ratio between the DM energy density as obtained with and without bound-state effects (labeled with $(\Omega_{\hbox{\tiny DM}}h^2)_{\textrm{only SE}}$) for different values of the $\alpha_5/\alpha$ ratios.}
    \label{fig:Omega_versus_xi}
\end{figure}

In figure~\ref{fig:Omega_versus_xi}, left panel, we show the dark matter energy density as a function of the $\xi$ parameter for $\alpha_5/\alpha=10^{-2}$, $M = \SI{2}{\TeV}$ and two different choices of the scalar coupling $\alpha$. The solid curves correspond to $ \langle  \sigma_{\textrm{eff}} \, v_{\textrm{rel}} \rangle$ \textit{without} bound-state effects, namely the second term in \Eq\eqref{cross_section_eff_BE} is dropped. The dashed curves entail both non-perturbative effects.
One may see how all curves flatten for large $\xi$ and provide the smallest DM energy density for each choice of the parameters (mass, couplings, and $\xi$). This is because the Coulomb limit is approached and the non-perturbative effects are maximal. Bound-state effects are absent for $\xi \lesssim 3$ for this choice of the couplings. This is due to the non-existence of the ground state for $\xi< 0.84$ \cite{Shepherd:2009sa} and to the phase space suppression of the bound-state formation process (\cf\Eqs\eqref{pps_1S} and \eqref{bsf_matrix_element_1S}), which quickly fades away with increasing $\xi$ values. For the benchmark point $\alpha=0.13$ and $M=2$ TeV, bound-state effects are crucial to attain a DM energy density compatible with the observed value or, at least, to avoid overclosing the universe.

The right panel of figure~\ref{fig:Omega_versus_xi} offers a  complementary perspective. Here, we consider the ratio between the DM energy density as obtained by including or omitting the bound-state formation and decays (\ie the second term on the right-hand side of \Eq\eqref{cross_section_eff_BE}), in addition to the Sommerfeld factors for the pair annihilation. An interesting trend can be observed. Bound-state formation becomes relatively more important by decreasing $\alpha_5/\alpha$ ratios, see the orange-dashed, pink-dotted, and dot-dashed-brown curves. This effect is traced back to the different dependence on the scalar and pseudo-scalar couplings of the annihilation cross section of the scattering states $\sigma_{\textrm{ann}} \, v_{\textrm{rel}}$, see \Eq\eqref{ann_cross_pNRY5} and the matching coefficients \Eqs\eqref{match_coeff_1}-\eqref{match_coeff_4}, with respect to the bound-state formation cross section  $ \sigma^{\hbox{\scriptsize 1S}}_{\textrm{\tiny BSF}}\vrel $ in \Eq\eqref{sigma_bsf_1S}. On the one hand, $\sigma_{\textrm{ann}} \, v_{\textrm{rel}}$ depends on $\alpha_5$, which triggers important velocity-independent contributions. On the other hand, the bound-state formation is independent of $\alpha_5$ at the accuracy we have computed such a process in this work (the dominant contribution to the bound-state formation is induced by the scalar coupling for $g_5< g$). However, such a trend gets inverted for very small $\alpha_5$. This is because the bound-state effects also depend on $\alpha_5$ through the decay width $\Gamma^{\hbox{\scriptsize 1S}}_{\textrm{ann}}$, and the factor $\Gamma^{\hbox{\scriptsize 1S}}_{\textrm{ann}}/(\Gamma^{\hbox{\scriptsize 1S}}_{\textrm{ann}}+\gammaBSD^{\hbox{\scriptsize 1S}})$ entering the effective cross section in \Eq\eqref{cross_section_eff_BE}. By taking very small pseudo-scalar couplings, the efficiency factor $\Gamma^{\hbox{\scriptsize 1S}}_{\textrm{ann}}/(\Gamma^{\hbox{\scriptsize 1S}}_{\textrm{ann}}+\gammaBSD^{\hbox{\scriptsize 1S}})$ decreases sensibly and, for $\alpha_5/\alpha=10^{-4}$, the bound-state decay is rather suppressed (see the dashed-red curve in figure~\ref{fig:Omega_versus_xi}).\footnote{In the limit $\alpha_5 \to 0$ the ground state cannot decay and bound-state effects, at the accuracy we are treating them, would be absent, namely $\Gamma^{\hbox{\tiny 1S}}_{\textrm{ann}}/(\Gamma^{\hbox{\tiny 1S}}_{\textrm{ann}}+\gammaBSD^{\hbox{\tiny 1S}}) \to 0$ in \Eq\eqref{cross_section_eff_BE}. The ground state decay width could be made independent of $\alpha_5$ if one included three-body decays, such as diagrams with three scalar vertices or with a three-scalar-field vertex $\rho_\phi \phi^3$.} The relative importance of bound-state effects with respect to the Sommerfeld factors is largely independent of the DM mass (at least in the mass range relevant for this work), that we have fixed to $M=2$ TeV in the exemplary case shown in figure~\ref{fig:Omega_versus_xi}, right panel.

In our estimation of the pair annihilation via bound states, both for the DM energy density and indirect detection, we include the ground state only, namely  1S$=|100\rangle$.  In light of recent findings on the relevance of excited states for DM freeze-out \cite{Garny:2021qsr,Binder:2021vfo,Binder:2023ckj}, some comments are in order. These works focus on models with vector mediators from unbroken gauge symmetries. In this class of models, they find a significant logarithmic enhancement of the total bound state formation cross section if a large number of Coulombic states is included \cite{Binder:2023ckj}. This is caused in part by the proliferation of bound states with $l\leq n-1$ for large $n$. In this work, we are interested in the case of a scalar mediator with a finite mass. As we are dealing with a Yukawa potential, which supports only a finite number of bound states, and not with the Coulombic case, a strong enhancement from summing BSF cross sections up to large $n$ is not expected here\footnote{Requiring the binding energy of the Coulombic bound state to exceed the mediator mass leads to a conservative upper limit on the number bound states that can be produced at all velocities. In the parameter space considered in our phenomenological analysis, this yields $n\leq 22$.}. Nevertheless, we have estimated the correction from higher states to the effective cross section in eq.~\eqref{cross_section_eff_BE} by including, in the Coulomb limit, the states with $n=2$, namely $2S$ and $2P$ for four additional bound states ($P$ states with $l=1$ have three polarizations of the magnetic quantum number). Varying the various parameters of the model, i.e.~couplings and masses, we found that the corrections to the relic density are always below $10\%$. The dominant contribution comes from the 2S state because of two features of the model under study. First, the efficiency factor $\Gamma_{\textrm{ann}}^n/(\Gamma_{\textrm{ann}}^n+\gammaBSD^n)$ for $2P$ states is fairly smaller than that of the $2S$ state. This is due to the different scaling of the corresponding decay width and dissociation rate with the couplings $\alpha$ and $\alpha_5$. Second, the ultrasoft vertices of pNRY$_{\gamma^5}$ only allow for $\Delta l =0$ and $\Delta l =2$ transitions~\cite{Oncala:2018bvl,Biondini:2021ccr,Biondini:2021ycj}. Thus the $\Delta l =1$ transitions for $2P$ states to $2S$ state induced by the electric-dipole transitions of  a vector-like mediator are absent, and $2P$ states cannot be reprocessed into the more effectively decaying $2S$ state. A more in-depth study of the tower of bound states for the scalar mediator model goes beyond the scope of our work.


\section{Indirect detection}
\label{sec:indirectdetection}

As discussed previously, the rate of dark matter annihilations in the late universe is too low to affect the total dark matter abundance in a meaningful way.  Nevertheless, the effects of such residual annihilations on the flux of cosmic rays can be significant. This opens a powerful method to test annihilating dark matter known as indirect detection. In this work, we are chiefly interested in the ability of current and future gamma-ray telescopes to test our model. Concretely, we will focus on the existing bounds that can be derived from Fermi-LAT observations of dwarf spheroidal galaxies (dSphs) and the prospects for CTA. Note, that due to the limited field of view of Imaging Air Cherenkov Telescopes a different observation strategy is anticipated for CTA, and the best prospects are expected for observations of the center of the Milky Way, Galactic Center (GC) in short.  In addition, we also consider bounds on the energy injection around the era of CMB decoupling that have been derived by the Planck collaboration. Thus, we need to model DM annihilations in three distinct astrophysical/cosmological environments: the CMB era, dSphs, and the GC.

\begin{figure}
    \centering
    \includegraphics[width=0.8\textwidth]{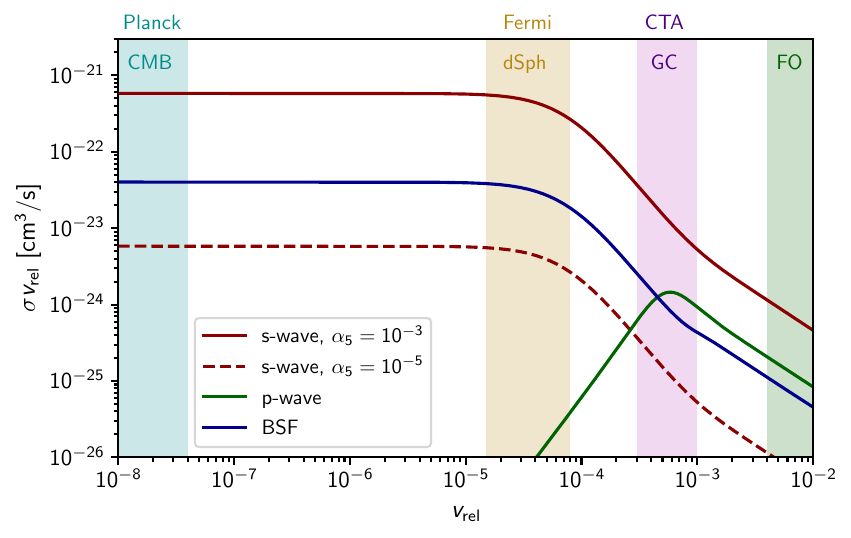}
    \caption{Leading order contributions to the DM annihilation cross section from $s$-wave and $p$-wave annihilation as well as BSF as a function of $\vrel$ for fixed $M=\SI{1}{\TeV}$, $m_\phi= \SI{1}{\GeV}$, $\alpha=0.1$ and different pseudo-scalar couplings $\alpha_5$ (note, that for the $p$-wave / BSF cross section, a change in $\alpha_5$ plays a subdominant / no role and is therefore not depicted). We also shaded the regions of typical DM velocities during freeze-out (FO), at times of the CMB decoupling as well as for dSph and the GC.}
    \label{fig:crossectionvrel}
\end{figure}

All these systems have in common that the typical velocities are much smaller than during freeze-out. However, there still remains a huge variation in their characteristic velocities with the CMB being sensitive to $\vrel=\mathcal{O}(10^{-8})$, while dSphs feature $\vrel=\mathcal{O}(10^{-5})$ and the Galactic Center even has $\vrel=\mathcal{O}(10^{-3})$. An illustration is shown in figure~\ref{fig:crossectionvrel}, where we depict the DM annihilation and BSF cross sections as a function of velocity for a representative set of parameters. We have highlighted the ranges of velocity that matter for the different environments for comparison. As can be seen, all our cross sections feature strong non-trivial velocity dependencies and the relative importance of annihilations in these systems has to be analyzed with care. Nevertheless, a few qualitative statements can be already made here. First off, since the velocities are significantly smaller than during freeze-out the velocity expansion of the cross sections can be truncated after the leading contribution. Therefore, $s$- and $p$-wave refer to the leading piece of the $l=0$ and $l=1$ cross sections in the following. Typically, the $s$-wave cross section for scattering states and the bound-state formation cross section are enhanced in the GC and dSphs compared to freeze-out since the velocities are smaller. Provided $m_\phi \ll \vrel M$, the same holds for the $p$-wave process since the Sommerfeld factor compensates for the velocity suppression usually associated with higher partial waves in this regime. Finally, the CMB is chiefly sensitive to the $s$-wave and bound state cross section since these tend to a constant value for small $\vrel$ whereas the $p$-wave cross section is suppressed at CMB velocities for all masses and couplings considered here. 

To make connection with observations one needs to refine the standard treatment to deal with the strong velocity dependence of the cross sections. In the case of the CMB, this is relatively simple since the universe is still highly homogeneous at that point. Thus, the rates can be determined by a simple thermal average as discussed in the previous section. However, the situation is more involved in the GC and in dSphs. Therefore, the following discussion concentrates on the signal at gamma-ray telescopes from these sources.

The key observable in gamma-ray telescopes is the photon flux from the target of interest. The contribution of DM annihilation to the differential photon flux is in general given by (\eg ref.~\cite{Ferrer:2013cla})
\begin{align}
    \dv{\Phi_\gamma}{E_\gamma}&=\frac{1}{16\pi M^2}\dv{N_\gamma}{E_\gamma}  \int_{\Delta\Omega}\dd\Omega\int_0^\infty\dd\psi \int \dd[3]v_1\int\dd[3]v_2 f_X(r(\psi,\Omega),\vec{v}_1) f_X(r(\psi,\Omega),\vec{v}_2) \, \sigma\vrel,
\end{align} 
for Dirac DM. Let us go through this expression piece by piece. The object $\dv{N_\gamma}{E_\gamma}$ describes the number of photons per energy per annihilation. If more annihilation channels are open it can be constructed from the photon numbers per energy of the different channels weighted by the branching fractions $B_f$ into the channels, \ie $\dv{N_\gamma}{E_\gamma}=\sum_f B_f\dv{N^{(f)}_\gamma}{E_\gamma}$ where the sum runs over all possible final states $f$. The first two integrals run over the observed solid angle $\Omega$ and the distance along the line of sight (l.o.s.) $\psi$. Finally, there are two integrals over the velocities of the two annihilating DM particles. The functions to be integrated consist of the product of the dark matter phase space distribution functions (DF) $f_X$ for the two particles and the in general velocity-dependent DM annihilation cross section $\sigma \vrel$. If $\sigma \vrel$ is independent of the  DM velocity, \eg for the leading contribution to $s$-wave annihilation without Sommerfeld enhancement, it can be pulled out of the integrals, and with
\begin{align}
    \int \dd[3]{v} f_X(\vec{r},\vec{v})=\rho_X(\vec{r})
\end{align}
one obtains the usual $J$-factor 
\begin{align}
    J_0=\int\dd{\Omega}\int\dd{\psi}\rho_X(\vec{r}(\psi,\Omega))^2\,.
\end{align}
In the following, we take the dark matter distribution in the dSphs and around the GC to be spherically symmetric and use $r=\abs{\vec{r}}$  to denote the distance from the center of the object under consideration. When non-perturbative effects are considered (or $p$-wave annihilations play a role), the velocity dependence of $\sigma \vrel$ is more complex and we need to perform the full velocity average. This requires knowledge of the DF. In the following, we assume that we can split up the cross section as 
\begin{equation}
    \sigma \vrel=\sum_{l,j=0}^{\infty}\sigmaann^{(l,j)}\vrel^{2(l+j)}\Sann^{(l)}(\zeta,\xi)+\sum_{n>l}\sigmaBSF^{(n,l)}\SBSF^{(n,l)}(\zeta,\xi)\,.
    \label{xsection_ID}
\end{equation} 
Ionization is negligible at late times such that the effective factor for BSF in \Eq\eqref{cross_section_eff_BE} can be neglected here. This motivates the definition of a generalized $J$-factor
\begin{equation}
    \label{Jsann}
    J_{\text{ann},\alpha}^{(l,j)}(\xi)\equiv\int_{\Delta\Omega}\dd\Omega\int_0^\infty\dd\psi \int \dd[3]v_1f_X(r(\psi,\Omega),\vec{v}_1)\int\dd[3]v_2f_X(r(\psi,\Omega),\vec{v}_2)\vrel^{2(l+j)}\Sann^{(l)}(\alpha/\vrel,\xi)
\end{equation}
and analogous to BSF. Regarding the annihilation cross section, we are mainly interested in the dominant $s$- and $p$-wave contributions which we identify as $\sigmaanns\equiv \sigmaann^{(0,0)} = {\rm{Im}}[f(^1S_0)]/M^2$ and $\sigmaannp\equiv \sigmaann^{(1,0)} =({\rm{Im}}[f(^3P_0)]+5{\rm{Im}}[f(^3P_2)])/12 M^2$. For the rest of the section, we limit ourselves to  BSF into the ground state  $\sigmaBSF\equiv \sigmaBSF^{(1,0)}=\pi \alpha^4/M^2$ and drop the labels $(n,l)$ in the following. Thus, the DM-induced photon flux can be compactly written  as
\begin{equation}
    \label{photonflux_Jfactors}
    \dv{\Phi_\gamma}{E_\gamma}=\frac{1}{16\pi M^2}\left(\sum_f B_f\dv{N^{(f)}_\gamma}{E_\gamma}\right)\left(\sigmaanns J_{\text{ann},\alpha}^{(0,0)}(\xi) +\sigmaannp J_{\text{ann},\alpha}^{(1,0)}(\xi)+\sigmaBSF J_{\text{\tiny BSF},\alpha}(\xi)\right).
\end{equation}
To make further progress here we need to determine the generalized $J$-factors and the photon number per energy per annihilation. 


\subsection[Distribution function $f_{X}$]{\protect\boldmath Distribution function $f_{X}$}
\label{subsec:fDM}

Unfortunately, the distribution function of dark matter in phase space is not directly accessible to observations. Therefore, additional information and/or approximations are required to make progress here. In the following, we use methods for the analysis of galactic dynamics that hold in collisionless systems in a quasi-equilibrium state, see ref.~\cite{Binneybook} for detailed discussion and derivation, and refs.~\cite{Ferrer:2013cla,Boddy:2017vpe,Lacroix:2018qqh} for earlier applications to DM halos. An alternative possibility which is not considered here is to sample $f_X$ directly from high-resolution numerical simulations of dark matter halos \cite{Board:2021bwj,McKeown:2021sob}.

We assume that we are dealing with isolated systems such that the gravitational potential $\Psi$ can be determined from the dark and baryonic matter distribution via a Poisson's equation from the local density
\begin{align}
    \Delta \Psi = - 4 \pi G \rho \, ,
\end{align}
where $G$ is Newton's constant and  $\rho$ the total matter density. Treating the DM halo as a spherically symmetric system of collisionless particles with density given by $\rho_X$, one can use the Eddington inversion method \cite{Binneybook} to obtain a unique and ergodic DF
\begin{equation}
    \label{EddingtonEquation}
    f_X(\epsilon)=\frac{1}{\sqrt{8}\pi^2}\left(\int_0^\epsilon\dv[2]{\rho_X}{\Psi}\frac{\dd{\Psi}}{\sqrt{\epsilon-\Psi}}-\frac{1}{\sqrt{-\epsilon}}\evalat{\dv{\rho_X}{\Psi}}{\Psi=0}\right)
\end{equation}
supported by this potential. Here $\epsilon=\Psi(r)-v^2/2$ is the energy per unit mass of the DM  in the gravitational potential. Setting the relative gravitational potential $\Psi(r)$ to $0$ at $r\to\infty$ removes the second term. Technically, this boundary condition is unphysical since the gravitational potential of a galaxy is naturally bounded by its neighbors. However, it was shown in ref.~\cite{Lacroix:2018qqh} that the effect of a large but finite boundary compared to an infinite one is rather small such that we will neglect it. Since $\Psi$ is a monotonic function in $r$, one can rewrite the Eddington equation directly as a function of $r$ and $v$ \cite{Ferrer:2013cla}
\begin{equation}
    \label{DF}
    f_X(r,v)=\frac{1}{\sqrt{8}\pi^2} \int_{r_{\text{min}}}^\infty \frac{\dd{r'}}{\sqrt{\Psi(r)-\Psi(r')-\frac{v^2}{2}}}\left(\dv{\Psi}{{r'}}\right)^{-1}\left[\dv[2]{\rho_X}{{r'}}-\left(\dv{\Psi}{{r'}}\right)^{-1}\dv[2]{\Psi}{{r'}}\dv{\rho_X}{{r'}}\right]
\end{equation}
where $r_{\text{min}}$ is given by $\Psi(r)-\Psi(r_{\text{min}})-\frac{v^2}{2}=0$.

Different functional forms for the dark matter halo are considered in the literature. In the following, we will use the NFW-profile \cite{Navarro:1995iw,Navarro:1996gj,Moore:1999nt} as our ansatz for dSphs. It is given by
\begin{equation}
    \rho_{\text{\tiny NFW}}= \frac{\rho_0}{\frac{r}{r_s}\left(1+\frac{r}{r_s}\right)^2}
\end{equation}
where $r_s$ is the scale radius, which corresponds to $\evalat{\dv{\log \rho_X}{\log r}}{r=r_s}=-2$, and $\rho_0$ fixes the normalization. For our analysis of classical dSphs, we use the best-fit values for $r_s$ and $\rho_0$ as reported in ref.~\cite{Boddy:2017vpe}. To stay as close as possible to the CTA sensitivity study for the Galactic Center, we model the halo with an Einasto profile in this case \cite{CTA:2020qlo}. It reads
\begin{equation}
    \label{eq:Einastoprofile}
    \rho_{\text{\tiny Ein}}= \rho_0 \,\mbox{Exp}\left[-\frac{2}{\gamma}\left[\left(\frac{r}{r_s}\right)^\gamma-1\right]\right]
\end{equation}
with $\gamma=0.17$. The associated gravitational potentials for both profiles can be computed analytically, see \eg ref.~\cite{Ferrer:2013cla}. It is more convenient to switch to dimensionless coordinates in the following. We use $x\equiv r/r_s$ and the scale mass  $M_s\equiv 4\pi\rho_0r_s^3 h$ with $h\equiv \int_0^1 \dd{x} x^2\rhochit(x)$ to define
\begin{equation}
    \rhochit(x)\equiv \frac{4\pi r_s^3}{M_s} h \rho_X(r),\quad \psit(x)\equiv\frac{r_s}{GM_s }\Psi(r),\quad \et(x,\vt)\equiv\frac{r_s}{G M_s }\epsilon(r,v)
\end{equation}
with $\vt\equiv \sqrt{\frac{r_s}{G M_s}} v$ (such that $\et=\psit-\vt^2/2$). The dimensionless DF follows 
\begin{equation}
    \label{ftildeDM}
    \fchit(x,\vt)=\int_{x_{\text{min}}}^\infty \frac{\dd{x'}}{\sqrt{\psit(x)-\psit(x')-\frac{\vt^2}{2}}}\left(\dv{\psit}{x'}\right)^{-1}\left[\dv[2]{\rhochit}{{x'}}-\left(\dv{\psit}{x'}\right)^{-1}\dv[2]{\psit}{{x'}}\dv{\rhochit}{{x'}}\right] \, ,
\end{equation}
with $x_{\text{min}}$ defined analogously to  $r_{\text{min}}$ and relates to the DF defined in \Eq\eqref{DF} via
\begin{equation}
    f_X(r,v)=\frac{1}{8\pi^3h\sqrt{2G^3 r_s^3 M_s}}\fchit(x,\vt).
\end{equation}
To compute the averages over the DFs we normalize them by the DM density and find position-dependent velocity distributions as follows
\begin{equation}
    P_r(v)\dd[3]v\equiv\frac{f_X(r,v)}{\rho_X(r)}\dd[3]v=\frac{1}{\sqrt{8}\pi^2}\frac{\fchit(x,\vt)}{\rhochit(x)}\dd[3]{\vt}\equiv P_x(\vt)\dd[3]{\vt} \, ,
\end{equation}
that are normalized to unity, \ie $4\pi\int_0^{\vesc}\dd{v} v^2P_r(v)=4\pi\int_0^{\vesct} \dd{v} \vt^2P_x(\vt)=1$ where $\vesc=\sqrt{2\Psi(r)}$ (and analogously $\vesct$) is the escape velocity.

For the dSphs, baryonic contributions to the gravitational potential can be neglected ($\Psi_{\text{\tiny dSph}}=\Psi_X$). However, in the  GC baryons contribute significantly to the potential and have to be taken into account ($\Psi_{\text{\tiny GC}}=\Psi_X+\Psi_b$). We will restrict ourselves in the following to model the baryonic bulge and stellar disk. To apply Eddingtons approach, we need the gravitational potential induced by baryons $\Psi_b$ to be spherically symmetric. This is clearly not the case (especially not for the disk). Following ref.~\cite{Ferrer:2013cla}, we use symmetrized models  
\begin{equation}
    \psit_{\text{bulge}}(x)=\frac{M_{\text{bulge}}}{M_s}\frac{1}{x+c_0/r_s},\qquad \psit_{\text{disk}}(x)=\frac{M_{\text{disk}}}{M_s}\frac{1-\exp\left[-\frac{r_s x}{b_{\text{disk}}}\right]}{x} \, ,
\end{equation}
where $M_{\text{bulge}}=\SI{1.5e10}{}M_\odot$ and $M_{\text{disk}}=\SI{5e10}{}M_\odot$ are the bulge and disk masses respectively, while $b_{\text{disk}}=\SI{4}{\kpc}$, $c_0=\SI{0.6}{\kpc}$ are model parameters that encode the spatial extent of these objects. Both are constructed to enclose the same mass as the flattened profiles.


\subsection[Generalized $J_s$ factors]{\protect\boldmath Generalized $J_s$ factors}
\label{subsec:Jsfactor}

We now turn to the effective $J$-factors as defined in \Eq\eqref{Jsann}. Three of the velocity integrals can be performed directly and only the magnitude of the velocities $v_1$ and $v_2$ and the angle $\phi$ between the annihilating particles remain. Since the integrand only depends on the relative velocity, it is convenient to trade this for the magnitude of the center of mass velocity $\vcm=|(\vec{v}_1+\vec{v}_2)/2|$, the magnitude of the relative velocity, $\vrel=|\vec{v}_1-\vec{v}_2|$ and $z=\cos\theta$, where $\theta$ denotes the angle between the two velocities.

Isolating the integration over $\vrel$ one gets
\begin{align}
     J_{\text{ann},\alpha}^{(l,j)}(\xi)=
     \int_{\Delta\Omega}\dd\Omega\int_0^\infty\dd\psi \rho_X(x(\psi,\Omega))^2K^{-2(l+j)}\int_0^{2\vesct}\dd{\vrelt}P_{x,\text{rel}}(\vrelt)\vrelt^{2(l+j)}\Sann^{(l)}(K\alpha/\vrelt,\xi)
\end{align}
where we introduced a unit conversion factor $K= \sqrt{\frac{r_s c^2}{G M_s}}$, and  restored factors of $c$ that are implicit in the computation of $\sigma \vrel$ in natural units. The integration over $\vcmt$ and $z$ is encapsulated in 
\begin{align}
    \label{Pxrel}
    P_{x,\text{rel}}(\vrelt)\equiv&\frac{2\vrelt^2}{\pi^2}\left\{\int_0^{\vesct-\frac{\vrelt}{2}}\dd{\vcmt}\vcmt^2\int_0^1\dd{z}+\int_{\vesct-\frac{\vrelt}{2}}^{\sqrt{\vesct^2-\frac{\vrelt^2}{4}}}\dd{\vcmt}\vcmt^2\int_0^{z_0}\dd{z}\right\}\nonumber\\
    & P_X(x,\sqrt{\vcmt^2+\vrelt^2/4+\vcmt\vrelt z})P_X(x,\sqrt{\vcmt^2+\vrelt^2/4-\vcmt\vrelt z})
\end{align}
where the limits of integration are set by the escape velocity and the kinematics with $z_0\equiv \frac{\vesct^2-\vcmt^2-\vrelt^2/4}{\vcmt\vrelt}$. In analogy to the thermal average, we define a velocity-averaged Sommerfeld factor. In a slight abuse of notation, we write it as 
\begin{equation}
    \langle\Sann^{(l,j)}(\zeta,\xi)\rangle(x)\equiv K^{-2(l+j)}\int_0^{2\vesct}\dd{\vrelt}P_{x,\text{rel}}(\vrelt)\vrelt^{2(l+j)}\Sann^{(l)}(K\alpha/\vrelt,\xi)
\end{equation}
which approaches unity for a decoupled mediator (\ie no SE) for the $s$-wave contribution as it should. Thus, we get
\begin{equation}
     J_{\text{ann},\alpha}^{(l,j)}(\xi)=\int_{\Delta\Omega}\dd\Omega\int_0^\infty\dd\psi \rho_X(x(\psi,\Omega))^2\langle\Sann^{(l,j)}(\zeta,\xi)\rangle(x)
\end{equation}
and analogously for $ J_{\text{\tiny BSF},\alpha}(\xi)$.

\begin{figure}
    \centering
    \includegraphics[width=0.8\textwidth]{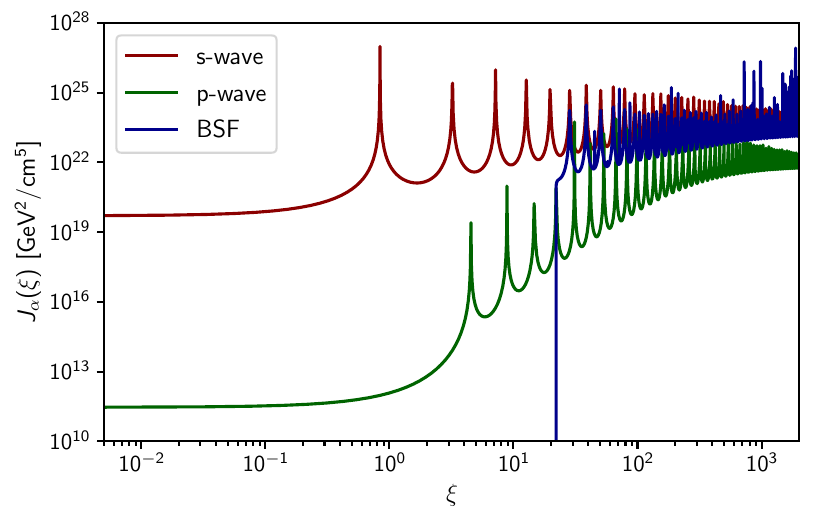}
    \includegraphics[width=0.8\textwidth]{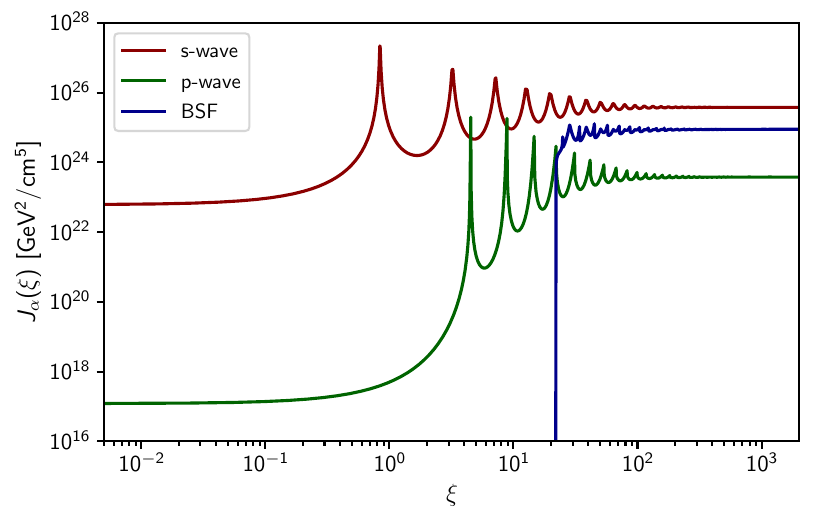}
    \caption{Effective $J$-factors for leading order $\vrel^0$ $s$-wave contribution $(l,m)=(0,0)$ (red), the leading order $\vrel^2$ $p$-wave annihilation $(l,m)=(1,0)$ (green) and BSF into the ground state $(n,l)=(1,0)$ as a function of $\xi$ for $\alpha=0.1$. 
    \textit{Top:} For the exemplary dSph Segue I,
    \textit{Bottom:}  For the Galactic Center ROI as defined in the text.}
    \label{fig:$J$-factors}
\end{figure}

This leaves the integration over the spatial coordinates. For dSph, we can safely assume that our solid angle $\Delta\Omega$ is large enough to cover the entire region where DM annihilation is relevant and exploit that the distance $D$ between us and the source is much bigger than its extent. Thus, we can change coordinates and integrate over a sphere centered around the considered dSph, \ie $\int\dd{\Omega}\int\dd{\psi}\to 4\pi/D^2\int r^2\dd{r}$. We consider the classical dwarfs Coma Berenices, Ursa Minor, Draco, and Sergue 1, and parameterize the halos as  NFW profiles. The parameters are taken from table 1 of ref.~\cite{Boddy:2017vpe}. We calculate the enhanced $J$-factors and present an exemplary set in figure~\ref{fig:$J$-factors} (top). As expected, the $J$-factors inherit the resonance structure of the Sommerfeld factors and we find a series of peaks once a smooth initial stage has been passed.  For large values of $\xi$ the enhancement is very pronounced for all contributions and the hierarchy between the $s$-wave and the $p$-wave is reduced considerably. For $m_\phi \lesssim E_b$  also BSF plays a role and the BSF $J$-factor rises swiftly with increasing $\xi$. It clearly exceeds the $p$-wave factor and is only mildly suppressed compared to the $s$-wave. Note, however, that due to the different dependence of the split-off prefactors on the couplings and the masses, this information is not sufficient to judge the relative importance of the different contributions for the present-day annihilation rates. 

For the GC, we take an Einasto profile and use the halo parameters $r_s= \SI{20}{\kpc}$, $\rho_s=\SI{0.081}{\GeV/\cm^3}$ as suggested by a dedicated prospect analysis for CTA \cite{CTA:2020qlo}. We consider the same region of interest (ROI) $l,b\in[-6^{\circ},6^{\circ}] $ where $l, b$ denote the galactic longitude and latitude, respectively, and use $D_{\odot}=\SI{8,5}{\kpc}$ as the distance of the Galactic Center to our sun.

We calculate the enhanced $J$-factors for the Galactic Center and present them in figure~\ref{fig:$J$-factors} (bottom) for the main contributions of s- and $p$-wave annihilation as well as BSF.  The broad structure of the $J$-factors is similar to the dSph case. However, as expected, the absolute values are larger since the Galactic Center exhibits the largest and closest concentration of the DM. It is also worth noting, that the resonance structure is a bit more washed out due to the higher average velocities compared to the dSphs. With the $J$-factors taken care of, we can now turn to the next piece needed to determine the gamma-ray flux.


\subsection{Photon spectrum}
\label{subsec:photonspectrum}

The photon spectrum $\dd N_\gamma/\dd E_\gamma$ accounts for the number of photons $N_\gamma$ produced by annihilating or decaying DM per photon energy $E_\gamma$. In our scenario, the annihilations produce pairs of scalars $\phi$, which then decay to SM particles via mixing with the Higgs. These SM particles in turn undergo decay and hadronization until only particles that are stable over astrophysical distances remain in the spectrum. Therefore, we need three ingredients to determine the final photon spectrum: i) the branching ratios of $\phi$ into the various SM final states, ii) the photon spectrum produced by a given SM final state, and iii) a prescription to take into account the large boost of the decaying $\phi$'s in the galactic rest frame.

In the following, we will collect the relevant branching ratios for the decay of the mediator into pairs of SM particles and determine the photon spectra for cascaded DM annihilation within two mass regimes for the mediator, namely $m_\phi \geq \SI{10}{\GeV}$ and $ 2 m_{\pi^0} \leq m_\phi \leq \, \SI{1}{\GeV}$, which we will refer to as the high and low mediator mass range. 


\subsubsection[Branching ratios for $\phi$]{\protect\boldmath Branching ratios for $\phi$}
\label{sec:BR_phi}

\begin{figure}
    \centering
    \includegraphics[width=0.8\textwidth]{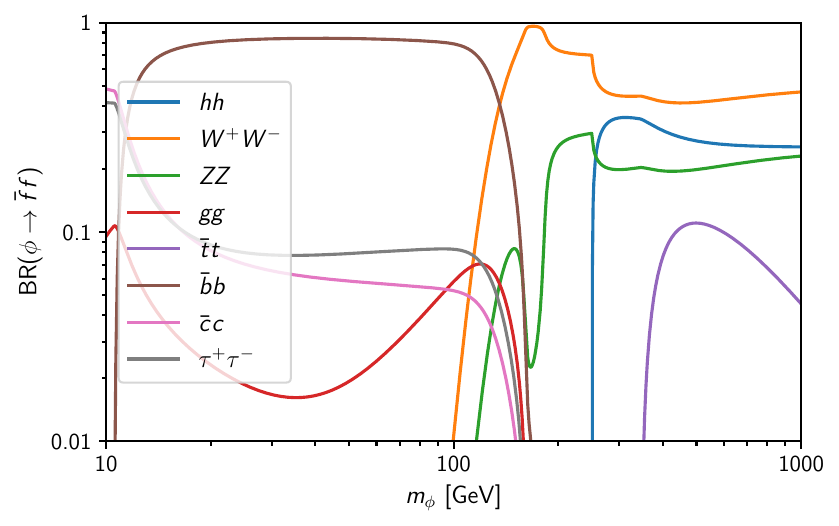}
    \caption{Branching ratios of the mediator $\phi$ decaying into SM particles as a function of the mediator mass. Assuming the mixing angle between the mediator and the Higgs is small enough, these are equivalent to the Branching ratios of a Higgs with arbitrary mass with the exception of the $\phi\to hh$ channel.}
    \label{fig:mediator_decay}
\end{figure}

The Higgs portal term in \Eq\eqref{L_portal} leads to a mixing between the mediator $\phi$ and the Higgs. From the point of indirect detection, the absolute value of the mixing is irrelevant, since even very small values lead to a decay of $\phi$ that is instantaneous on astrophysical scales. Thus, the only question is what $\phi$ decays into. For $m_\phi\lesssim 2 m_h$, the branching ratio of the mediator into SM particles equals the one of a Higgs with the same mass. For $m_\phi>2m_h$ the decay channel of the mediator in pairs of Higgs particles opens. The contribution of this channel to the total decay width modifies the branching rations slightly in this regime and we explicitly take it into account. We compute the branching ratios of a hypothetical Higgs with  $m_h \gtrsim\SI{2}{\GeV}$ using the expressions compiled in \cite{Djouadi:2005gi}. The running of couplings and masses is taken from \cite{Chetyrkin:2000yt}. Above the threshold, the additional decay rate into two Higgs particles is given by 
\begin{equation}
	\Gamma_{\phi\to hh}=\frac{\left(m_\phi^2+2m_h^2\right)^2}{32\pi m_\phi v^2}\sqrt{1-\frac{4m_h^2}{m_\phi^2}}\sin^2\delta
\end{equation}
where
\begin{equation}
    \label{eq:sindeltaapprox}
 \sin\delta\approx \frac{\mu_{\phi h} v}{|m_h^2 -m_\phi^2|}\,.
\end{equation}
The BR of the $\phi\to hh$ decay becomes comparable to the $W$ and $Z$ contributions above the mass threshold. A display of the branching ratios of all major decay channels within the whole mediator mass range is given in figure~\ref{fig:mediator_decay}. For the analysis of the high mediator mass range, we consider the decay channels $\phi\to\{hh, W^+W^-, ZZ, gg, \bar{t}t, \bar{b}b, \bar{c}c, \tau^+\tau^-\}$.

This method breaks down close to the confinement scale. For  $m_\phi\lesssim\SI{2}{\GeV}$ we therefore use the branching ratios derived by \cite{Winkler:2018qyg}. It is interesting to note here, that in this regime decays into pion pairs play a large role and completely dominate the total width below the kaon threshold. Since $\pi^0$s decay to a photon pair with a branching ratio of $99\%$, a very large fraction of the energy goes directly to hard gamma rays in this mass range. Hence, we focus on this decay channel here and neglect photons produced in other decays of light mediators. We will only be interested in the continuum flux in setting the limits or prospects and, therefore, this approach is conservative.


\subsubsection{Cascade annihilations}
\label{sec:cascadeannihilation}

We are interested in the photon spectrum produced by DM annihilations. In our case, this process can be broken into two pieces: the annihilation into pairs of $\phi$ and the decay of $\phi$ into SM particles. For center of mass energies of the SM pair, \ie $\phi$ masses, larger than a few GeV, the second part can be solved with the help of event generators for high-energy physics collision such as Pythia \cite{Bierlich:2022pfr} or Herwig \cite{Bellm:2015jjp}. Tabulated photon spectra as a function of the center of mass energy $E_{\text{\tiny{CM}}}$ for all relevant SM final states have been published in \eg  \cite{Belanger:2001fz,Cirelli:2010xx,Bringmann:2018lay}. In the following we use the results of \cite{Cirelli:2010xx}. For $m_\phi\leq \mbox{a few  GeV}$ this approach is unreliable since non-perturbative effects in QCD become large. There have been attempts to derive photon spectra for lighter vector mediators using vector meson dominance combined with data from $e^+ e^-$ to hadrons \cite{Plehn:2019jeo,Coogan:2022cdd}. Unfortunately, this is not applicable to scalar mediators and, therefore, we exclude the range $1 \, \mbox{GeV} \leq m_\phi \leq 10$ GeV from our analysis. Lighter scalars decay dominantly into mesons and thus the spectra can be constructed from those of meson decays, see below for more details.

This leaves us with the task of translating the photons spectrum computed in the $\phi$ rest frame to the galactic rest frame. For an isotropically decaying scalar, the spectrum in the center of mass frame of the annihilations is given by \cite{Mardon:2009rc}
\begin{equation}
	\dv{N_\gamma}{E_1}=\int_{-1}^{1}\dd{\cos\theta}\int_0^{m_\phi/2}\dd{E_0}\dv{N_\gamma}{E_0}\delta(E_1-E_1^*(E_0))
\end{equation}
where $E_0$ and $E_1$ denote the photon energy in the restframe of the decay and the center of mass frame of the annihilation, respectively. The $\delta$ function incorporates the Lorentz boost constraint $E_1^*(E_0)=E_0/\epsilon_\phi\,(1+\sqrt{1-\epsilon_\phi^2}\cos\theta)$ with $\theta$ the angle between the direction of the boost and the direction of the photon in the scalar rest frame and $\epsilon_\phi=m_\phi/M$.

Mediators in a mass range $2 m_{\pi^0} \leq  m_\phi \leq 1\, \si{\GeV}$ decay predominantly into pions which makes the derivation of the photon spectrum easy.\footnote{Even lighter mediators decay into electron-positron pairs and the most stringent limits come from the positron flux measured AMS-02 \cite{AMS:2015tnn}, see \eg\cite{Elor:2015bho} for a DM interpretation. We do not include these final states since we want to focus on gamma rays.}  The considered process is $\bar{X}+X\to 2\phi\to 4\pi^0\to8\gamma$, which is an example of a two-step cascade annihilation and can be dealt with using the method outlined in \cite{Elor:2015tva}. Since the spectrum of pion decay, $\dd{N_{\gamma}}/\dd{x_0}=2\delta(x_0-1)$ with $x_0=2E_0/m_{\pi^0}$, is particularly simple, the cascade spectrum can be computed analytically. In the limit  $\epsilon_{\phi,\pi^0}\ll1$, where $\epsilon_{\pi^0}=2m_{\pi^0}/m_\phi$, the result is
\begin{align}
    \dv{N_\gamma}{x}=8 \ln(1/x)\,.
\end{align}
where we defined $x=E_1/M$. However, these mass ratios are not reached in practice. Therefore, the full result has to be used which reads
\begin{equation}
	\dv{N_\gamma}{x}=\frac{8}{\sqrt{1-\epsilon_{\pi^0}^2}\sqrt{1-\epsilon_\phi^2}}
	\begin{cases}
		\ln\left(x/K_+^-\right) & K_+^-\leq x<\min\{K_+^+,K_-^-\} \\
		\ln\left(\min\{K_+^+,K_-^-\} /K_+^-\right) & \min\{K_+^+,K_-^-\}\leq x < \max\{K_+^+,K_-^-\} \\
		\ln\left(K_-^+/x\right) &  \max\{K_+^+,K_-^-\} \leq x<K_-^+\\
	\end{cases}
\end{equation}
and $0$ elsewhere, with
\begin{equation}
	K_\pm^\pm=\frac{\left(1\pm\sqrt{1-\epsilon_{\pi^0}^2}\right)\epsilon_\phi^2}{4\left(1\pm\sqrt{1-\epsilon_\phi^2}\right)}.
\end{equation}
An interesting feature of this spectrum is that it is considerably harder than the power spectra expected for most astrophysical backgrounds. Therefore, the signal is most significant at relatively high energies and has the potential to induce a spectral feature that can boost the sensitivity in a dedicated analysis \cite{Bringmann:2011ye,Bringmann:2012vr}. We do not pursue this idea further here and rely on results from a strategy for continuum spectra.


\subsection{Current bounds and prospects}
\label{sec:boundsandprospects}

We now make connections with observations and interpret bounds derived by the Fermi-LAT and Planck collaborations in our scenario. Prospects for CTA are then included in a similar way.
\begin{figure}
    \centering
    \includegraphics[width=0.8\textwidth]{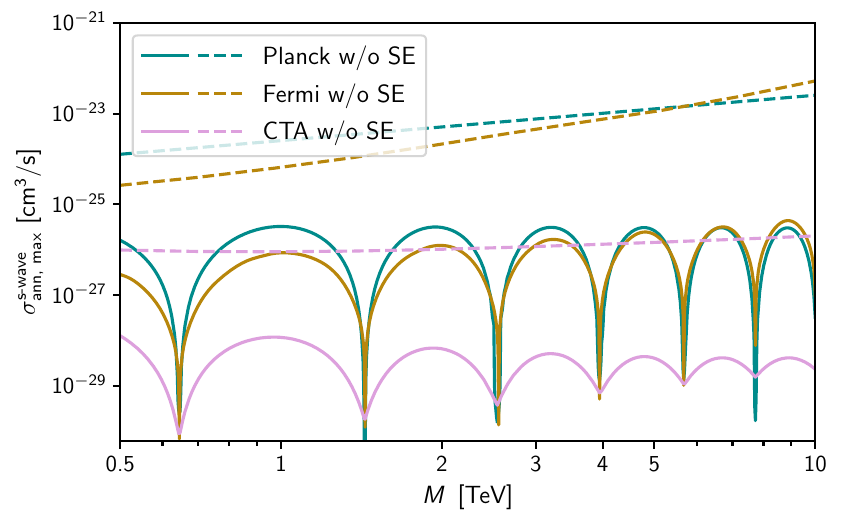}
    \includegraphics[width=0.8\textwidth]{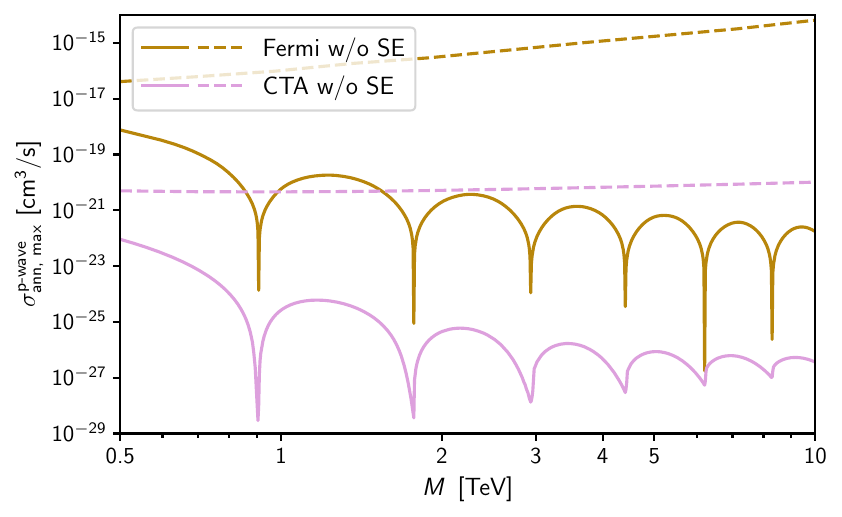}
    \caption{Indirect detection limits on the $s$-wave (top) and $p$-wave (bottom) dominated DM annihilation scenarios as a function of the DM mass $M$ for a benchmark point of $m_\phi=\SI{10}{\GeV}$ and $\alpha=0.1$. Turquoise lines correspond to existing CMB limits from Planck data, golden lines from an analysis of Fermi data from dSph, whereas the pink lines depict prospects for the upcoming CTA measurement in the Galactic Center. Dotted lines give limits without SE, whereas for the solid lines SE has been taken into account.}
    \label{fig:crosssectionlimit}
\end{figure}
\vspace{0.5cm}

{\bf Fermi-LAT limits:} We derive our limits from the supplementary material of ref.~\cite{Fermi-LAT:2015att}, which is based on six years of Fermi-LAT observations of the classical dwarfs Coma Berenices, Ursa Minor, Draco, and Sergue 1. In addition to the limits for benchmark annihilation channels into pure SM final states, which are not applicable in our case, the collaboration published the energy-bin by energy-bin likelihood as a function of the photon flux \cite{Fermi_logL}. This allows to derive approximate likelihood functions for a general DM annihilation spectrum and puts an upper limit on the annihilation rate. We follow the prescription outlined by the Fermi-LAT collaboration and use their data in combination with our photon spectra to derive upper limits on the annihilation rate which can be related to limits on the model parameters using our generalized $J$-factors. As a check, we also derived limits for the benchmark cases analyzed in \cite{Fermi-LAT:2015att} following the same procedure. They are in good agreement with the official limits.\footnote{Exact agreement cannot be expected since this statistical treatment is simplified compared to the full method used by the collaboration.}

If only one annihilation channel contributes significantly to the photon flux, then we can also place a limit on the individual cross section. An illustration of such a case for a representative set of parameters is shown in figure~\ref{fig:crosssectionlimit}. As can be seen, the limit on the $s$-wave  cross section is rather modest if Sommerfeld enhancement is not included. It lies broadly in the same range as the limits on pure SM final states for these DM masses. The inclusion of the Sommerfeld effect strengthens the bounds by several orders of magnitude everywhere. As expected, the strongest impact is found around the resonances we already observed in the $J$-factors, which can push the limit even below $10^{-29}\mbox{cm}^3 \mbox{s}^{-1}$. A similar analysis for the $p$-wave case is shown in the lower panel of figure~\ref{fig:crosssectionlimit}. Here, the limits are considerably weaker due to the velocity suppression of the annihilation rate. Without Sommerfeld enhancement the limits are very far from the values relevant for a thermal relic and even with the Sommerfeld effect only the resonantly enhanced parts dip below $10^{-23}\mbox{cm}^3 \mbox{s}^{-1}$ in our example. The overall limit improves with higher masses since we are keeping $m_\phi$ fixed here which leads to stronger Sommerfeld enhancement as $M$ is increased.
\vspace{0.5cm}

{\bf Planck limits: } The CMB is sensitive to dark matter annihilations since exotic energy injection can affect the recombination history and change the optical depth of the CMB.  Concretely, the additional energetic particles lead to higher ionization levels at redshifts $z\approx600 -1000$, which modifies the CMB anisotropies and the polarization \cite{Padmanabhan:2005es,Galli:2011rz}. The most stringent limits on such a non-standard energy injection have been derived by the Planck collaboration  \cite{Planck:2018vyg}. Assuming that the redshift dependence of the energy injection rate is controlled entirely by the change of  the DM density, which implies that the cross section is velocity independent, they place an upper limit on the combination
\begin{align}
    p_{\text{ann}}=f_{\text{eff}} \frac{\langle \sigma \vrel \rangle}{M}\leq 3.5 \times 10^{-28} \mbox{cm}^3 \mbox{s}^{-1} \mbox{GeV}^{-1}
\end{align}
where $f_{\text{eff}}$ is an efficiency factor that encodes how efficiently the released energy is absorbed by the intergalactic medium. We use a conservative estimate of $f_{\text{eff}}\simeq 0.137$ for all channels based on the results of ref.~\cite{Slatyer:2015jla}. Thus, we find
\begin{equation}
    \label{CMBbound}
    \langle\sigma \vrel\rangle<\SI{2.5e-24}{\frac{\cm^3}{s}}\left(\frac{M}{\si{\TeV}}\right) \, ,
\end{equation}
where $\langle\sigma \vrel\rangle$ is the sum of all thermally averaged cross sections that scale as $\vrel^0$. In our case, this happens for the $s$-wave and the BSF cross section since the Sommerfeld and BSF factors have reached a plateau at the relevant velocities (compare figure~\ref{fig:crossectionvrel} for an illustration). In contrast, $p$-wave annihilation is highly suppressed in this regime and does not contribute. The limits on the $s$-wave cross section for a representative set of parameters are shown in figure~\ref{fig:crosssectionlimit}. As can be seen, the inclusion of the Sommerfeld effect strengthens the Planck limit on the cross section considerably and we find an improvement of several orders of magnitude throughout the mass range. Interestingly, the Planck limit turns out the be relatively similar to the one derived from Fermi-LAT observations. At high DM masses, Planck even outperforms the gamma-ray limits. This can be qualitatively understood based on the scaling of the relevant rates with $M$. Fermi-LAT is sensitive to the photon flux, which scales as $\langle \sigma v \rangle \frac{\rho^2}{M^2}$, whereas the CMB is sensitive to the energy release rate which scales as  $M \langle \sigma v \rangle \frac{\rho^2}{M^2}$ and thus has a weaker dependence on the DM mass.
\vspace{0.5cm}

{\bf CTA prospects: }
The dark matter prospects for CTA observations of the Galactic Center have been analyzed carefully by the Cherenkov Telescope Array Consortium in \cite{CTA:2020qlo}. Similar to the Fermi-LAT analysis discussed above, this publication comes with tabulated bin-by-bin likelihoods \cite{bringmann_torsten_2020_4057987}. These can be used to estimate the expected upper limit on the dark matter annihilation rate for any spectrum in a way that is analogous to the one for the Fermi-LAT dwarf limits. An illustration of the prospects that can be derived in this way is presented in figure~\ref{fig:crosssectionlimit}. As expected, CTA has the potential to outperform both Fermi-LAT and Planck limits by orders of magnitude in the range of masses considered here.

Before closing this section, we would like to comment briefly on the dependence of limits and prospects from observations of the galactic center on the halo profile. As the matter in the galactic center is dominated by the baryonic component, the $J$-factor is not constrained as well as in the case of dSphs and the interpretation of the (prospective) observations is less robust. In particular, cuspy profiled tend to predict larger fluxes compared to cored ones. In this context, we would like to point out that the Einasto profile does not necessarily fall into the category of a cuspy profile. The inner slope of the profile is controlled by the input parameter $\gamma$. With the factor $\gamma=0.17$ adopted in this work, the profile is in-between a cuspy and a cored profile. Nevertheless, the dependence of the prospects on this choice still warrants study. In order to estimate the effect of a more clearly cored profile we again follow the suggestion of the CTA-collaboration and consider an artificially cored Einasto profile \cite{CTA:2020qlo}. Here the density is taken to be constant in the inner part of the halo (below $\SI{1}{\kpc}$). Note that this ad hoc ansatz does not allow for the application of Eddington inversion and we need to determine the impact of the density profile on the velocity dependent $J$-factors differently. We compute the usual velocity independent $J$-factors for s-wave annihilations in both halos and assume that the ratio between the velocity dependent $J$-factors is the same. Using this procedure we find that the cored $J$-factor is about $50\%$ smaller which translates to a weaker limit by the same ratio. The comparably small change in the $J$-factor between different density models in contrast to previous studies has also been addressed in \cite{CTA:2020qlo} and can be explained by the larger ROI considered here. Naturally, one may wonder how robust this estimate is. In order to check our approach we consider a NFW profile which allows the use of the Eddington inversion method. Computing the ratio of the velocity dependent $J$-factors, we find that the ratio between the Einasto and the NFW $J$-factors is largely velocity independent and agrees to $10\%$ with the ratio estimated using the  velocity independent (s-wave without Sommerfeld) $J$-factors. Therefore, we believe that the  method outline above is robust. These results are also similar to those of Ref. \cite{Rinchiuso:2020skh}, which studies the CTA sensitivity to Wino and Higgsino dark matter. Comparing an Einasto model to a cored Einasto model (with core size 1 kpc) and using a slightly smaller ROI they find a factor of 3 reduction of the prospect compared to our factor of 2. Finally, we would like to caution, that there are other sources of uncertainties which can affect the limits apart from the $J$-factors. Taking these into account requires a full analysis of the data and cannot be done based on the likelihood tables provided by \cite{bringmann_torsten_2020_4057987}. Therefore, we do not consider them here, see \cite{CTA:2020qlo} for a detailed discussion of these effects.


\section{Complementary searches and bounds}
\label{sec:other_exp_limit}
In this section we address the connection between the DM model and experimental searches that are complementary to indirect detection. We consider direct detection, BBN limits, thermalization of the dark sector and electric dipole moments.


\subsection{Direct detection and BBN limits}
\label{sec:DD_BBN}
The mixing between the Higgs and the mediator allows for the elastic scattering of DM on SM particles. The rate of this process is strongly constrained by direct detection experiments (\eg \cite{PandaX-4T:2021bab,LZ:2022lsv,XENON:2023cxc}) The spin-independent scattering cross section on a nucleon $N$ is given by \cite{Duerr:2016tmh}
\begin{align}
    \sigma_{SI}= \frac{g^2}{\pi} \frac{\mu^2_{X N} m^2_N f_N^2}{v^2}\sin^2{\delta} \cos^2{\delta} \left(\frac{1}{m^2_\phi}-\frac{1}{m_h^2}
    \right)^2
\end{align}
where $m_N$ is the nucleon mass, $\mu_{X N}$ the reduced mass of the dark matter-nucleon system and $f_N\approx 0.35$ the effective coupling of the Higgs to the nucleon \footnote{The coupling $g_5$ can be neglected here since it leads to a momentum dependent direct detection cross section \cite{Anand:2013yka} which is constrained much more weakly.}. For $m_\phi\ll m_h$ and $M \gg m_N$ this leads to a simple  upper limit on the  mixing angle 
\begin{equation}
    (\sin\delta)_{\text{max}} \simeq \frac{m_\phi^2\,v}{2\, f_N \,m_N^2}\sqrt{\frac{\sigma_{\text{SI}}(M)}{\alpha}}.
\end{equation}
Currently the most stringent upper bound on the spin-independent DM nucleus scattering cross section comes from the LZ experiment \cite{LZ:2022lsv}. We always assume that $\sin \delta$ is small enough not to be in conflict with the direct detection data. As the exact value is not relevant for indirect detection this does not have an effect on the indirect detection phenomenology discussed above.

Note, however, that this limit can be combined with limits from BBN to exclude low mass mediators \cite{Wise:2014jva}. In order to leave the abundances of the primordial elements unaffected one needs to ensure that the mediators decay before the onset of BBN. Therefore we demand that the lifetime of $\phi$, $\tau=1/\Gamma_{\phi}$, is shorter than the age of the Universe at the onset of BBN which we take to be $T=1$ MeV. This yields
\begin{equation}
    T_{\text{dec}}\approx \sqrt{0.3  \,g_{\textrm{eff}}^{-1/2}M_{\textrm{Pl}}\Gamma_\phi}\,.
\end{equation}
As BBN leads to a lower limit on $\sin{\delta}$ while direct detection provides an upper one, the combination of this argument with the LZ limits allows to exclude a range of scalar masses. Numerically, we find that an unsuppressed partial width into muons is crucial to ensure a short enough lifetime. Therefore, masses of $m_\phi$ slightly above $2m_\mu$ are excluded due to the onset of the kinematic threshold.


\subsection{Thermal contact between the SM and the dark sector}
\label{sec:thermalization}
Our relic density computation assumes that the SM and the dark sector share a common temperature and remain in thermal contact during the freeze-out. While a deviation from this assumption is not a problem as such, it would require a more sophisticated analysis of the freeze-out process and the accuracy of the relic density prediction will vary depending on the details of the thermal decoupling between the sectors. Therefore, it is important to check that our assumptions hold. We follow standard arguments in the literature, see e.g.~\cite{Lebedev:2021xey}, to estimate the couplings required for efficient thermal contact between the two sectors and we summarize the main steps in the following.

\begin{figure}[t!]
    \centering
    \includegraphics[scale=0.5]{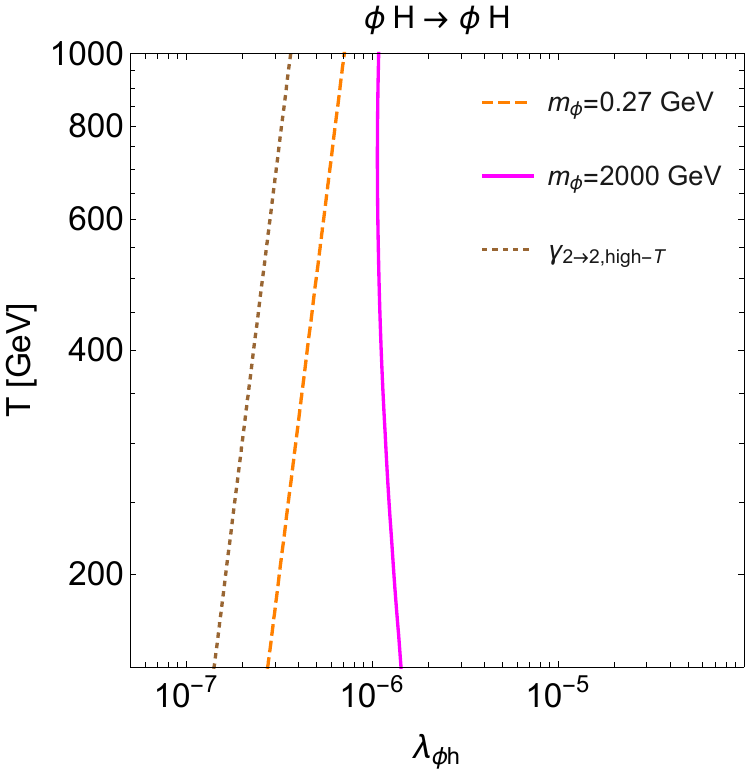}
    \hspace{0.2 cm}
    \includegraphics[scale=0.5]{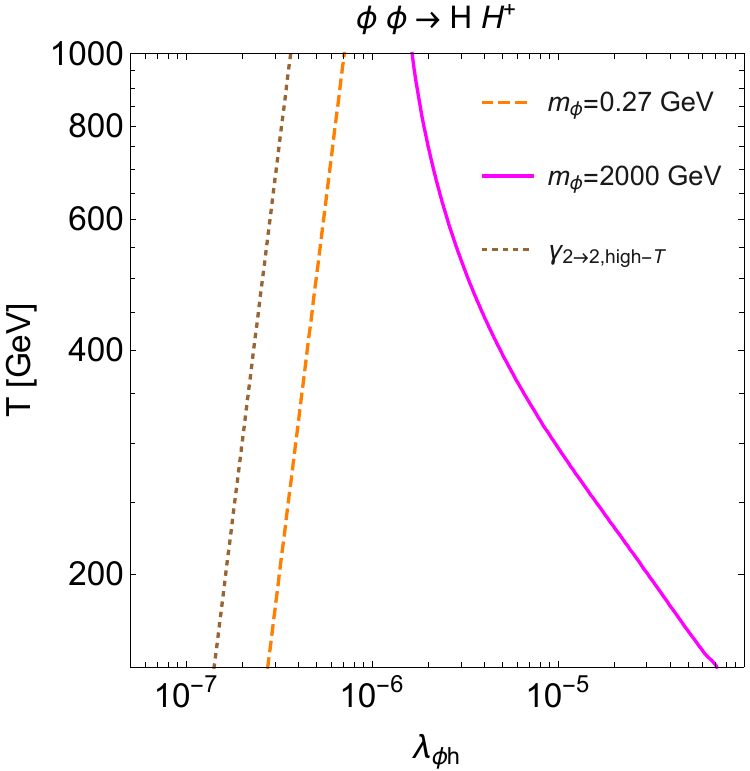}
      \caption{Contours for the condition $\gamma_{2 \to 2}/(3H n_{\phi,\textrm{eq}}) =1$.  The dashed-orange and solid-magenta curves stand for the condition thermally averaged cross section as with full account of the scalar mass $m_\phi$ (smallest and largest values respectively); the dotted-brown line stands for the estimation based on dimensional analysis only.}
    \label{fig:Gamma_Hubble_symmetric}
\end{figure}

In the symmetric SM phase, the dominant processes are found to be $2 \to 2 $ scatterings of the form $\phi H \leftrightarrow \phi H$ and $\phi \phi \leftrightarrow H H^\dagger$. As usual, we want the interactions to be efficient and faster than the expansion rate of the universe, which translates to the condition $\gamma_{2 \to 2} \geq 3 \, H(T) \, n_{\phi,\textrm{eq}}$, where $\gamma_{2 \to 2}$ is the rate density (i.e. the right-hand side of the Boltzmann equation) and $H(T)$ denotes the Hubble rate, which was already defined in sec.~\ref{sec:relic_density},  and  $n_{\phi,\textrm{eq}}$ is the equilibrium number density of the dark scalar. For an order of magnitude estimate, it is sufficient to consider the leading-$T$ behavior of the rates density $\gamma$ entering the Boltzmann equation, namely $\gamma_{2 \to 2}^{\textrm{high}-T} \propto \lambda^2_{\phi h}T^4/(32 \pi^2)$. By using $n_{\phi,\textrm{eq}} \approx T^3/\pi^2$, one finds the condition to be satisfied for 
\begin{equation}
    \lambda_{\phi h} \geq 1.3 \times 10 \, \left( \frac{ T \, \sqrt{g_{\textrm{eff}}(T)} }{M_{\textrm{Pl}}} \right)^{1/2} \, ,
\end{equation}
where $g_{\textrm{eff}}(T)$ denotes the effective number of SM degrees of freedom. For $T \in [150, 10^3]$~GeV one finds $\lambda_{\phi h} \simeq [ 10^{-7}, 5 \times 10^{-7}]$. A more accurate extraction based on full thermally averaged cross sections that accounts for the finite mass of the dark scalar leads to the results shown in Fig.~\ref{fig:Gamma_Hubble_symmetric}. The more efficient elastic scattering $\phi H \leftrightarrow \phi H$ allows for  $\lambda_{\phi h} \simeq 10^{-6}$ for the whole range of scalar masses considered in our study. 

At lower temperatures, the dominant thermal connection between the two sectors is provided by the Higgs-scalar mixing. In this situation the most relevant processes for establishing thermal contact between the SM and the dark sector are $f \phi \leftrightarrow f V$ scattering processes, where $f$ denotes a SM fermion and $V$ is a gauge boson, see e.g. appendix A.2 of \cite{Evans:2017kti} for a detailed discussion. The values of $\sin \delta$ required to achieve thermal contact between the sectors as a function of the freeze-out temperature $T_{\text{fo}}$ can be read off from Fig.~2 of the same reference mentioned above \cite{Evans:2017kti}. In the regime of interest to us, i.e. $T_{\text{fo}}\gtrsim 25$ GeV, one finds that $\sin \delta \gtrsim 10^{-6} - 10^{-7}$ is required to establish thermal contact. In the bulk of the parameter space considered in this work, these values are comfortably below the experimental limits. Note, however, that for very light scalars with $m_\phi\lesssim 500$ MeV the direct detection constrains become very tight and start to exclude the mixing angles needed to establish thermal contact. This region largely overlaps with the combined direct detection-BBN bound.


\subsection{Electric dipole moments}
\label{sec:EDM}
\begin{figure}[t!]
    \centering
    \includegraphics[scale=0.6]{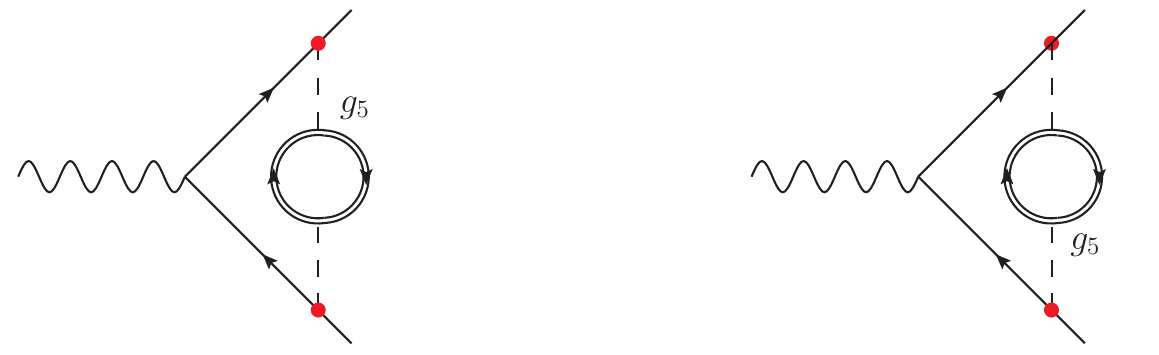}
    \caption{Two loop diagrams that involve a non-trivial complex phase from the pseudo-scalar vertex in the dark sector. Double solid lines stand for the dark matter fermion $X$, dashed lines the dark scalar $\phi$, solid lines the SM fermions, and the red vertex denotes the mixing interaction $\sin\delta \bar{f} f$.}
    \label{fig:EDM}
\end{figure}

The pseudo-scalar interaction in \Eq\eqref{lag_mod_relativistic} violates CP in the dark sector. Through portal interactions CP violation can be transferred to the SM sector and induce electric dipole moments \cite{Bernreuther:1990jx,Chupp:2017rkp,Alarcon:2022ero}. In the model of our paper, the mixing with the SM Higgs boson does not produce a pseudo-scalar interaction among the dark scalar $\phi$ and the SM fermions at leading order, rather one finds only a scalar interaction of the form $\mathcal{L}_{\textrm{int}} \supset - \sin\delta\bar{f}f \phi$. Hence, there is no contribution to one-loop topologies. The first contribution to EDM of SM fermions can arise at two-loop level. The relevant diagrams are shown in figure~\ref{fig:EDM}, where one scalar and one pseudo-scalar vertex and a DM fermion loop enter.

The most stringent EDM bounds are in place for the electron, namely $|d_e|<1.1 \times 10^{-29}  \; \text{cm} \, e$ \cite{ACME:2018yjb}. One can estimate the two-loop contribution to the EDM by
\begin{equation}
    d_e \approx \frac{e}{(4 \pi^2)^2} 4\pi \sqrt{\alpha \alpha_5} \; y_e^2 \; \sin^2 \delta \frac{1}{m_e} \times \left( 1, \frac{m_e^2}{m_\phi^2},  \frac{m_e^2}{M^2}\right),
\end{equation}
where we have indicated the possible relevant combinations from the scales running in the loop, the electron mass being the smallest scale in the problem. Even taking the least suppressed contribution from the loop, and by using $m_e \simeq 0.5 \, \text{MeV} \simeq 2 \times 10^{10} \, \text{cm}^{-1} $, $y_e \simeq 2 \times 10^{-6}$, $\alpha=0.1$, $\alpha_5=\alpha/10$ and $\sin\delta=10^{-2}$, we obtain $|d_e| \simeq 10^{-30} \; \text{cm} \, e$, which is smaller than the experimental limit. Note that the value for the mixing angle used in the estimation are larger that those we consider in our study. In addition, the heavy scales in the problem are expected to push the EDM to much smaller values than found in this conservative estimate. Therefore, the electron EDM bound is irrelevant in this model.

One may wonder if the situation is more promising if the electric dipole moment of the muon is considered instead but it turns out that the larger Yukawa coupling is not sufficient to compensate for the larger mass and the weaker experimental limit. 


\section{Parameter space of thermal dark matter}
\label{sec:parameterspace}
It is of key interest to understand where the limits and prospects of indirect searches stand compared to the theoretical expectation for thermally produced dark matter. Therefore, we now focus on the parameter space of our model that allows us to explain the full relic density via the freeze-out mechanism. In general, the model is characterized by five parameters: two masses ($M$ and $m_\phi$), two interaction strengths ($\alpha$ and $\alpha_5$), and the mixing angle between the mediator $\phi$ and the SM Higgs. As long as the mixing angle is small enough to avoid limits from direct detection and Higgs decays, its exact value is largely irrelevant to the phenomenology. This reduces the number of effective parameters to four. The relic density is measured with percent level accuracy \cite{Planck:2018vyg}. Therefore, it is possible to reduce the free parameters further by imposing that the correct DM abundance is produced in the early universe. We opt to fix $\alpha$ in this way while varying the two masses and $\alpha_5$. With only three parameters left, it is possible to show slices through the parameter space for two of them by fixing the third. As the phenomenology is mainly driven by the masses of the involved particles, we take $M$ and $m_\phi$ as variables in our analysis and show fixed ratios of $\alpha_5/\alpha$ to explore that direction.
\begin{figure}
    \centering
    \includegraphics[width=0.45\textwidth]{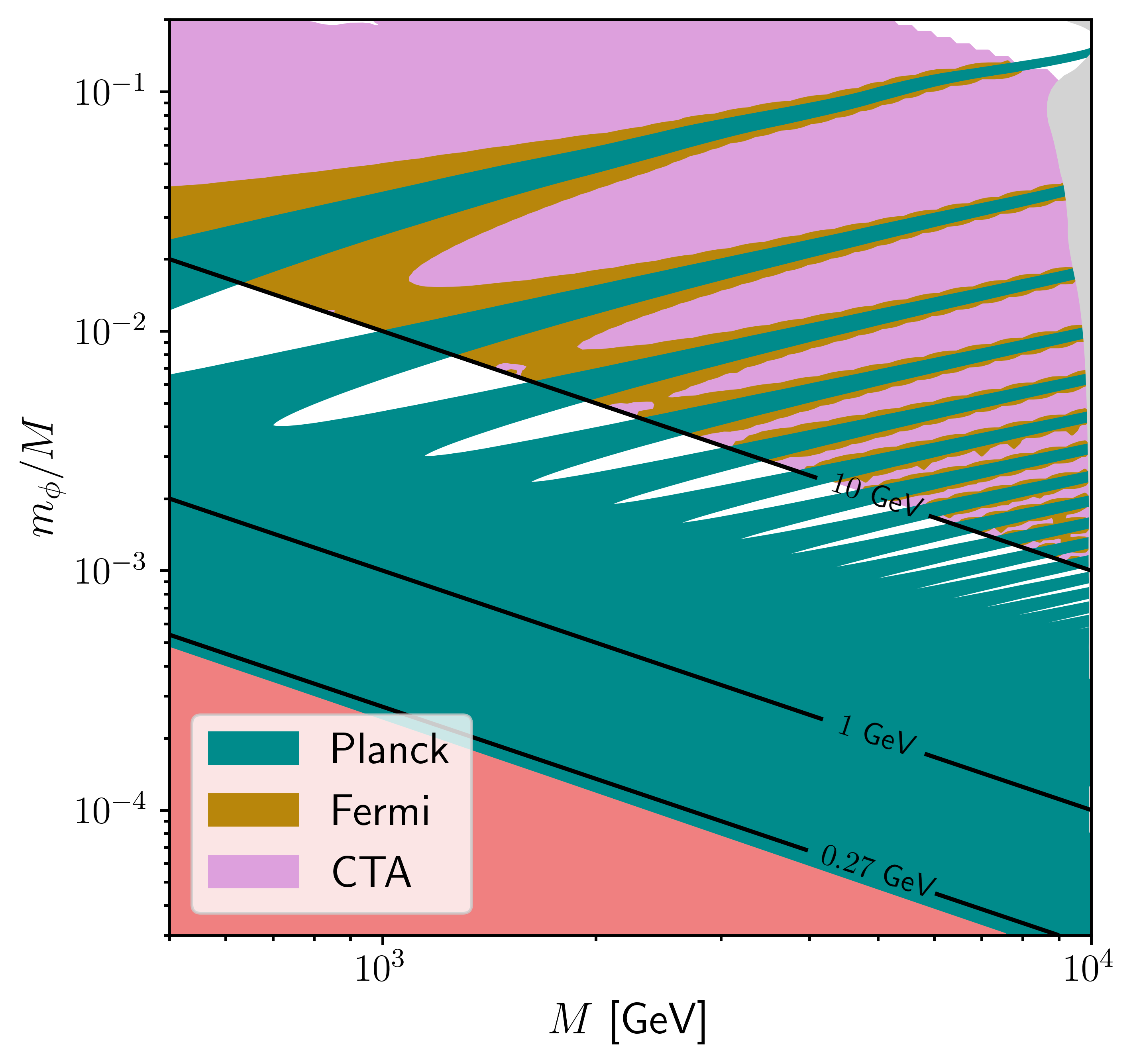}
    \includegraphics[width=0.45\textwidth]{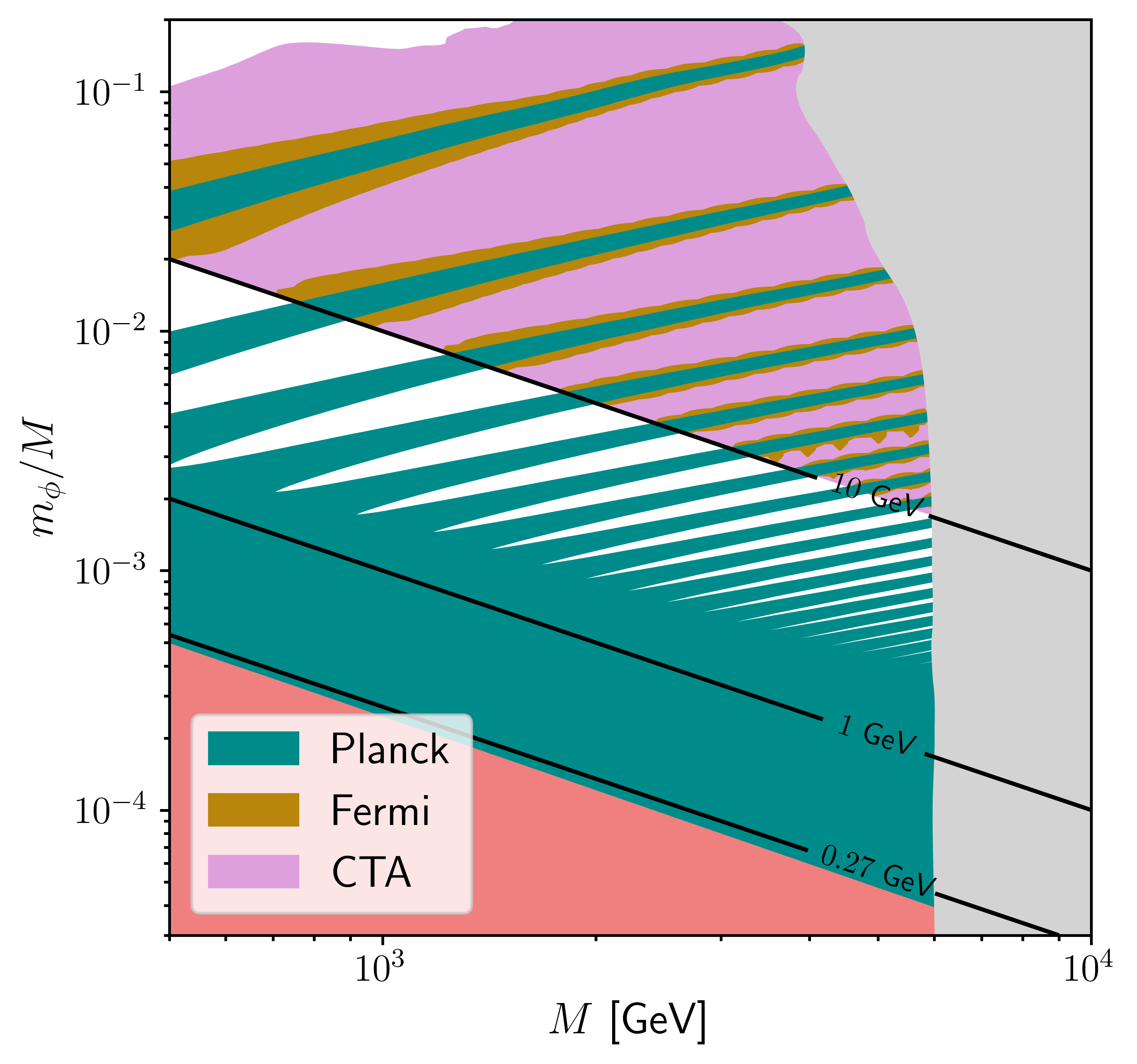}
    \includegraphics[width=0.45\textwidth]{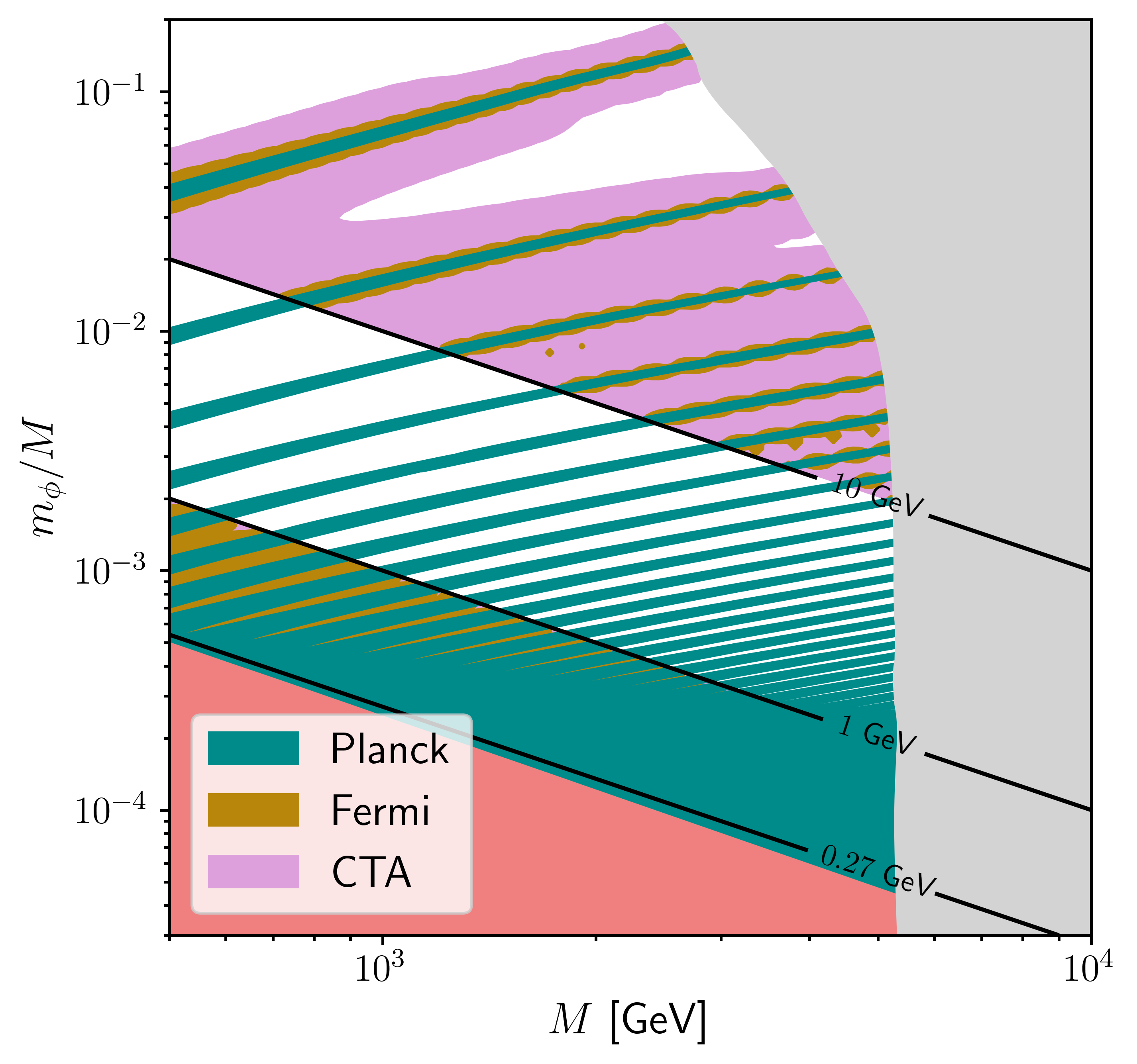}
    \includegraphics[width=0.45\textwidth]{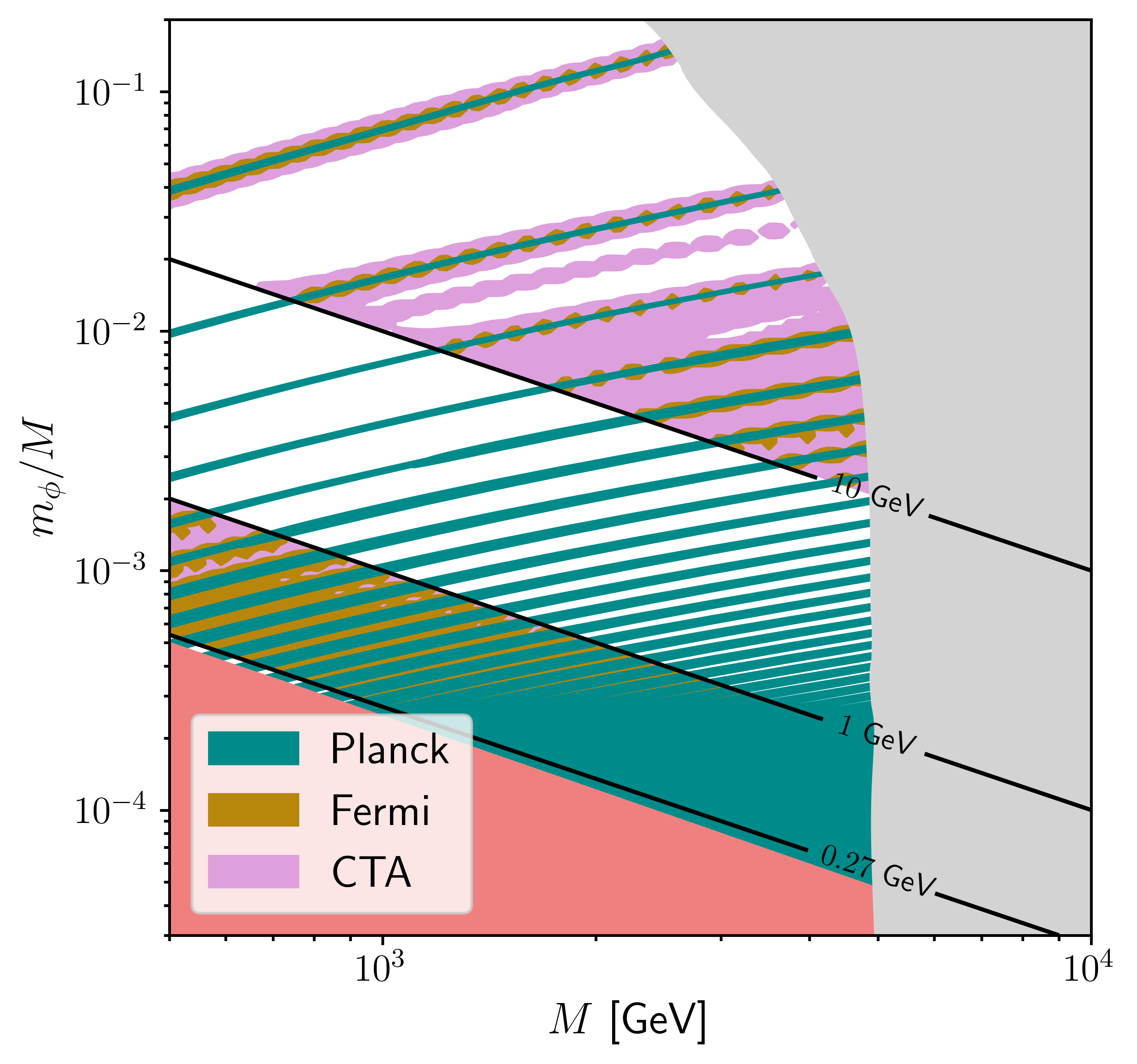}
    \caption{Slices through the cosmologically preferred parameter space in the $M$-$m_\phi/M$ plane for four different representative values of $\alpha_5/\alpha$ ($10^{-1}, 10^{-2}$, $10^{-3}$ and $10^{-4}$). The value of $\alpha$ is fixed by the relic density requirement. The regions excluded by Planck are shown in turquoise and the Fermi-LAT exclusion in gold. Prospects for CTA are shown in pink. The red region depicts mediator masses, which are excluded by the combined BBN and direct detection limit on the mixing angle. In the gray region the relic density requires $\alpha>0.25$. The black lines indicate the characteristic values of $m_\phi$ that delineate the different gamma ray production regimes.}
    \label{fig:parameter_space}
\end{figure}

Representative examples of slices through the cosmologically preferred parameter space are shown in figure~\ref{fig:parameter_space}. We have limited the analysis  to the region $\SI{0.5}{\TeV}\leq M\leq \SI{10}{\TeV}$ and  $ m_\phi \leq \alpha M$,  where the upper limit on $m_\phi$ ensures the validity of the NREFT employed in our derivation of the cross sections and induces long-range interactions. Gamma ray constraints and prospects are evaluated for $m_\phi \geq 2 m_{\pi^0}$ while the CMB bounds are not restricted by that condition. However, the combined limit by BBN and direct detection searches excludes mediator masses slightly below this threshold (red). Furthermore, we do not consider regions where the relic density computations point towards $\alpha \gtrsim 0.25$ (gray). In this regime corrections that are not included in our derivations become large and an analysis of the phenomenology based on our computations can lead to misleading conclusions. These include $\alpha$ and $\alpha_5$ corrections to the matching coefficients \eqref{match_coeff_1}-\eqref{match_coeff_4}, and corrections to the binding energy of the bound states. As can be seen, the indirect detection limits are already quite constraining if $\alpha_5$ is not much smaller than $\alpha$. This is expected since larger values of $\alpha_5$ increase the $s$-wave cross section which is not velocity suppressed at any $m_\phi$. Together with the Sommerfeld effect, this ensures that essentially all configurations with $m_\phi \lesssim \SI{3}{\GeV}$ are excluded by the Planck measurements for $\alpha_5/\alpha= 0.1$. At higher mediator masses, CMB exclusions rely on the presence of resonances in the Sommerfeld enhanced $J$-factors and we find stripes of excluded parameter space that extend up to very high values of $M$ and $m_\phi$. As the overall cross sections decrease for higher masses these become thinner as the mass increases since the resonance condition needs to be fulfilled more and more precisely. In this regime, the gamma-ray telescopes add more information. As one can see, the structure of the Fermi-LAT limits is comparable to the CMB ones. However, due to the better sensitivity of Fermi, they require a less pronounced Sommerfeld enhancement and, therefore,  it has broader stripes that encompass the CMB ones for low $M$. Towards the high end of the considered mass range, this effect becomes less prominent due to the different scaling of the sensitivity with $M$ as noted previously. Fermi does not add significantly to the CMB bounds for $M\approx 10$ TeV. It is very interesting to note, that CTA can be a game changer. With its significantly stronger exclusion prospects, it does not rely on the resonances. This will close the gaps between the stripes and allows to probe almost the whole parameter space. Only very high DM masses and large $m_\phi$ can evade the prospects here. Note that gamma rays also have something to add at low $m_\phi$. The strip with $2 m_{\pi^0} \leq m_\phi \leq \SI{1}{\GeV}$, the low mass-region from now on, is also excluded by Fermi-LAT. These limits are just hidden below the CMB ones to increase the readability of the plot.

For smaller $\alpha_5/\alpha$ ratios the $p$-wave and BSF contributions become more important. For $\alpha_5/\alpha=10^{-2}$ the overall picture remains qualitatively similar; new feature become visible for $\alpha_5/\alpha=10^{-3}$. Here the limits from Planck measurements decrease substantially in the low mediator mass regime such that configurations with $m_\phi \lesssim \SI{1}{\GeV}$ cannot all be reached by CMB limits alone. However, Fermi-LAT limits are still relatively strong here due to the large number of energetic photons produced in the pion cascade regime. Therefore, most of this parameter space is still ruled out by current experiments. For $m_\phi > 10$ GeV only quite narrow stripes with a strong resonant enhancement are excluded by present experiments. Prospects for CTA are still excellent and most of the gaps between the enhanced stripes can be closed if its sensitivity fulfills the expectations. Note that the granularity of the gamma-ray limits which becomes noticeable compared to the Planck limits in this figure is a numerical artifact caused by the lower resolution of our Fermi/CTA scans. The determination of the $J$-factors required for the gamma-ray limits and prospects is computationally intensive and increasing the resolution to the same level as the Planck limits was not feasible with the available computing power. Finally, at $\alpha_5/\alpha=10^{-4}$ the signals are dominated by $p$-wave and BSF. Here, existing limits are rather weak and CTA is already needed to close the gaps in the low-mass region. At $m_\phi\geq 10$ GeV most of the parameter space is currently unconstrained. CTA is still able to contribute here but for $m_\phi/M\gtrsim 10^{-2}$, it also relies on resonances in the Sommerfeld factors. Some further strips become visible in this figure. These are driven by the resonances of the $p$-wave Sommerfeld factors which do not coincide with those of the s-wave. All considered, indirect detection is highly effective at testing the model in the parts of parameter space where $s$-wave annihilation is relevant. In the $p$-wave and BSF-dominated regime, there are already some interesting limits and good prospects for parts of the mass range. However, a definitive test of the model will remain challenging even with future instruments.


\section{Conclusion}
\label{sec:conclusion}

The nature and origin of DM remain two of the most pressing problems in physics today. One particularly attractive possibility is that DM was produced thermally by its interactions with SM particles during the first second after the Big Bang. Despite the simplicity of the basic idea, this scenario allows for a very diverse phenomenology which has inspired a huge range of experimental searches. Among these, indirect searches for the annihilation products of DM in cosmic rays stand out since the relevant observables are directly connected with the processes that drive cosmological production.

Studying such indirect signatures for DM annihilations is particularly timely since new and more sensitive instruments such as the CTA will provide new data in the near future. Therefore, it is important to understand the sensitivity of the CTA to thermal DM and compare its abilities in a fair manner with existing bounds. Since there is a huge range of possible observables, we choose the rather robust limits from Planck and Fermi-LAT observations of dSphs as our reference points.

In this work, we focused on a model of fermionic DM interacting with the SM via a (pseudo)-scalar interaction with a massive scalar mediator. This model is largely unconstrained from direct detection and the LHC, and thus constitutes an interesting benchmark for indirect detection. When the mediator is light compared to the DM mass, special care is required in the derivation of the annihilation cross section. Due to the long-range force mediated by $\phi$ exchange, non-perturbative corrections commonly known as Sommerfeld enhancement become relevant and the formation of bound states is also possible. Bound fermion-antifermion pairs possess a total energy below that of a free pair and are, therefore, thermally preferred because of a lifting of the Boltzmann suppression. Hence, they can have an important effect on the relic density even though the formation cross section is suppressed by additional powers of the coupling. Both Sommerfeld enhancement and BSF are of crucial importance for light mediators and can drastically alter the relevant rates in the early universe and astrophysical environments. We employ NREFT and pNREFT techniques, which allow us to take these effects into account systematically in the computation of the relevant cross section, and we derive the DM relic density with the usual Boltzmann equation method.

As a consequence of the non-perturbative effects, the cross sections of interest exhibit a strong and non-trivial dependence on the relative DM velocity. This breaks the common lore of indirect detection that only the leading velocity-independent contribution to $s$-wave annihilation is detectable in a realistic experiment. Therefore, we employ an extended version of the standard formalism to compute the appropriate velocity averaged $J$-factors. This requires position-dependent velocity distributions for the targets under consideration. We derive these using Eddington inversion from the density profiles for four dSphs and the Galactic Center. This allows us to predict the photon flux from DM annihilation from the particle physics parameters and impose upper limits (identify the prospects).

Combining this with our predictions for the relic density allows us to assess the status of thermal DM in this model. If the suppression of $\alpha_5$ relative to $\alpha$ is modest, $s$-wave annihilations play a large role and there are strong constraints on light mediators irrespective of the DM mass. At higher $m_\phi$ the resonant enhancement due to the Yukawa Sommerfeld factors becomes important and allows to exclude the regions close to the resonance with Planck and Fermi-LAT observations. Interestingly, the gaps that remain in the testable parameter space can be closed by CTA in this case. Testing the model becomes more difficult for smaller $\alpha_5/\alpha$. Here, the sensitivity of Planck and Fermi to the cosmologically preferred parameter space diminishes and CTA is crucial in this regime since it will allow testing large regions of the parameter space for the first time. 


\acknowledgments

J.B. and S.V. acknowledge the support of the Research Training Group (RTG) 2044 funded by the German Research Foundation (DFG). The work of S.V. was supported by the DFG with an Individual Research Grant (project number 496940663). The work of S.B. is supported by the Swiss National Science Foundation (SNSF) under the Ambizione grant PZ00P2\_185783. We thank M. Winkler for providing the branching ratios of a light dark Higgs in machine-readable form.


\newpage
\bibliographystyle{JHEP2.bst}    
\bibliography{references}

\providecommand{\href}[2]{#2}\begingroup\raggedright\begin{thebibliography}{100}

\bibitem{Planck:2018vyg}
{\scshape Planck} collaboration, \emph{{Planck 2018 results. VI. Cosmological
  parameters}},
  \href{https://doi.org/10.1051/0004-6361/201833910}{\emph{Astron. Astrophys.}
  {\bfseries 641} (2020) A6}
  [\href{https://arxiv.org/abs/1807.06209}{{\ttfamily 1807.06209}}].

\bibitem{Bertone:2004pz}
G.~Bertone, D.~Hooper and J.~Silk, \emph{{Particle dark matter: Evidence,
  candidates and constraints}},
  \href{https://doi.org/10.1016/j.physrep.2004.08.031}{\emph{Phys. Rept.}
  {\bfseries 405} (2005) 279}
  [\href{https://arxiv.org/abs/hep-ph/0404175}{{\ttfamily hep-ph/0404175}}].

\bibitem{Feng:2010gw}
J.L.~Feng, \emph{{Dark Matter Candidates from Particle Physics and Methods of
  Detection}},
  \href{https://doi.org/10.1146/annurev-astro-082708-101659}{\emph{Ann. Rev.
  Astron. Astrophys.} {\bfseries 48} (2010) 495}
  [\href{https://arxiv.org/abs/1003.0904}{{\ttfamily 1003.0904}}].

\bibitem{Kolb:1990vq}
E.W.~Kolb and M.S.~Turner, \emph{{The Early Universe}}, vol.~69 (1990),
  \href{https://doi.org/10.1201/9780429492860}{10.1201/9780429492860}.

\bibitem{MarrodanUndagoitia:2015veg}
T.~Marrod\'an~Undagoitia and L.~Rauch, \emph{{Dark matter direct-detection
  experiments}}, \href{https://doi.org/10.1088/0954-3899/43/1/013001}{\emph{J.
  Phys. G} {\bfseries 43} (2016) 013001}
  [\href{https://arxiv.org/abs/1509.08767}{{\ttfamily 1509.08767}}].

\bibitem{Klasen:2015uma}
M.~Klasen, M.~Pohl and G.~Sigl, \emph{{Indirect and direct search for dark
  matter}}, \href{https://doi.org/10.1016/j.ppnp.2015.07.001}{\emph{Prog. Part.
  Nucl. Phys.} {\bfseries 85} (2015) 1}
  [\href{https://arxiv.org/abs/1507.03800}{{\ttfamily 1507.03800}}].

\bibitem{Gaskins:2016cha}
J.M.~Gaskins, \emph{{A review of indirect searches for particle dark matter}},
  \href{https://doi.org/10.1080/00107514.2016.1175160}{\emph{Contemp. Phys.}
  {\bfseries 57} (2016) 496}
  [\href{https://arxiv.org/abs/1604.00014}{{\ttfamily 1604.00014}}].

\bibitem{Arcadi:2017kky}
G.~Arcadi, M.~Dutra, P.~Ghosh, M.~Lindner, Y.~Mambrini, M.~Pierre et~al.,
  \emph{{The waning of the WIMP? A review of models, searches, and
  constraints}},
  \href{https://doi.org/10.1140/epjc/s10052-018-5662-y}{\emph{Eur. Phys. J. C}
  {\bfseries 78} (2018) 203}
  [\href{https://arxiv.org/abs/1703.07364}{{\ttfamily 1703.07364}}].

\bibitem{Kahlhoefer:2017dnp}
F.~Kahlhoefer, \emph{{Review of LHC Dark Matter Searches}},
  \href{https://doi.org/10.1142/S0217751X1730006X}{\emph{Int. J. Mod. Phys. A}
  {\bfseries 32} (2017) 1730006}
  [\href{https://arxiv.org/abs/1702.02430}{{\ttfamily 1702.02430}}].

\bibitem{Fermi-LAT:2015att}
{\scshape Fermi-LAT} collaboration, \emph{{Searching for Dark Matter
  Annihilation from Milky Way Dwarf Spheroidal Galaxies with Six Years of Fermi
  Large Area Telescope Data}},
  \href{https://doi.org/10.1103/PhysRevLett.115.231301}{\emph{Phys. Rev. Lett.}
  {\bfseries 115} (2015) 231301}
  [\href{https://arxiv.org/abs/1503.02641}{{\ttfamily 1503.02641}}].

\bibitem{Fermi-LAT:2016uux}
{\scshape Fermi-LAT, DES} collaboration, \emph{{Searching for Dark Matter
  Annihilation in Recently Discovered Milky Way Satellites with Fermi-LAT}},
  \href{https://doi.org/10.3847/1538-4357/834/2/110}{\emph{Astrophys. J.}
  {\bfseries 834} (2017) 110}
  [\href{https://arxiv.org/abs/1611.03184}{{\ttfamily 1611.03184}}].

\bibitem{Fermi-LAT:2017opo}
{\scshape Fermi-LAT} collaboration, \emph{{The Fermi Galactic Center GeV Excess
  and Implications for Dark Matter}},
  \href{https://doi.org/10.3847/1538-4357/aa6cab}{\emph{Astrophys. J.}
  {\bfseries 840} (2017) 43}
  [\href{https://arxiv.org/abs/1704.03910}{{\ttfamily 1704.03910}}].

\bibitem{HESS:2016mib}
{\scshape H.E.S.S.} collaboration, \emph{{Search for dark matter annihilations
  towards the inner Galactic halo from 10 years of observations with H.E.S.S}},
  \href{https://doi.org/10.1103/PhysRevLett.117.111301}{\emph{Phys. Rev. Lett.}
  {\bfseries 117} (2016) 111301}
  [\href{https://arxiv.org/abs/1607.08142}{{\ttfamily 1607.08142}}].

\bibitem{VERITAS:2017tif}
{\scshape VERITAS} collaboration, \emph{{Dark Matter Constraints from a Joint
  Analysis of Dwarf Spheroidal Galaxy Observations with VERITAS}},
  \href{https://doi.org/10.1103/PhysRevD.95.082001}{\emph{Phys. Rev. D}
  {\bfseries 95} (2017) 082001}
  [\href{https://arxiv.org/abs/1703.04937}{{\ttfamily 1703.04937}}].

\bibitem{Hoof:2018hyn}
S.~Hoof, A.~Geringer-Sameth and R.~Trotta, \emph{{A Global Analysis of Dark
  Matter Signals from 27 Dwarf Spheroidal Galaxies using 11 Years of Fermi-LAT
  Observations}},
  \href{https://doi.org/10.1088/1475-7516/2020/02/012}{\emph{JCAP} {\bfseries
  02} (2020) 012} [\href{https://arxiv.org/abs/1812.06986}{{\ttfamily
  1812.06986}}].

\bibitem{AMS:2015tnn}
{\scshape AMS} collaboration, \emph{{Precision Measurement of the Proton Flux
  in Primary Cosmic Rays from Rigidity 1 GV to 1.8 TV with the Alpha Magnetic
  Spectrometer on the International Space Station}},
  \href{https://doi.org/10.1103/PhysRevLett.114.171103}{\emph{Phys. Rev. Lett.}
  {\bfseries 114} (2015) 171103}
  [\href{https://arxiv.org/abs/1612.08562}{{\ttfamily 1612.08562}}].

\bibitem{Cuoco:2016eej}
A.~Cuoco, M.~Kr\"amer and M.~Korsmeier, \emph{{Novel Dark Matter Constraints
  from Antiprotons in Light of AMS-02}},
  \href{https://doi.org/10.1103/PhysRevLett.118.191102}{\emph{Phys. Rev. Lett.}
  {\bfseries 118} (2017) 191102}
  [\href{https://arxiv.org/abs/1610.03071}{{\ttfamily 1610.03071}}].

\bibitem{Heisig:2020nse}
J.~Heisig, M.~Korsmeier and M.W.~Winkler, \emph{{Dark matter or correlated
  errors: Systematics of the AMS-02 antiproton excess}},
  \href{https://doi.org/10.1103/PhysRevResearch.2.043017}{\emph{Phys. Rev.
  Res.} {\bfseries 2} (2020) 043017}
  [\href{https://arxiv.org/abs/2005.04237}{{\ttfamily 2005.04237}}].

\bibitem{PAMELA:2008gwm}
{\scshape PAMELA} collaboration, \emph{{An anomalous positron abundance in
  cosmic rays with energies 1.5-100 GeV}},
  \href{https://doi.org/10.1038/nature07942}{\emph{Nature} {\bfseries 458}
  (2009) 607} [\href{https://arxiv.org/abs/0810.4995}{{\ttfamily 0810.4995}}].

\bibitem{AMS:2019rhg}
{\scshape AMS} collaboration, \emph{{Towards Understanding the Origin of
  Cosmic-Ray Positrons}},
  \href{https://doi.org/10.1103/PhysRevLett.122.041102}{\emph{Phys. Rev. Lett.}
  {\bfseries 122} (2019) 041102}.

\bibitem{John:2021ugy}
I.~John and T.~Linden, \emph{{Cosmic-Ray Positrons Strongly Constrain
  Leptophilic Dark Matter}},
  \href{https://doi.org/10.1088/1475-7516/2021/12/007}{\emph{JCAP} {\bfseries
  12} (2021) 007} [\href{https://arxiv.org/abs/2107.10261}{{\ttfamily
  2107.10261}}].

\bibitem{ANTARES:2015vis}
{\scshape ANTARES} collaboration, \emph{{Search of Dark Matter Annihilation in
  the Galactic Centre using the ANTARES Neutrino Telescope}},
  \href{https://doi.org/10.1088/1475-7516/2015/10/068}{\emph{JCAP} {\bfseries
  10} (2015) 068} [\href{https://arxiv.org/abs/1505.04866}{{\ttfamily
  1505.04866}}].

\bibitem{IceCube:2022clp}
{\scshape IceCube} collaboration, \emph{{Searches for Connections between Dark
  Matter and High-Energy Neutrinos with IceCube}},
  \href{https://arxiv.org/abs/2205.12950}{{\ttfamily 2205.12950}}.

\bibitem{IceCube:2021xzo}
{\scshape IceCube} collaboration, \emph{{Search for GeV-scale dark matter
  annihilation in the Sun with IceCube DeepCore}},
  \href{https://doi.org/10.1103/PhysRevD.105.062004}{\emph{Phys. Rev. D}
  {\bfseries 105} (2022) 062004}
  [\href{https://arxiv.org/abs/2111.09970}{{\ttfamily 2111.09970}}].

\bibitem{CTAConsortium:2017dvg}
{\scshape CTA Consortium} collaboration, B.S.~Acharya et~al., \emph{{Science
  with the Cherenkov Telescope Array}}, WSP (11, 2018),
  \href{https://doi.org/10.1142/10986}{10.1142/10986},
  [\href{https://arxiv.org/abs/1709.07997}{{\ttfamily 1709.07997}}].

\bibitem{Pospelov:2007mp}
M.~Pospelov, A.~Ritz and M.B.~Voloshin, \emph{{Secluded WIMP Dark Matter}},
  \href{https://doi.org/10.1016/j.physletb.2008.02.052}{\emph{Phys. Lett. B}
  {\bfseries 662} (2008) 53} [\href{https://arxiv.org/abs/0711.4866}{{\ttfamily
  0711.4866}}].

\bibitem{Pospelov:2008jd}
M.~Pospelov and A.~Ritz, \emph{{Astrophysical Signatures of Secluded Dark
  Matter}}, \href{https://doi.org/10.1016/j.physletb.2008.12.012}{\emph{Phys.
  Lett. B} {\bfseries 671} (2009) 391}
  [\href{https://arxiv.org/abs/0810.1502}{{\ttfamily 0810.1502}}].

\bibitem{PandaX-4T:2021bab}
{\scshape PandaX-4T} collaboration, \emph{{Dark Matter Search Results from the
  PandaX-4T Commissioning Run}},
  \href{https://doi.org/10.1103/PhysRevLett.127.261802}{\emph{Phys. Rev. Lett.}
  {\bfseries 127} (2021) 261802}
  [\href{https://arxiv.org/abs/2107.13438}{{\ttfamily 2107.13438}}].

\bibitem{LZ:2022lsv}
{\scshape LZ} collaboration, \emph{{First Dark Matter Search Results from the
  LUX-ZEPLIN (LZ) Experiment}},
  \href{https://doi.org/10.1103/PhysRevLett.131.041002}{\emph{Phys. Rev. Lett.}
  {\bfseries 131} (2023) 041002}
  [\href{https://arxiv.org/abs/2207.03764}{{\ttfamily 2207.03764}}].

\bibitem{XENON:2023cxc}
{\scshape XENON} collaboration, \emph{{First Dark Matter Search with Nuclear
  Recoils from the XENONnT Experiment}},
  \href{https://doi.org/10.1103/PhysRevLett.131.041003}{\emph{Phys. Rev. Lett.}
  {\bfseries 131} (2023) 041003}
  [\href{https://arxiv.org/abs/2303.14729}{{\ttfamily 2303.14729}}].

\bibitem{ATLAS:2019wdu}
{\scshape ATLAS} collaboration, \emph{{Constraints on mediator-based dark
  matter and scalar dark energy models using $\sqrt s = 13$ TeV $pp$ collision
  data collected by the ATLAS detector}},
  \href{https://doi.org/10.1007/JHEP05(2019)142}{\emph{JHEP} {\bfseries 05}
  (2019) 142} [\href{https://arxiv.org/abs/1903.01400}{{\ttfamily
  1903.01400}}].

\bibitem{ATLAS:2021kxv}
{\scshape ATLAS} collaboration, \emph{{Search for new phenomena in events with
  an energetic jet and missing transverse momentum in $pp$ collisions at $\sqrt
  {s}$ =13 TeV with the ATLAS detector}},
  \href{https://doi.org/10.1103/PhysRevD.103.112006}{\emph{Phys. Rev. D}
  {\bfseries 103} (2021) 112006}
  [\href{https://arxiv.org/abs/2102.10874}{{\ttfamily 2102.10874}}].

\bibitem{CMS:2017jdm}
{\scshape CMS} collaboration, \emph{{Search for dark matter produced with an
  energetic jet or a hadronically decaying W or Z boson at $ \sqrt{s}=13 $
  TeV}}, \href{https://doi.org/10.1007/JHEP07(2017)014}{\emph{JHEP} {\bfseries
  07} (2017) 014} [\href{https://arxiv.org/abs/1703.01651}{{\ttfamily
  1703.01651}}].

\bibitem{CMS:2018mgb}
{\scshape CMS} collaboration, \emph{{Search for narrow and broad dijet
  resonances in proton-proton collisions at $ \sqrt{s}=13 $ TeV and constraints
  on dark matter mediators and other new particles}},
  \href{https://doi.org/10.1007/JHEP08(2018)130}{\emph{JHEP} {\bfseries 08}
  (2018) 130} [\href{https://arxiv.org/abs/1806.00843}{{\ttfamily
  1806.00843}}].

\bibitem{Hisano:2003ec}
J.~Hisano, S.~Matsumoto and M.M.~Nojiri, \emph{{Explosive dark matter
  annihilation}},
  \href{https://doi.org/10.1103/PhysRevLett.92.031303}{\emph{Phys. Rev. Lett.}
  {\bfseries 92} (2004) 031303}
  [\href{https://arxiv.org/abs/hep-ph/0307216}{{\ttfamily hep-ph/0307216}}].

\bibitem{Hisano:2004ds}
J.~Hisano, S.~Matsumoto, M.M.~Nojiri and O.~Saito, \emph{{Non-perturbative
  effect on dark matter annihilation and gamma ray signature from galactic
  center}}, \href{https://doi.org/10.1103/PhysRevD.71.063528}{\emph{Phys. Rev.
  D} {\bfseries 71} (2005) 063528}
  [\href{https://arxiv.org/abs/hep-ph/0412403}{{\ttfamily hep-ph/0412403}}].

\bibitem{Arkani-Hamed:2008hhe}
N.~Arkani-Hamed, D.P.~Finkbeiner, T.R.~Slatyer and N.~Weiner, \emph{{A Theory
  of Dark Matter}},
  \href{https://doi.org/10.1103/PhysRevD.79.015014}{\emph{Phys. Rev. D}
  {\bfseries 79} (2009) 015014}
  [\href{https://arxiv.org/abs/0810.0713}{{\ttfamily 0810.0713}}].

\bibitem{March-Russell:2008klu}
J.D.~March-Russell and S.M.~West, \emph{{WIMPonium and Boost Factors for
  Indirect Dark Matter Detection}},
  \href{https://doi.org/10.1016/j.physletb.2009.04.010}{\emph{Phys. Lett. B}
  {\bfseries 676} (2009) 133}
  [\href{https://arxiv.org/abs/0812.0559}{{\ttfamily 0812.0559}}].

\bibitem{vonHarling:2014kha}
B.~von Harling and K.~Petraki, \emph{{Bound-state formation for thermal relic
  dark matter and unitarity}},
  \href{https://doi.org/10.1088/1475-7516/2014/12/033}{\emph{JCAP} {\bfseries
  1412} (2014) 033} [\href{https://arxiv.org/abs/1407.7874}{{\ttfamily
  1407.7874}}].

\bibitem{Feng:2010zp}
J.L.~Feng, M.~Kaplinghat and H.-B.~Yu, \emph{{Sommerfeld Enhancements for
  Thermal Relic Dark Matter}},
  \href{https://doi.org/10.1103/PhysRevD.82.083525}{\emph{Phys. Rev. D}
  {\bfseries 82} (2010) 083525}
  [\href{https://arxiv.org/abs/1005.4678}{{\ttfamily 1005.4678}}].

\bibitem{Slatyer:2011kg}
T.R.~Slatyer, N.~Toro and N.~Weiner, \emph{{Sommerfeld-enhanced annihilation in
  dark matter substructure: Consequences for constraints on cosmic-ray
  excesses}}, \href{https://doi.org/10.1103/PhysRevD.86.083534}{\emph{Phys.
  Rev. D} {\bfseries 86} (2012) 083534}
  [\href{https://arxiv.org/abs/1107.3546}{{\ttfamily 1107.3546}}].

\bibitem{Abazajian:2011ak}
K.N.~Abazajian and J.P.~Harding, \emph{{Constraints on WIMP and
  Sommerfeld-Enhanced Dark Matter Annihilation from HESS Observations of the
  Galactic Center}},
  \href{https://doi.org/10.1088/1475-7516/2012/01/041}{\emph{JCAP} {\bfseries
  01} (2012) 041} [\href{https://arxiv.org/abs/1110.6151}{{\ttfamily
  1110.6151}}].

\bibitem{Lu:2017jrh}
B.-Q.~Lu, Y.-L.~Wu, W.-H.~Zhang and Y.-F.~Zhou, \emph{{Constraints on the
  Sommerfeld-enhanced dark matter annihilation from the gamma rays of subhalos
  and dwarf galaxies}},
  \href{https://doi.org/10.1088/1475-7516/2018/04/035}{\emph{JCAP} {\bfseries
  04} (2018) 035} [\href{https://arxiv.org/abs/1711.00749}{{\ttfamily
  1711.00749}}].

\bibitem{Ando:2021jvn}
S.~Ando and K.~Ishiwata, \emph{{Sommerfeld-enhanced dark matter searches with
  dwarf spheroidal galaxies}},
  \href{https://doi.org/10.1103/PhysRevD.104.023016}{\emph{Phys. Rev. D}
  {\bfseries 104} (2021) 023016}
  [\href{https://arxiv.org/abs/2103.01446}{{\ttfamily 2103.01446}}].

\bibitem{Petraki:2015hla}
K.~Petraki, M.~Postma and M.~Wiechers, \emph{{Dark-matter bound states from
  Feynman diagrams}},
  \href{https://doi.org/10.1007/JHEP06(2015)128}{\emph{JHEP} {\bfseries 06}
  (2015) 128} [\href{https://arxiv.org/abs/1505.00109}{{\ttfamily
  1505.00109}}].

\bibitem{Beneke:2016ync}
M.~Beneke, A.~Bharucha, F.~Dighera, C.~Hellmann, A.~Hryczuk, S.~Recksiegel
  et~al., \emph{{Relic density of wino-like dark matter in the MSSM}},
  \href{https://doi.org/10.1007/JHEP03(2016)119}{\emph{JHEP} {\bfseries 03}
  (2016) 119} [\href{https://arxiv.org/abs/1601.04718}{{\ttfamily
  1601.04718}}].

\bibitem{Ellis:2015vna}
J.~Ellis, J.L.~Evans, F.~Luo and K.A.~Olive, \emph{{Scenarios for Gluino
  Coannihilation}}, \href{https://doi.org/10.1007/JHEP02(2016)071}{\emph{JHEP}
  {\bfseries 02} (2016) 071}
  [\href{https://arxiv.org/abs/1510.03498}{{\ttfamily 1510.03498}}].

\bibitem{Liew:2016hqo}
S.P.~Liew and F.~Luo, \emph{{Effects of QCD bound states on dark matter relic
  abundance}}, \href{https://doi.org/10.1007/JHEP02(2017)091}{\emph{JHEP}
  {\bfseries 02} (2017) 091}
  [\href{https://arxiv.org/abs/1611.08133}{{\ttfamily 1611.08133}}].

\bibitem{Mitridate:2017izz}
A.~Mitridate, M.~Redi, J.~Smirnov and A.~Strumia, \emph{{Cosmological
  Implications of Dark Matter Bound States}},
  \href{https://doi.org/10.1088/1475-7516/2017/05/006}{\emph{JCAP} {\bfseries
  1705} (2017) 006} [\href{https://arxiv.org/abs/1702.01141}{{\ttfamily
  1702.01141}}].

\bibitem{Cirelli:2016rnw}
M.~Cirelli, P.~Panci, K.~Petraki, F.~Sala and M.~Taoso, \emph{{Dark Matter's
  secret liaisons: phenomenology of a dark U(1) sector with bound states}},
  \href{https://doi.org/10.1088/1475-7516/2017/05/036}{\emph{JCAP} {\bfseries
  05} (2017) 036} [\href{https://arxiv.org/abs/1612.07295}{{\ttfamily
  1612.07295}}].

\bibitem{Beneke:2016jpw}
M.~Beneke, A.~Bharucha, A.~Hryczuk, S.~Recksiegel and P.~Ruiz-Femenia,
  \emph{{The last refuge of mixed wino-Higgsino dark matter}},
  \href{https://doi.org/10.1007/JHEP01(2017)002}{\emph{JHEP} {\bfseries 01}
  (2017) 002} [\href{https://arxiv.org/abs/1611.00804}{{\ttfamily
  1611.00804}}].

\bibitem{Harz:2018csl}
J.~Harz and K.~Petraki, \emph{{Radiative bound-state formation in unbroken
  perturbative non-Abelian theories and implications for dark matter}},
  \href{https://doi.org/10.1007/JHEP07(2018)096}{\emph{JHEP} {\bfseries 07}
  (2018) 096} [\href{https://arxiv.org/abs/1805.01200}{{\ttfamily
  1805.01200}}].

\bibitem{Biondini:2018pwp}
S.~Biondini and M.~Laine, \emph{{Thermal dark matter co-annihilating with a
  strongly interacting scalar}}, {\emph{JHEP} {\bfseries 04} (2018) 072}
  [\href{https://arxiv.org/abs/1801.05821}{{\ttfamily 1801.05821}}].

\bibitem{Oncala:2019yvj}
R.~Oncala and K.~Petraki, \emph{{Dark matter bound state formation via emission
  of a charged scalar}},
  \href{https://doi.org/10.1007/JHEP02(2020)036}{\emph{JHEP} {\bfseries 02}
  (2020) 036} [\href{https://arxiv.org/abs/1911.02605}{{\ttfamily
  1911.02605}}].

\bibitem{Oncala:2021tkz}
R.~Oncala and K.~Petraki, \emph{{Bound states of WIMP dark matter in
  Higgs-portal models. Part I. Cross-sections and transition rates}},
  \href{https://doi.org/10.1007/JHEP06(2021)124}{\emph{JHEP} {\bfseries 06}
  (2021) 124} [\href{https://arxiv.org/abs/2101.08666}{{\ttfamily
  2101.08666}}].

\bibitem{Biondini:2021ycj}
S.~Biondini and V.~Shtabovenko, \emph{{Bound-state formation, dissociation and
  decays of darkonium with potential non-relativistic Yukawa theory for scalar
  and pseudoscalar mediators}},
  \href{https://doi.org/10.1007/JHEP03(2022)172}{\emph{JHEP} {\bfseries 03}
  (2022) 172} [\href{https://arxiv.org/abs/2112.10145}{{\ttfamily
  2112.10145}}].

\bibitem{Garny:2021qsr}
M.~Garny and J.~Heisig, \emph{{Bound-state effects on dark matter
  coannihilation: Pushing the boundaries of conversion-driven freeze-out}},
  \href{https://doi.org/10.1103/PhysRevD.105.055004}{\emph{Phys. Rev. D}
  {\bfseries 105} (2022) 055004}
  [\href{https://arxiv.org/abs/2112.01499}{{\ttfamily 2112.01499}}].

\bibitem{Binder:2019erp}
T.~Binder, K.~Mukaida and K.~Petraki, \emph{{Rapid bound-state formation of
  Dark Matter in the Early Universe}},
  \href{https://doi.org/10.1103/PhysRevLett.124.161102}{\emph{Phys. Rev. Lett.}
  {\bfseries 124} (2020) 161102}
  [\href{https://arxiv.org/abs/1910.11288}{{\ttfamily 1910.11288}}].

\bibitem{Binder:2021vfo}
T.~Binder, A.~Filimonova, K.~Petraki and G.~White, \emph{{Saha equilibrium for
  metastable bound states and dark matter freeze-out}},
  \href{https://doi.org/10.1016/j.physletb.2022.137323}{\emph{Phys. Lett. B}
  {\bfseries 833} (2022) 137323}
  [\href{https://arxiv.org/abs/2112.00042}{{\ttfamily 2112.00042}}].

\bibitem{Laha:2015yoa}
R.~Laha, \emph{{Directional detection of dark matter in universal bound
  states}}, \href{https://doi.org/10.1103/PhysRevD.92.083509}{\emph{Phys. Rev.
  D} {\bfseries 92} (2015) 083509}
  [\href{https://arxiv.org/abs/1505.02772}{{\ttfamily 1505.02772}}].

\bibitem{Asadi:2016ybp}
P.~Asadi, M.~Baumgart, P.J.~Fitzpatrick, E.~Krupczak and T.R.~Slatyer,
  \emph{{Capture and Decay of Electroweak WIMPonium}},
  \href{https://doi.org/10.1088/1475-7516/2017/02/005}{\emph{JCAP} {\bfseries
  1702} (2017) 005} [\href{https://arxiv.org/abs/1610.07617}{{\ttfamily
  1610.07617}}].

\bibitem{Coskuner:2018are}
A.~Coskuner, D.M.~Grabowska, S.~Knapen and K.M.~Zurek, \emph{{Direct Detection
  of Bound States of Asymmetric Dark Matter}},
  \href{https://doi.org/10.1103/PhysRevD.100.035025}{\emph{Phys. Rev. D}
  {\bfseries 100} (2019) 035025}
  [\href{https://arxiv.org/abs/1812.07573}{{\ttfamily 1812.07573}}].

\bibitem{Chu:2018faw}
X.~Chu, C.~Garcia-Cely and H.~Murayama, \emph{{Finite-size dark matter and its
  effect on small-scale structure}},
  \href{https://doi.org/10.1103/PhysRevLett.124.041101}{\emph{Phys. Rev. Lett.}
  {\bfseries 124} (2020) 041101}
  [\href{https://arxiv.org/abs/1901.00075}{{\ttfamily 1901.00075}}].

\bibitem{Bottaro:2021srh}
S.~Bottaro, A.~Strumia and N.~Vignaroli, \emph{{Minimal Dark Matter bound
  states at future colliders}},
  \href{https://doi.org/10.1007/JHEP06(2021)143}{\emph{JHEP} {\bfseries 06}
  (2021) 143} [\href{https://arxiv.org/abs/2103.12766}{{\ttfamily
  2103.12766}}].

\bibitem{An:2016gad}
H.~An, M.B.~Wise and Y.~Zhang, \emph{{Effects of Bound States on Dark Matter
  Annihilation}}, \href{https://doi.org/10.1103/PhysRevD.93.115020}{\emph{Phys.
  Rev. D} {\bfseries 93} (2016) 115020}
  [\href{https://arxiv.org/abs/1604.01776}{{\ttfamily 1604.01776}}].

\bibitem{Pearce:2015zca}
L.~Pearce, K.~Petraki and A.~Kusenko, \emph{{Signals from dark atom formation
  in halos}}, \href{https://doi.org/10.1103/PhysRevD.91.083532}{\emph{Phys.
  Rev. D} {\bfseries 91} (2015) 083532}
  [\href{https://arxiv.org/abs/1502.01755}{{\ttfamily 1502.01755}}].

\bibitem{Baldes:2020hwx}
I.~Baldes, F.~Calore, K.~Petraki, V.~Poireau and N.L.~Rodd, \emph{{Indirect
  searches for dark matter bound state formation and level transitions}},
  \href{https://doi.org/10.21468/SciPostPhys.9.5.068}{\emph{SciPost Phys.}
  {\bfseries 9} (2020) 068} [\href{https://arxiv.org/abs/2007.13787}{{\ttfamily
  2007.13787}}].

\bibitem{Kahlhoefer:2017umn}
F.~Kahlhoefer, K.~Schmidt-Hoberg and S.~Wild, \emph{{Dark matter
  self-interactions from a general spin-0 mediator}},
  \href{https://doi.org/10.1088/1475-7516/2017/08/003}{\emph{JCAP} {\bfseries
  08} (2017) 003} [\href{https://arxiv.org/abs/1704.02149}{{\ttfamily
  1704.02149}}].

\bibitem{Kaplinghat:2013yxa}
M.~Kaplinghat, S.~Tulin and H.-B.~Yu, \emph{{Direct Detection Portals for
  Self-interacting Dark Matter}},
  \href{https://doi.org/10.1103/PhysRevD.89.035009}{\emph{Phys. Rev. D}
  {\bfseries 89} (2014) 035009}
  [\href{https://arxiv.org/abs/1310.7945}{{\ttfamily 1310.7945}}].

\bibitem{DeSimone:2016fbz}
A.~De~Simone and T.~Jacques, \emph{{Simplified models vs. effective field
  theory approaches in dark matter searches}},
  \href{https://doi.org/10.1140/epjc/s10052-016-4208-4}{\emph{Eur. Phys. J. C}
  {\bfseries 76} (2016) 367}
  [\href{https://arxiv.org/abs/1603.08002}{{\ttfamily 1603.08002}}].

\bibitem{Kahlhoefer:2015bea}
F.~Kahlhoefer, K.~Schmidt-Hoberg, T.~Schwetz and S.~Vogl, \emph{{Implications
  of unitarity and gauge invariance for simplified dark matter models}},
  \href{https://doi.org/10.1007/JHEP02(2016)016}{\emph{JHEP} {\bfseries 02}
  (2016) 016} [\href{https://arxiv.org/abs/1510.02110}{{\ttfamily
  1510.02110}}].

\bibitem{Duerr:2016tmh}
M.~Duerr, F.~Kahlhoefer, K.~Schmidt-Hoberg, T.~Schwetz and S.~Vogl, \emph{{How
  to save the WIMP: global analysis of a dark matter model with two s-channel
  mediators}}, \href{https://doi.org/10.1007/JHEP09(2016)042}{\emph{JHEP}
  {\bfseries 09} (2016) 042}
  [\href{https://arxiv.org/abs/1606.07609}{{\ttfamily 1606.07609}}].

\bibitem{Wise:2014jva}
M.B.~Wise and Y.~Zhang, \emph{{Stable Bound States of Asymmetric Dark Matter}},
  \href{https://doi.org/10.1103/PhysRevD.90.055030}{\emph{Phys. Rev. D}
  {\bfseries 90} (2014) 055030}
  [\href{https://arxiv.org/abs/1407.4121}{{\ttfamily 1407.4121}}].

\bibitem{Oncala:2018bvl}
R.~Oncala and K.~Petraki, \emph{{Dark matter bound states via emission of
  scalar mediators}},
  \href{https://doi.org/10.1007/JHEP01(2019)070}{\emph{JHEP} {\bfseries 01}
  (2019) 070} [\href{https://arxiv.org/abs/1808.04854}{{\ttfamily
  1808.04854}}].

\bibitem{Kainulainen:2015sva}
K.~Kainulainen, K.~Tuominen and V.~Vaskonen, \emph{{Self-interacting dark
  matter and cosmology of a light scalar mediator}},
  \href{https://doi.org/10.1103/PhysRevD.93.015016}{\emph{Phys. Rev. D}
  {\bfseries 93} (2016) 015016}
  [\href{https://arxiv.org/abs/1507.04931}{{\ttfamily 1507.04931}}].

\bibitem{Shepherd:2009sa}
W.~Shepherd, T.M.P.~Tait and G.~Zaharijas, \emph{{Bound states of weakly
  interacting dark matter}},
  \href{https://doi.org/10.1103/PhysRevD.79.055022}{\emph{Phys. Rev. D}
  {\bfseries 79} (2009) 055022}
  [\href{https://arxiv.org/abs/0901.2125}{{\ttfamily 0901.2125}}].

\bibitem{Luke:1997ys}
M.E.~Luke and M.J.~Savage, \emph{{Power counting in dimensionally regularized
  NRQCD}}, \href{https://doi.org/10.1103/PhysRevD.57.413}{\emph{Phys. Rev. D}
  {\bfseries 57} (1998) 413}
  [\href{https://arxiv.org/abs/hep-ph/9707313}{{\ttfamily hep-ph/9707313}}].

\bibitem{Biondini:2021ccr}
S.~Biondini and V.~Shtabovenko, \emph{{Non-relativistic and potential
  non-relativistic effective field theories for scalar mediators}},
  \href{https://doi.org/10.1007/JHEP08(2021)114}{\emph{JHEP} {\bfseries 08}
  (2021) 114} [\href{https://arxiv.org/abs/2106.06472}{{\ttfamily
  2106.06472}}].

\bibitem{Luke:1996hj}
M.E.~Luke and A.V.~Manohar, \emph{{Bound states and power counting in effective
  field theories}}, \href{https://doi.org/10.1103/PhysRevD.55.4129}{\emph{Phys.
  Rev. D} {\bfseries 55} (1997) 4129}
  [\href{https://arxiv.org/abs/hep-ph/9610534}{{\ttfamily hep-ph/9610534}}].

\bibitem{Biondini:2023zcz}
S.~Biondini, N.~Brambilla, G.~Qerimi and A.~Vairo, \emph{{Effective field
  theories for dark matter pairs in the early universe: cross sections and
  widths}},  \href{https://arxiv.org/abs/2304.00113}{{\ttfamily 2304.00113}}.

\bibitem{Bodwin:1994jh}
G.T.~Bodwin, E.~Braaten and G.P.~Lepage, \emph{{Rigorous QCD analysis of
  inclusive annihilation and production of heavy quarkonium}},
  \href{https://doi.org/10.1103/PhysRevD.55.5853,
  10.1103/PhysRevD.51.1125}{\emph{Phys. Rev.} {\bfseries D51} (1995) 1125}
  [\href{https://arxiv.org/abs/hep-ph/9407339}{{\ttfamily hep-ph/9407339}}].

\bibitem{Caswell:1985ui}
W.E.~Caswell and G.P.~Lepage, \emph{{Effective Lagrangians for Bound State
  Problems in QED, QCD, and Other Field Theories}},
  \href{https://doi.org/10.1016/0370-2693(86)91297-9}{\emph{Phys. Lett.}
  {\bfseries 167B} (1986) 437}.

\bibitem{Pineda:1997bj}
A.~Pineda and J.~Soto, \emph{{Effective field theory for ultrasoft momenta in
  NRQCD and NRQED}},
  \href{https://doi.org/10.1016/S0920-5632(97)01102-X}{\emph{Nucl. Phys. Proc.
  Suppl.} {\bfseries 64} (1998) 428}
  [\href{https://arxiv.org/abs/hep-ph/9707481}{{\ttfamily hep-ph/9707481}}].

\bibitem{Brambilla:2002nu}
N.~Brambilla, D.~Eiras, A.~Pineda, J.~Soto and A.~Vairo, \emph{{Inclusive
  decays of heavy quarkonium to light particles}},
  \href{https://doi.org/10.1103/PhysRevD.67.034018}{\emph{Phys. Rev. D}
  {\bfseries 67} (2003) 034018}
  [\href{https://arxiv.org/abs/hep-ph/0208019}{{\ttfamily hep-ph/0208019}}].

\bibitem{Brambilla:2004jw}
N.~Brambilla, A.~Pineda, J.~Soto and A.~Vairo, \emph{{Effective Field Theories
  for Heavy Quarkonium}},
  \href{https://doi.org/10.1103/RevModPhys.77.1423}{\emph{Rev. Mod. Phys.}
  {\bfseries 77} (2005) 1423}
  [\href{https://arxiv.org/abs/hep-ph/0410047}{{\ttfamily hep-ph/0410047}}].

\bibitem{Sommerfeld}
A.~Sommerfeld, \emph{{\"Uber die Beugung und Bremsung der Elektronen}},
  {\emph{Ann. Phys.(1931)} {\bfseries 403} (1931) }.

\bibitem{Iengo:2009ni}
R.~Iengo, \emph{{Sommerfeld enhancement: General results from field theory
  diagrams}}, \href{https://doi.org/10.1088/1126-6708/2009/05/024}{\emph{JHEP}
  {\bfseries 05} (2009) 024} [\href{https://arxiv.org/abs/0902.0688}{{\ttfamily
  0902.0688}}].

\bibitem{Cassel:2009wt}
S.~Cassel, \emph{{Sommerfeld factor for arbitrary partial wave processes}},
  \href{https://doi.org/10.1088/0954-3899/37/10/105009}{\emph{J. Phys. G}
  {\bfseries 37} (2010) 105009}
  [\href{https://arxiv.org/abs/0903.5307}{{\ttfamily 0903.5307}}].

\bibitem{Biondini:2017ufr}
S.~Biondini and M.~Laine, \emph{{Re-derived overclosure bound for the inert
  doublet model}}, \href{https://doi.org/10.1007/JHEP08(2017)047}{\emph{JHEP}
  {\bfseries 08} (2017) 047}
  [\href{https://arxiv.org/abs/1706.01894}{{\ttfamily 1706.01894}}].

\bibitem{Gondolo:1990dk}
P.~Gondolo and G.~Gelmini, \emph{{Cosmic abundances of stable particles:
  Improved analysis}},
  \href{https://doi.org/10.1016/0550-3213(91)90438-4}{\emph{Nucl. Phys.}
  {\bfseries B360} (1991) 145}.

\bibitem{Brambilla:2008cx}
N.~Brambilla, J.~Ghiglieri, A.~Vairo and P.~Petreczky, \emph{{Static
  quark-antiquark pairs at finite temperature}},
  \href{https://doi.org/10.1103/PhysRevD.78.014017}{\emph{Phys. Rev.}
  {\bfseries D78} (2008) 014017}
  [\href{https://arxiv.org/abs/0804.0993}{{\ttfamily 0804.0993}}].

\bibitem{Beneke:1997zp}
M.~Beneke and V.A.~Smirnov, \emph{{Asymptotic expansion of Feynman integrals
  near threshold}},
  \href{https://doi.org/10.1016/S0550-3213(98)00138-2}{\emph{Nucl. Phys. B}
  {\bfseries 522} (1998) 321}
  [\href{https://arxiv.org/abs/hep-ph/9711391}{{\ttfamily hep-ph/9711391}}].

\bibitem{Ellis:2015vaa}
J.~Ellis, F.~Luo and K.A.~Olive, \emph{{Gluino Coannihilation Revisited}},
  \href{https://doi.org/10.1007/JHEP09(2015)127}{\emph{JHEP} {\bfseries 09}
  (2015) 127} [\href{https://arxiv.org/abs/1503.07142}{{\ttfamily
  1503.07142}}].

\bibitem{Binder:2023ckj}
T.~Binder, M.~Garny, J.~Heisig, S.~Lederer and K.~Urban, \emph{{Excited bound
  states and their role in dark matter production}},
  \href{https://arxiv.org/abs/2308.01336}{{\ttfamily 2308.01336}}.

\bibitem{Ferrer:2013cla}
F.~Ferrer and D.R.~Hunter, \emph{{The impact of the phase-space density on the
  indirect detection of dark matter}},
  \href{https://doi.org/10.1088/1475-7516/2013/09/005}{\emph{JCAP} {\bfseries
  09} (2013) 005} [\href{https://arxiv.org/abs/1306.6586}{{\ttfamily
  1306.6586}}].

\bibitem{Binneybook}
J.~{Binney} and S.~{Tremaine}, \emph{{Galactic Dynamics: Second Edition}}
  (2008).

\bibitem{Boddy:2017vpe}
K.K.~Boddy, J.~Kumar, L.E.~Strigari and M.-Y.~Wang, \emph{{Sommerfeld-Enhanced
  $J$-Factors For Dwarf Spheroidal Galaxies}},
  \href{https://doi.org/10.1103/PhysRevD.95.123008}{\emph{Phys. Rev. D}
  {\bfseries 95} (2017) 123008}
  [\href{https://arxiv.org/abs/1702.00408}{{\ttfamily 1702.00408}}].

\bibitem{Lacroix:2018qqh}
T.~Lacroix, M.~Stref and J.~Lavalle, \emph{{Anatomy of Eddington-like inversion
  methods in the context of dark matter searches}},
  \href{https://doi.org/10.1088/1475-7516/2018/09/040}{\emph{JCAP} {\bfseries
  09} (2018) 040} [\href{https://arxiv.org/abs/1805.02403}{{\ttfamily
  1805.02403}}].

\bibitem{Board:2021bwj}
E.~Board, N.~Bozorgnia, L.E.~Strigari, R.J.J.~Grand, A.~Fattahi, C.S.~Frenk
  et~al., \emph{{Velocity-dependent J-factors for annihilation radiation from
  cosmological simulations}},
  \href{https://doi.org/10.1088/1475-7516/2021/04/070}{\emph{JCAP} {\bfseries
  04} (2021) 070} [\href{https://arxiv.org/abs/2101.06284}{{\ttfamily
  2101.06284}}].

\bibitem{McKeown:2021sob}
D.~McKeown, J.S.~Bullock, F.J.~Mercado, Z.~Hafen, M.~Boylan-Kolchin, A.~Wetzel
  et~al., \emph{{Amplified J-factors in the Galactic Centre for
  velocity-dependent dark matter annihilation in FIRE simulations}},
  \href{https://doi.org/10.1093/mnras/stac966}{\emph{Mon. Not. Roy. Astron.
  Soc.} {\bfseries 513} (2022) 55}
  [\href{https://arxiv.org/abs/2111.03076}{{\ttfamily 2111.03076}}].

\bibitem{Navarro:1995iw}
J.F.~Navarro, C.S.~Frenk and S.D.M.~White, \emph{{The Structure of cold dark
  matter halos}}, \href{https://doi.org/10.1086/177173}{\emph{Astrophys. J.}
  {\bfseries 462} (1996) 563}
  [\href{https://arxiv.org/abs/astro-ph/9508025}{{\ttfamily
  astro-ph/9508025}}].

\bibitem{Navarro:1996gj}
J.F.~Navarro, C.S.~Frenk and S.D.M.~White, \emph{{A Universal density profile
  from hierarchical clustering}},
  \href{https://doi.org/10.1086/304888}{\emph{Astrophys. J.} {\bfseries 490}
  (1997) 493} [\href{https://arxiv.org/abs/astro-ph/9611107}{{\ttfamily
  astro-ph/9611107}}].

\bibitem{Moore:1999nt}
B.~Moore, S.~Ghigna, F.~Governato, G.~Lake, T.R.~Quinn, J.~Stadel et~al.,
  \emph{{Dark matter substructure within galactic halos}},
  \href{https://doi.org/10.1086/312287}{\emph{Astrophys. J. Lett.} {\bfseries
  524} (1999) L19} [\href{https://arxiv.org/abs/astro-ph/9907411}{{\ttfamily
  astro-ph/9907411}}].

\bibitem{CTA:2020qlo}
{\scshape CTA} collaboration, \emph{{Sensitivity of the Cherenkov Telescope
  Array to a dark matter signal from the Galactic centre}},
  \href{https://doi.org/10.1088/1475-7516/2021/01/057}{\emph{JCAP} {\bfseries
  01} (2021) 057} [\href{https://arxiv.org/abs/2007.16129}{{\ttfamily
  2007.16129}}].

\bibitem{Djouadi:2005gi}
A.~Djouadi, \emph{{The Anatomy of electro-weak symmetry breaking. I: The Higgs
  boson in the standard model}},
  \href{https://doi.org/10.1016/j.physrep.2007.10.004}{\emph{Phys. Rept.}
  {\bfseries 457} (2008) 1}
  [\href{https://arxiv.org/abs/hep-ph/0503172}{{\ttfamily hep-ph/0503172}}].

\bibitem{Chetyrkin:2000yt}
K.G.~Chetyrkin, J.H.~Kuhn and M.~Steinhauser, \emph{{RunDec: A Mathematica
  package for running and decoupling of the strong coupling and quark masses}},
  \href{https://doi.org/10.1016/S0010-4655(00)00155-7}{\emph{Comput. Phys.
  Commun.} {\bfseries 133} (2000) 43}
  [\href{https://arxiv.org/abs/hep-ph/0004189}{{\ttfamily hep-ph/0004189}}].

\bibitem{Winkler:2018qyg}
M.W.~Winkler, \emph{{Decay and detection of a light scalar boson mixing with
  the Higgs boson}},
  \href{https://doi.org/10.1103/PhysRevD.99.015018}{\emph{Phys. Rev. D}
  {\bfseries 99} (2019) 015018}
  [\href{https://arxiv.org/abs/1809.01876}{{\ttfamily 1809.01876}}].

\bibitem{Bierlich:2022pfr}
C.~Bierlich et~al., \emph{{A comprehensive guide to the physics and usage of
  PYTHIA 8.3}},  \href{https://arxiv.org/abs/2203.11601}{{\ttfamily
  2203.11601}}.

\bibitem{Bellm:2015jjp}
J.~Bellm et~al., \emph{{Herwig 7.0/Herwig++ 3.0 release note}},
  \href{https://doi.org/10.1140/epjc/s10052-016-4018-8}{\emph{Eur. Phys. J. C}
  {\bfseries 76} (2016) 196}
  [\href{https://arxiv.org/abs/1512.01178}{{\ttfamily 1512.01178}}].

\bibitem{Belanger:2001fz}
G.~Belanger, F.~Boudjema, A.~Pukhov and A.~Semenov, \emph{{MicrOMEGAs: A
  Program for calculating the relic density in the MSSM}},
  \href{https://doi.org/10.1016/S0010-4655(02)00596-9}{\emph{Comput. Phys.
  Commun.} {\bfseries 149} (2002) 103}
  [\href{https://arxiv.org/abs/hep-ph/0112278}{{\ttfamily hep-ph/0112278}}].

\bibitem{Cirelli:2010xx}
M.~Cirelli, G.~Corcella, A.~Hektor, G.~Hutsi, M.~Kadastik, P.~Panci et~al.,
  \emph{{PPPC 4 DM ID: A Poor Particle Physicist Cookbook for Dark Matter
  Indirect Detection}},
  \href{https://doi.org/10.1088/1475-7516/2012/10/E01}{\emph{JCAP} {\bfseries
  03} (2011) 051} [\href{https://arxiv.org/abs/1012.4515}{{\ttfamily
  1012.4515}}].

\bibitem{Bringmann:2018lay}
T.~Bringmann, J.~Edsj\"o, P.~Gondolo, P.~Ullio and L.~Bergstr\"om,
  \emph{{DarkSUSY 6 : An Advanced Tool to Compute Dark Matter Properties
  Numerically}},
  \href{https://doi.org/10.1088/1475-7516/2018/07/033}{\emph{JCAP} {\bfseries
  07} (2018) 033} [\href{https://arxiv.org/abs/1802.03399}{{\ttfamily
  1802.03399}}].

\bibitem{Plehn:2019jeo}
T.~Plehn, P.~Reimitz and P.~Richardson, \emph{{Hadronic Footprint of GeV-Mass
  Dark Matter}},
  \href{https://doi.org/10.21468/SciPostPhys.8.6.092}{\emph{SciPost Phys.}
  {\bfseries 8} (2020) 092} [\href{https://arxiv.org/abs/1911.11147}{{\ttfamily
  1911.11147}}].

\bibitem{Coogan:2022cdd}
A.~Coogan, L.~Morrison, T.~Plehn, S.~Profumo and P.~Reimitz, \emph{{Hazma meets
  HERWIG4DM: precision gamma-ray, neutrino, and positron spectra for light dark
  matter}}, \href{https://doi.org/10.1088/1475-7516/2022/11/033}{\emph{JCAP}
  {\bfseries 11} (2022) 033}
  [\href{https://arxiv.org/abs/2207.07634}{{\ttfamily 2207.07634}}].

\bibitem{Mardon:2009rc}
J.~Mardon, Y.~Nomura, D.~Stolarski and J.~Thaler, \emph{{Dark Matter Signals
  from Cascade Annihilations}},
  \href{https://doi.org/10.1088/1475-7516/2009/05/016}{\emph{JCAP} {\bfseries
  05} (2009) 016} [\href{https://arxiv.org/abs/0901.2926}{{\ttfamily
  0901.2926}}].

\bibitem{Elor:2015bho}
G.~Elor, N.L.~Rodd, T.R.~Slatyer and W.~Xue, \emph{{Model-Independent Indirect
  Detection Constraints on Hidden Sector Dark Matter}},
  \href{https://doi.org/10.1088/1475-7516/2016/06/024}{\emph{JCAP} {\bfseries
  06} (2016) 024} [\href{https://arxiv.org/abs/1511.08787}{{\ttfamily
  1511.08787}}].

\bibitem{Elor:2015tva}
G.~Elor, N.L.~Rodd and T.R.~Slatyer, \emph{{Multistep cascade annihilations of
  dark matter and the Galactic Center excess}},
  \href{https://doi.org/10.1103/PhysRevD.91.103531}{\emph{Phys. Rev. D}
  {\bfseries 91} (2015) 103531}
  [\href{https://arxiv.org/abs/1503.01773}{{\ttfamily 1503.01773}}].

\bibitem{Bringmann:2011ye}
T.~Bringmann, F.~Calore, G.~Vertongen and C.~Weniger, \emph{{On the Relevance
  of Sharp Gamma-Ray Features for Indirect Dark Matter Searches}},
  \href{https://doi.org/10.1103/PhysRevD.84.103525}{\emph{Phys. Rev. D}
  {\bfseries 84} (2011) 103525}
  [\href{https://arxiv.org/abs/1106.1874}{{\ttfamily 1106.1874}}].

\bibitem{Bringmann:2012vr}
T.~Bringmann, X.~Huang, A.~Ibarra, S.~Vogl and C.~Weniger, \emph{{Fermi LAT
  Search for Internal Bremsstrahlung Signatures from Dark Matter
  Annihilation}},
  \href{https://doi.org/10.1088/1475-7516/2012/07/054}{\emph{JCAP} {\bfseries
  07} (2012) 054} [\href{https://arxiv.org/abs/1203.1312}{{\ttfamily
  1203.1312}}].

\bibitem{Fermi_logL}
{Fermi-LAT Collaboration}, ``Supplementary material for {Fermi-LAT} dwarf
  spheroidal paper.'' \url{http://www-glast.stanford.edu/pub_data/1048/}, 2015.

\bibitem{Padmanabhan:2005es}
N.~Padmanabhan and D.P.~Finkbeiner, \emph{{Detecting dark matter annihilation
  with CMB polarization: Signatures and experimental prospects}},
  \href{https://doi.org/10.1103/PhysRevD.72.023508}{\emph{Phys. Rev. D}
  {\bfseries 72} (2005) 023508}
  [\href{https://arxiv.org/abs/astro-ph/0503486}{{\ttfamily
  astro-ph/0503486}}].

\bibitem{Galli:2011rz}
S.~Galli, F.~Iocco, G.~Bertone and A.~Melchiorri, \emph{{Updated CMB
  constraints on Dark Matter annihilation cross-sections}},
  \href{https://doi.org/10.1103/PhysRevD.84.027302}{\emph{Phys. Rev. D}
  {\bfseries 84} (2011) 027302}
  [\href{https://arxiv.org/abs/1106.1528}{{\ttfamily 1106.1528}}].

\bibitem{Slatyer:2015jla}
T.R.~Slatyer, \emph{{Indirect dark matter signatures in the cosmic dark ages.
  I. Generalizing the bound on s-wave dark matter annihilation from Planck
  results}}, \href{https://doi.org/10.1103/PhysRevD.93.023527}{\emph{Phys. Rev.
  D} {\bfseries 93} (2016) 023527}
  [\href{https://arxiv.org/abs/1506.03811}{{\ttfamily 1506.03811}}].

\bibitem{bringmann_torsten_2020_4057987}
T.~Bringmann, C.~Eckner, A.~Sokolenko, L.~Yang and G.~Zaharijas, ``{Likelihoods
  for the CTA sensitivity to a dark matter signal from the Galactic centre}.''
  \url{https://doi.org/10.5281/zenodo.4057987}, 2020.

\bibitem{Rinchiuso:2020skh}
L.~Rinchiuso, O.~Macias, E.~Moulin, N.L.~Rodd and T.R.~Slatyer,
  \emph{{Prospects for detecting heavy WIMP dark matter with the Cherenkov
  Telescope Array: The Wino and Higgsino}},
  \href{https://doi.org/10.1103/PhysRevD.103.023011}{\emph{Phys. Rev. D}
  {\bfseries 103} (2021) 023011}
  [\href{https://arxiv.org/abs/2008.00692}{{\ttfamily 2008.00692}}].

\bibitem{Anand:2013yka}
N.~Anand, A.L.~Fitzpatrick and W.C.~Haxton, \emph{{Weakly interacting massive
  particle-nucleus elastic scattering response}},
  \href{https://doi.org/10.1103/PhysRevC.89.065501}{\emph{Phys. Rev. C}
  {\bfseries 89} (2014) 065501}
  [\href{https://arxiv.org/abs/1308.6288}{{\ttfamily 1308.6288}}].

\bibitem{Lebedev:2021xey}
O.~Lebedev, \emph{{The Higgs portal to cosmology}},
  \href{https://doi.org/10.1016/j.ppnp.2021.103881}{\emph{Prog. Part. Nucl.
  Phys.} {\bfseries 120} (2021) 103881}
  [\href{https://arxiv.org/abs/2104.03342}{{\ttfamily 2104.03342}}].

\bibitem{Evans:2017kti}
J.A.~Evans, S.~Gori and J.~Shelton, \emph{{Looking for the WIMP Next Door}},
  \href{https://doi.org/10.1007/JHEP02(2018)100}{\emph{JHEP} {\bfseries 02}
  (2018) 100} [\href{https://arxiv.org/abs/1712.03974}{{\ttfamily
  1712.03974}}].

\bibitem{Bernreuther:1990jx}
W.~Bernreuther and M.~Suzuki, \emph{{The electric dipole moment of the
  electron}}, \href{https://doi.org/10.1103/RevModPhys.63.313}{\emph{Rev. Mod.
  Phys.} {\bfseries 63} (1991) 313}.

\bibitem{Chupp:2017rkp}
T.~Chupp, P.~Fierlinger, M.~Ramsey-Musolf and J.~Singh, \emph{{Electric dipole
  moments of atoms, molecules, nuclei, and particles}},
  \href{https://doi.org/10.1103/RevModPhys.91.015001}{\emph{Rev. Mod. Phys.}
  {\bfseries 91} (2019) 015001}
  [\href{https://arxiv.org/abs/1710.02504}{{\ttfamily 1710.02504}}].

\bibitem{Alarcon:2022ero}
R.~Alarcon et~al., \emph{{Electric dipole moments and the search for new
  physics}},  in \emph{{Snowmass 2021}}, 3, 2022
  [\href{https://arxiv.org/abs/2203.08103}{{\ttfamily 2203.08103}}].

\bibitem{ACME:2018yjb}
{\scshape ACME} collaboration, \emph{{Improved limit on the electric dipole
  moment of the electron}},
  \href{https://doi.org/10.1038/s41586-018-0599-8}{\emph{Nature} {\bfseries
  562} (2018) 355}.

\end{thebibliography}\endgroup

\end{document}